\documentclass{IEEEtran}

\usepackage{cite}
\usepackage{amsmath,amssymb,amsfonts}
\usepackage{algorithmic}
\usepackage{graphicx,color}
\usepackage{textcomp}
\def\BibTeX{{\rm B\kern-.05em{\sc i\kern-.025em b}\kern-.08em
		T\kern-.1667em\lower.7ex\hbox{E}\kern-.125emX}}
\AtBeginDocument{\definecolor{ojcolor}{cmyk}{0.93,0.59,0.15,0.02}}
    
\usepackage{url,hyperref,lineno,microtype,subcaption,siunitx,booktabs}
\usepackage[table]{xcolor}
\usepackage{tabularray,multirow,ulem}
\usepackage{tcolorbox}
\usepackage{hhline}
\usepackage{forest,inputenc}
\usepackage{tikz}
\usepackage{longtable}
\usetikzlibrary{trees}
\usepackage{float}
\usepackage{multicol}
\usepackage{nomencl,soul}
\definecolor{NavyBlue}{RGB}{0, 41, 142}     % Dark blue (Used in both)
\definecolor{ReviewerColor}{RGB}{80, 80, 80} % Dark Gray for comments
\definecolor{SteelBlue}{RGB}{10, 80, 180}
\definecolor{SlateGray}{RGB}{112, 128, 144} % Muted gray
\definecolor{DarkGreen}{RGB}{0, 102, 51}    % Dark green
\definecolor{Burgundy}{RGB}{128, 0, 32}      % Deep red
\definecolor{LightGray}{RGB}{220, 220, 220} % Light gray
\makeatletter
\def\ps@IEEEtitlepagestyle{%
	\def\@oddfoot{\mycopyrightnotice}%
}
\def\mycopyrightnotice{%
	\begin{minipage}{\textwidth}
		\centering \scriptsize
		\vspace{10pt}
		\copyright 2026 IEEE. Personal use of this material is permitted. Permission from IEEE must be obtained for all other uses, in any current or future media, including reprinting/republishing this material for advertising or promotional purposes, creating new collective works, for resale or redistribution to servers or lists, or reuse of any copyrighted component of this work in other works.
	\end{minipage}
}
\makeatother

\begin{document}
	
	%\twocolumn
	%\newpage
	%\setcounter{figure}{0}
	%\setcounter{table}{0}

	\title{Corridor Design and Separation Definition in Advanced Air Mobility: Systematic Literature Review}
	
	\author{\IEEEauthorblockN{Evgenii~Vinogradov$^{1}$,
		Debashisha~Mishra$^{1}$,
		Mariam~Ali~Askar~Alobeidli$^{2}$,
		Jamal~Khaled~Al~Ali$^{2}$,
		Ahmed~Saleh~Alshehhi$^{2}$,
		Jennifer~Simonjan$^{1}$,
		and~Enrico~Natalizio$^{1}$}
	\IEEEauthorblockA{$^1$\textit{Technology Innovation Institute (TII), United Arab Emirates. Emails: firstname.lastname@tii.ae}
	\\ $^2$\textit{General Civil Aviation Authority (GCAA), United Arab Emirates. Emails: \{Maalobeidli,jalali,ashehhi\}@gcaa.gov.ae}}
}
	
%	\markboth{Corridor Design and Separation Definition in Advanced Air Mobility: Systematic Literature Review}{Vinogradov \textit{et al.}}
	
		\maketitle
	
	\begin{abstract}
		Advanced Air Mobility (AAM) uses electric vertical take-off and landing (eVTOL) vehicles to address urban congestion and emissions. However, corridor design, operation management, and separation standards remain underexamined for safe high-density operations. This paper applies the Preferred Reporting Items for Systematic Reviews and Meta-Analyses (PRISMA) guidelines to systematically review relevant literature from IEEE Xplore and Web of Science, focusing on publications from 2010 to 2024. A Context, Intervention, Mechanism, and Outcome (CIMO) framework guided the development of research questions. After screening 2,039 journal and conference papers, 62 articles met the inclusion criteria. The findings reveal a lack of integrated corridor design approaches, limited operational strategies, and reliance on standards originally designed for conventional aviation. A unified corridor design and separation definition frameworks and taxonomies are proposed to address these shortcomings, informing future investigations and operational frameworks for safe, efficient eVTOL operation deployment in urban settings.
		\vspace{1em}
	\end{abstract}

	\section{Introduction}
	\IEEEPARstart{U}rbanization is increasing pressure on existing transportation infrastructure. As cities continue to expand, traditional transportation systems struggle to keep up with increasing demand. Advanced Air Mobility (AAM) is the next-generation passenger and cargo {air} transportation (e.g., flying taxi) paradigm incorporating such advancements as remotely piloted, autonomous, or vertical take-off and landing (VTOL) aircraft. This includes those powered by electric (i.e., eVTOLs) or hybrid-electric propulsion. {The AAM ``corridor" is defined as an air route within the airspace linking two vertiports and designed for safe and efficient movement of eVTOLs. The corridor can be geometrically represented as three dimensional airspace regions with defined horizontal and vertical limits within which the movement of eVTOLs are permitted.} Among the most compelling advantages, AAM can reduce congestion, lower emissions, and enhance sustainability in urban environments, offering a cleaner and more effective alternative to traditional surface-based transport. The government and commercial sectors are investing in AAM to enhance the overall mobility of urban environments that are becoming increasingly congested and to shift the development towards smart, sustainable cities.
	
	The recent research focus on AAM resulted in several survey papers exploring various dimensions of its implementation and integration into existing transportation systems. 
	Based on an extensive analysis of papers, we have categorized these studies into following themes: Air Mobility Overview and Evolution, Market Demand and Economic Analysis, Airspace and Corridor Organization, Ground Infrastructure and Vertiports, and Safety\&Security Concerns.
	
	Several contributions provide comprehensive overviews of AAM. Straubinger \textit{et al.} \cite{Straubinger2020scene} offer a broad overview of Urban Air Mobility (UAM - a subset of AAM), discussing vehicle-related aspects, operational concepts, market structures, and public acceptance. Similarly, Cohen \textit{et al.} \cite{Cohen2021market} describe the evolution of AAM, emphasizing phased development and potential barriers such as regulatory environments and public acceptance. The integration of UAM into multimodal transportation systems is discussed through research initiatives like the German Aerospace Center’s HorizonUAM project \cite{Pak2024dlr}.
	
	Understanding market demand and economic viability is essential for assessing the potential success of AAM services as evidenced by demand analyses \cite{Goyal2021demand,Long2023demand}, meta-analyses comparing AAM with Electric and Autonomous Vehicles~\cite{Garrow2021ev}, and comprehensive reviews of AAM ecosystems \cite{Ahmed2024challenges}. Moreover, the authors of \cite{Wiedemann2024regulations} provide an evaluation of AAM policies from four different countries highlighting that, while AAM policy development happening globally is in its infancy, the United Arab Emirates has the most developed regulatory framework.
	
	Airspace design studies examine the structural aspects necessary for efficient and safe AAM operations. A multi-layered air corridor structure and traffic flow rules (intersections, engagement rules) are explored in \cite{Muna2021corridor}. Bauranov and Rakas~\cite{Bauranov2021airspace} analyze urban airspace design concepts. Additionally, Nithya \textit{et al.} \cite{Nithya2024wind} address operational complexities in urban airspace related to wind flow influencing the corridor design.
	
	Ground infrastructure, particularly vertiports, is critical for the successful implementation of AAM systems. Several works review literature on UAM ground infrastructure, focusing on the guidelines for vertiport design \cite{Schweiger2022vertiport,Brunelli2023vertiport}. Marvraj \textit{et al.} \cite{Marvraj2022_SLR_ground} offer an overview of AAM ground-based infrastructure, highlighting take-off and landing sites, maintenance facilities, energy supply, and regulatory frameworks, highlighting the need for a holistic perspective on infrastructure challenges.
	
	The systematic literature review in~\cite{Ferrao2022_SLR_security} focuses on air taxi safety and security, and reviews advances in techniques and architectures to ensure safe and autonomous operations. Bauranov \cite{Bauranov2021airspace} also touches upon safety factors within airspace design, though not as comprehensively as in \cite{Ferrao2022_SLR_security}.
	
	\textbf{Limitations of the State of the Art:} Despite the extensive coverage provided by the surveys, only two systematic literature reviews \cite{Marvraj2022_SLR_ground,Ferrao2022_SLR_security} have been conducted using the recognized Preferred Reporting Items for Systematic Reviews and Meta-Analyses (PRISMA) framework. The majority of other papers, while insightful, use methodologies that i) do not permit the reproduction of results and ii) have a higher risk of bias, limiting their utility for systematic analysis of AAM and consequent regulatory advice.
	
	Beyond these methodological concerns, several gaps exist in the current research literature:
	
	\begin{enumerate}
		\item \textbf{Lack of an Integrated Framework:} Existing studies often focus on individual aspects of AAM, such as vehicle-related factors, airspace design, or ground infrastructure, without offering a unified framework that integrates vertiport network design, corridor formulation, operation management, and separation definitions. This fragmentation hinders a holistic understanding of the interactions between these components.
		
		\item \textbf{Insufficient Taxonomies and Structured Approaches:} While some literature provides overviews, there is a lack of comprehensive taxonomies that cover all essential factors for AAM corridor design. This limitation restricts the ability to systematically address the challenges involved in deploying AAM systems.
		
		\item \textbf{Inadequate Exploration of Operational Management Strategies:} 
		Existing studies often overlook advanced strategies for dynamically aligning flight demand with available capacity, such as corridor and vertiport management techniques, which are crucial for maintaining operational efficiency and reliability.
		
		\item \textbf{Underdeveloped Separation Definition Factors:} Current research frequently considers existing separation standards defined for commercial and general aviation. This oversight can lead to conservative safety standards hindering higher-density AAM operations.
	\end{enumerate}

	\textbf{Beyond the State of the Art:} {We focus on AAM corridor design and separation definition as the core research dimensions because they connect ground infrastructure to daily flight operations. While vertiports act as the isolated nodes, the corridors form the physical links (requiring complex spatial and environmental considerations), and separation definitions define the operational capacity within those links. Without a systematic understanding of how to structure these routes and safely distance the vehicles within them, the holistic integration of vertiports and flight operations remains impossible.}
	Given {these interconnected challenges}, this Systematic Literature Review (SLR) is both timely and essential. To address {the identified research gaps}, we employed a PRISMA-based methodology, drawing on the Context, Intervention, Mechanism, and Outcome (CIMO) framework to formulate research questions that target the key technological, environmental, and societal factors shaping these airspace structures. This approach aims to minimize bias, ensure reproducibility, and synthesize a broad range of recent studies. Our review seeks to establish comprehensive taxonomies for AAM design factors and propose an integrated framework that unifies vertiport network design, corridor formulation, operation management, and separation standards. The complete methodology, including search strategies, selection criteria, and quality assessment protocols, is detailed in Section~\ref{sec:methodology}. The review results are presented in Section~\ref{sec:results} while Section~\ref{sec:discussion} provides insights on missing factors and under-investigated areas.
	
	\section{Literature review Methodology}\label{sec:methodology}
	This study presents a Systematic Literature Review conducted in accordance with the PRISMA 2020 guidelines \cite{Page2021} to ensure transparency, replicability, and comprehensive reporting. The objective of this SLR is to systematically identify, evaluate, and synthesize research findings related to the design and separation methodologies of AAM corridors for passenger and freight eVTOLs (flying cars) in urban environments. In the following text, we (i) provide a high-level overview of the methodology and (ii) detail the PRISMA-based selection procedure.
	
	\subsection{Methodology overview}\label{sec:methodology_detailed}
	\subsubsection{Overview of the Systematic Review Process}
	\begin{figure}[]
		\centering
		\includegraphics[width=0.7\columnwidth]{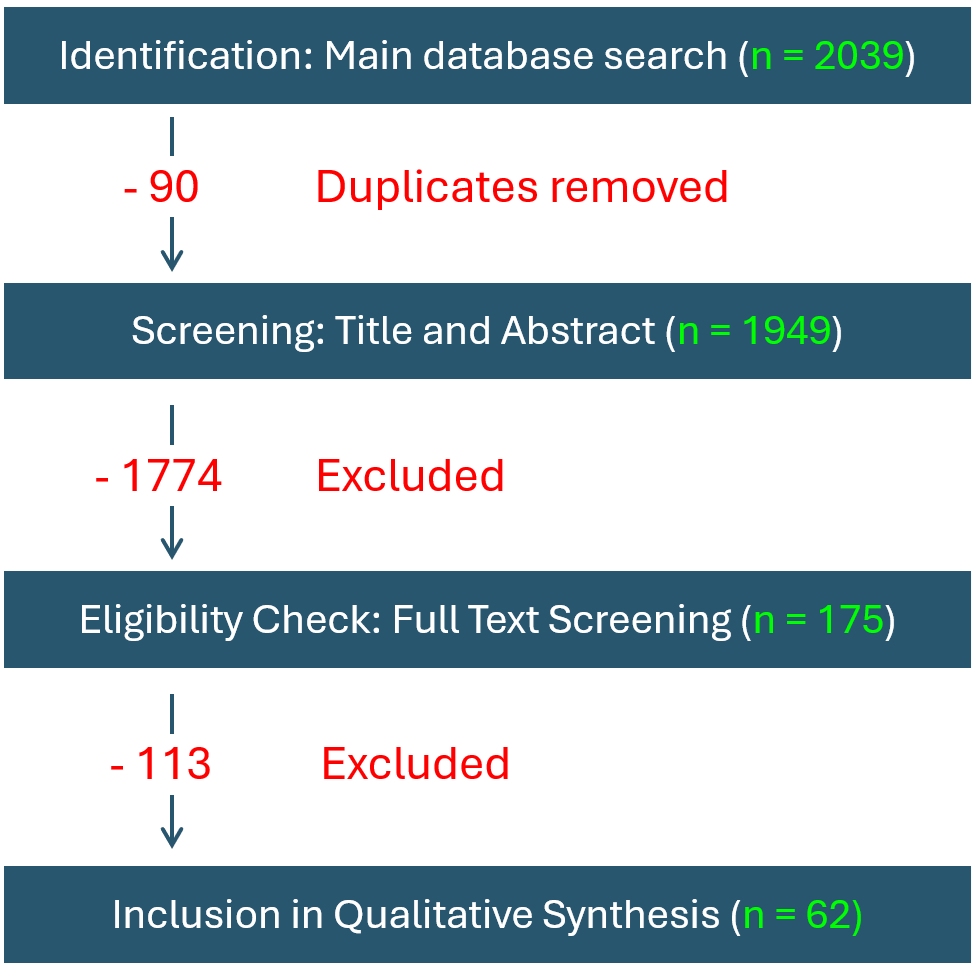}
		\caption{Flow Diagram for the selection of the literature reviewed.}
		\label{fig:PRISMA}
		\vspace{-0.5cm}
	\end{figure}
	
	The SLR process (Fig. \ref{fig:PRISMA}) comprises three main phases: (i) Identification, (ii) Screening, and (iii) Eligibility. Upon the formulation of research questions, the identification phase defines the search strategy, including the choice of data sources and extraction methods for collecting relevant papers. The screening and eligibility phases outline the inclusion and exclusion criteria aligned with the specific requirements and scope of the review, where papers are filtered based on titles and abstracts (screening) and full-text (eligibility). The answers to the research questions are then synthesized, while the challenges, opportunities, and limitations are highlighted.
	
	\begin{figure*}[ht]
		\begin{subfigure}[t]{0.4\textwidth}
			\includegraphics[width=1\linewidth]{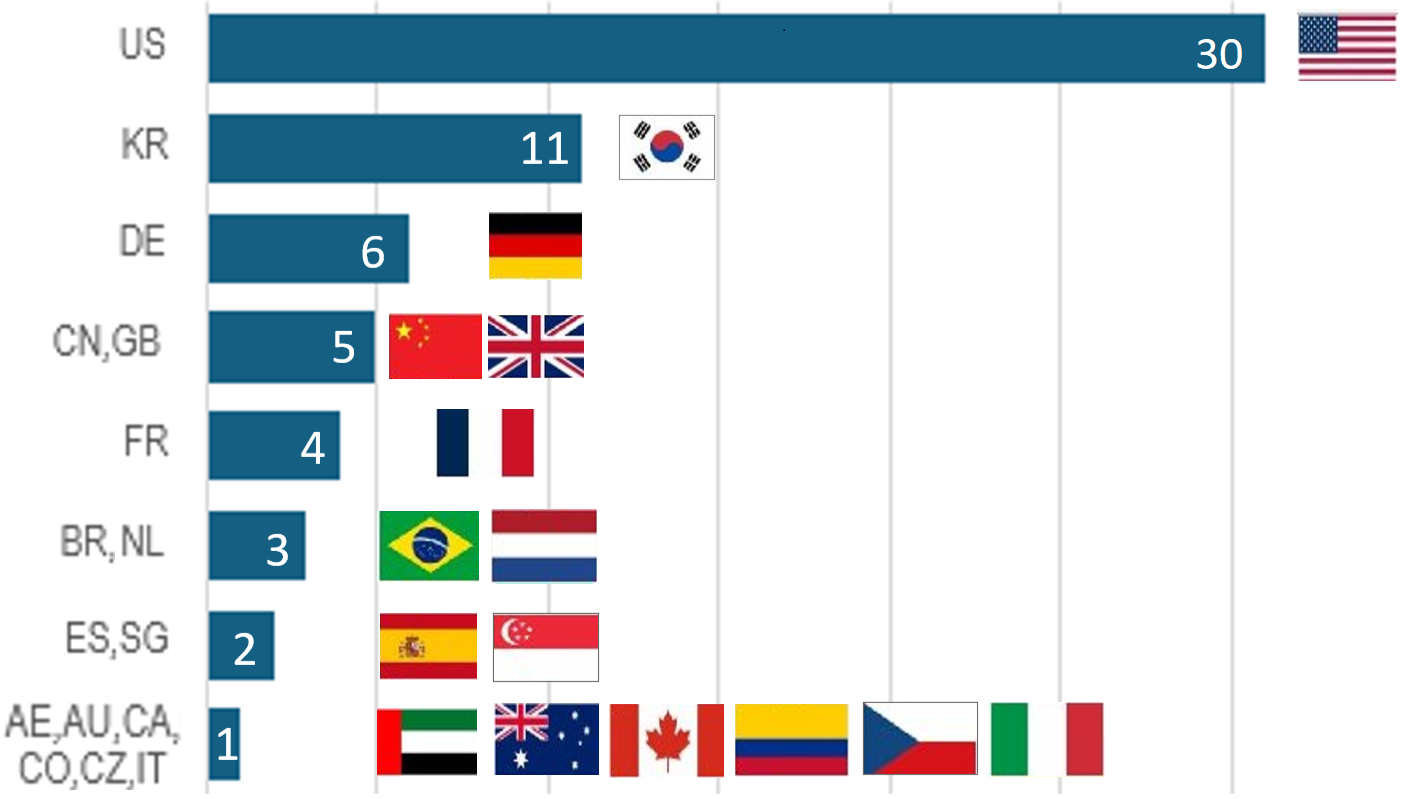}
			\caption{By author affiliations}\label{fig:auth-affl}
		\end{subfigure}
		\begin{subfigure}[t]{0.29\textwidth}
			\includegraphics[width=1\linewidth]{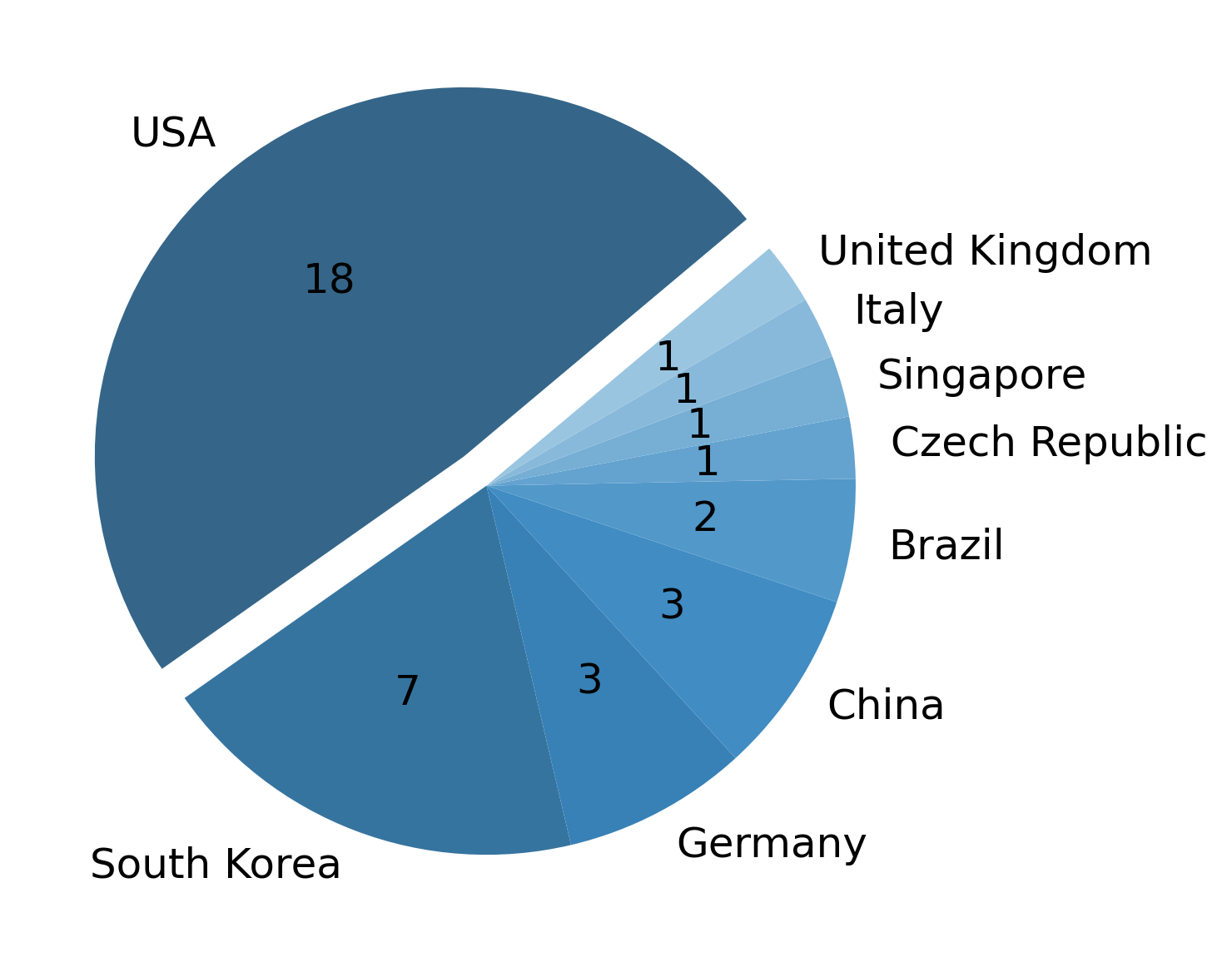}
			\caption{By case study countries}\label{fig:case-study-country}
		\end{subfigure}
		\begin{subfigure}[t]{0.29\textwidth}
			\includegraphics[width=1\linewidth]{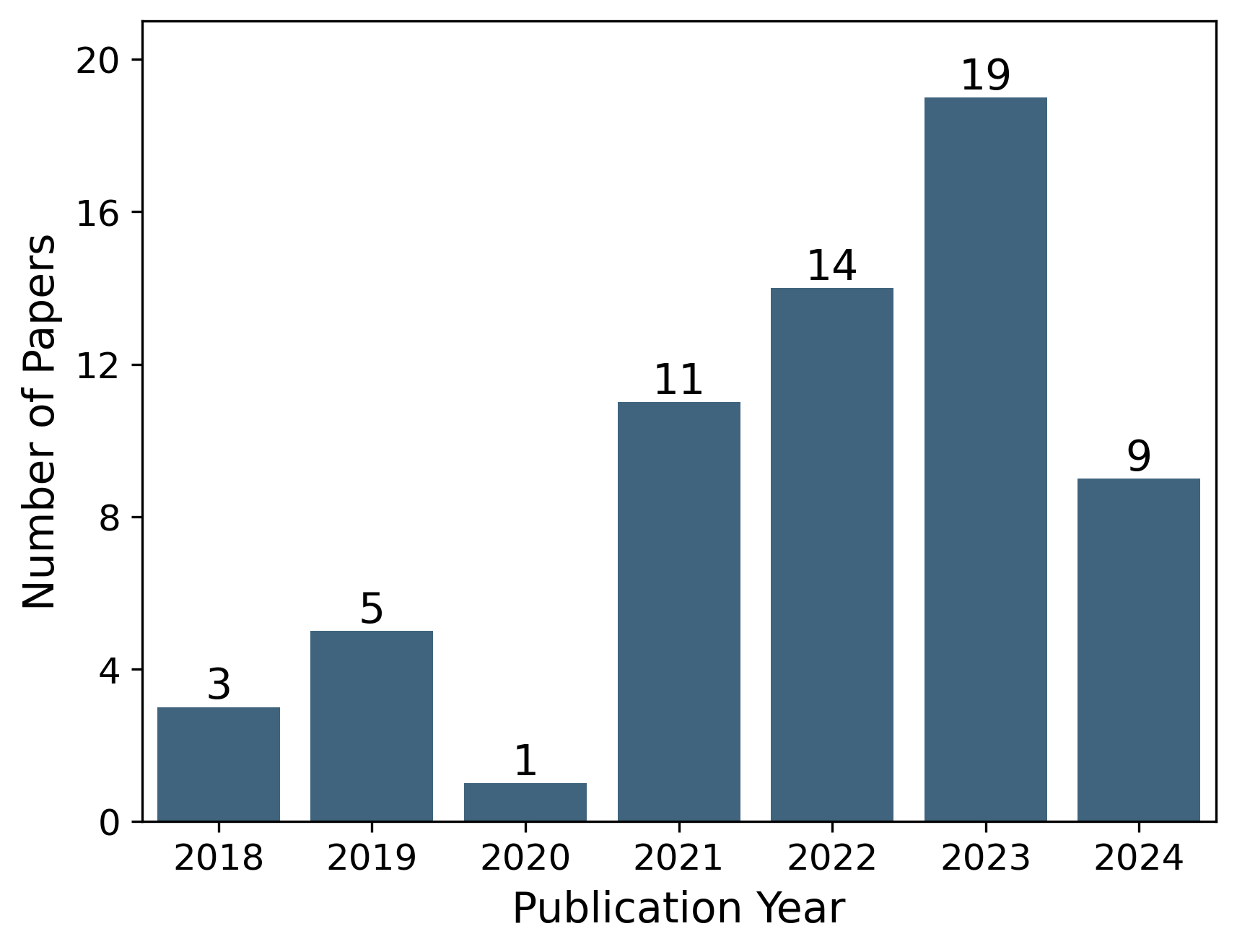}
			\caption{By publication year}\label{fig:publication-year}
		\end{subfigure}
		
		\caption{Distribution of included papers}
		\label{fig:included-papers}
		\vspace{-0.6cm}
	\end{figure*}
	
	The search process represented by the flowchart in Fig. \ref{fig:PRISMA} resulted in a total of \textit{2039} articles from IEEE Xplore and Web of Science databases. After the removal of \textit{90} duplicate records, \textit{1949} unique papers remained. Out of these, \textit{1774} studies were excluded during the title and abstract screening phase as they did not fulfill the inclusion criteria, primarily due to irrelevance to the research questions, focus on non-urban areas, or not addressing passenger or heavy freight eVTOL-related topics\footnote{Note that papers related to small drones were excluded due to the significant difference in operational and safety requirements. Moreover, the drone topic has been extensively covered in literature~\cite{Huang2019uavCollisions,Vinogradov2020safe,Rezaee2024uavCollisions,Asghari2025uavSLR, hamissi2024comprehensive}.}. This screening process retained \textit{175} studies for full-text review. Subsequently, \textit{113} of these were excluded after inspecting full texts. Thus, finally, only \textit{62} studies {(see the full list in Appendix~\ref{Appendix:overview})} have been selected for inclusion in the current review and are summarized in the following sections. The selection process took approximately three months to complete.

	\vspace{-0.1cm}
	\subsubsection{Geographical Distribution of Author Affiliations}
	Fig.~\ref{fig:included-papers} presents statistics of the selected papers. In total, researchers from 16 countries contributed to 62 selected papers (Fig.~\ref{fig:auth-affl}). Thirty papers were authored by researchers based in the USA, while 11 papers were written by South Korean researchers. Germany, the UK, China, and France each contributed between 4 to 6 papers. South America is represented by Brazil (3) and Colombia (1), the Middle East by the UAE (1), ASEAN by Singapore (2) and Australia (1).
	
	{Approximately one half of the papers (37) presented a realistic case study of an AAM deployment in a city. As observed in Fig.~\ref{fig:case-study-country}, most of the case studies considered the USA (18) followed by countries like South Korea (7), Germany (3), and China (3), underlining the extensive geographical coverage of AAM research.}
	
	The relevant papers have been published from 2018 onwards, as illustrated in Fig.~\ref{fig:publication-year}. The USA, South Korea, and Germany have consistently published studies since 2018, reflecting sustained interest and investment in AAM technologies. In contrast, China only began contributing actively starting in 2023, indicating a recent surge in research activity within the country. The highest number of papers was published in 2023, while a relatively large number of related publications have been recorded for 2024. The trend of increasing interest is evident, despite only one paper being published in 2020, which can be attributed to the COVID-19 pandemic's impact on research activities.

	\subsubsection{Research Themes}
	Thematic analysis of the selected papers reveals that the majority focus on various aspects of AAM operation design, divided into three primary phases: vertiport location definition, corridor formulation, and operational design (including scheduling and dynamic conflict management). These themes account for approximately three fourths of the studies reviewed. Conversely, the investigation of separation distances between eVTOLs is less prevalent, constituting about one fourth of the papers. This indicates a potential area for future research to ensure the safety and efficiency of advanced air mobility systems.
	
	\subsubsection{Publication Venues}
	
	Of the 62 selected studies, 24 were published in peer-reviewed journals, while 38 were presented at conferences. The most prominent journal is \textit{Transportation Research Part C: Emerging Technologies} by Elsevier, which accounts for 4 papers. Among the conferences, the \textit{IEEE/AIAA Digital Avionics Systems Conference (DASC)} is the leading venue with 19 papers, followed by the \textit{Integrated Communications, Navigation and Surveillance Conference} with 8 papers.

	\subsection{Review Steps}
	\begin{figure}[b!]
		\centering
		\vspace{-0.5cm}
		\includegraphics[width=0.7\columnwidth]{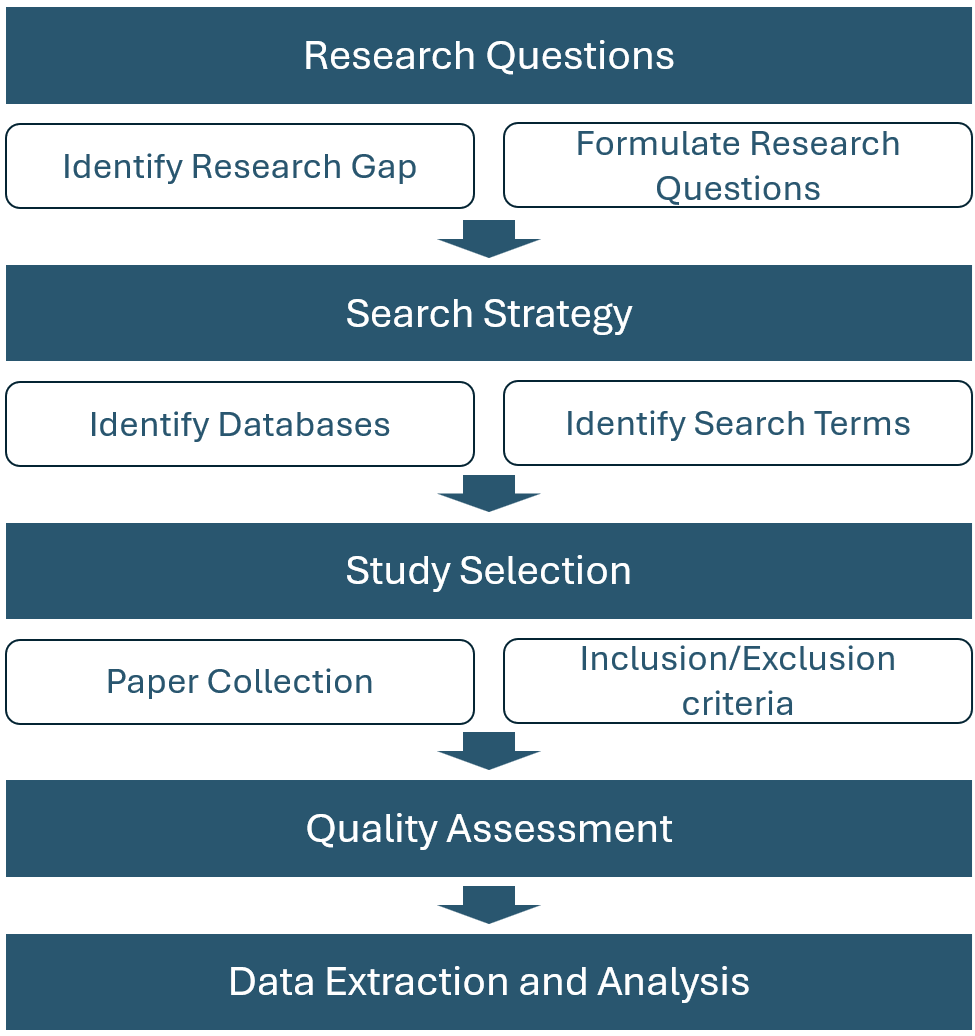}
		\caption{Methodology of the systematic review process}
		\label{fig:method}
		\vspace{-0.6cm}
	\end{figure}
	Fig.~\ref{fig:method} shows the systematic review steps: specifying research questions, designing the search strategy, screening studies, assessing quality, extracting data, and analyzing findings. Each of these blocks is detailed in the text below where we discuss the procedures and decisions made at every stage.
	\subsubsection{Research Questions (RQs)}
	
	The SLR methodology requires well-formulated research questions. In this work, we rely on the CIMO framework (Context, Intervention, Mechanism, and Output) to define the research questions for our methodological study. Therefore, the research questions devised for this systematic review are:
	
	\begin{itemize}
		\item \textbf{RQ1:} In urban areas (Context), what technical, environmental, and societal factors (Mechanism) influence the design (Intervention) of AAM corridors (Outcome) for eVTOLs (flying cars)?
		\item \textbf{RQ2:} In urban areas (Context), what technical, environmental, and societal factors (Mechanism) influence the definition (Intervention) of separation distances (Outcome) for eVTOLs (flying cars)?
		\item \textbf{RQ3:} In urban areas with eVTOL operations (Context), what quantitative methodologies (Mechanism) are used to define (Intervention) safe separation distances (Outcome) for eVTOLs (flying cars)?
	\end{itemize}
	
	These key questions have been formulated to guide the systematic review to understand the factors influencing design and operational parameters of AAM corridors for passenger or freight transportation in urban settings. Moreover, this paper seeks to present a comprehensive taxonomy of the factors, highlighting gaps and opportunities for future research.
	
	\subsubsection{Search Strategy}
	
	To address the research questions, we conducted a comprehensive search using two scientific databases: IEEE Xplore and Web of Science (WoS). These databases were selected due to their extensive coverage of peer-reviewed journals and conference proceedings in the fields of engineering, technology, and applied sciences, which are important to the study of AAM. 
	
	The search for relevant publications was initiated on 26\textsuperscript{th} of May 2024, with updates for any additional papers finalized on 19\textsuperscript{th} of June 2024. Using the research questions, a set of keywords and queries were formulated tailored to the search syntax and requirements of each database. A combination of keywords '\textit{VTOL}', '\textit{Vertical Take-Off and Landing}', '\textit{air mobility}', '\textit{UAM}', '\textit{U-SPACE}', '\textit{UTM}', '\textit{AAM}', '\textit{urban airspace}', '\textit{air taxi}', '\textit{flying car}', '\textit{corridor}', '\textit{separation distance}', '\textit{separation}', '\textit{airspace design}', '\textit{traffic management}', '\textit{flight path}', '\textit{route}', '\textit{trajectory}', '\textit{flight corridor}', '\textit{air traffic management}', '\textit{conflict}', '\textit{collision}', and '\textit{navigation}' was employed. 
	
	All searches were limited to the "journal" and "conference" document types. After completing the first draft of the search strings, pilot searches were conducted in each database to evaluate their effectiveness in retrieving relevant studies. Effectiveness was assessed based on the relevance and number of retrieved papers. Given that WoS and IEEE Xplore support different wildcard characters ('*' for WoS and '?' for IEEE), the search strings were slightly adjusted to conform to each database's syntax. Through iterative testing and refinement, the placement and combination of wildcards were optimized to increase the number of papers, ensuring a comprehensive collection of relevant literature. The final search time frame covered publications from January 2010 to June 2024.
	
	The resulting search string used in WoS:
	\begin{tcolorbox}[width=\linewidth, colback=white!95!black]
		TS=( ("*VTOL*" OR "Vertical Take-Off and Landing" OR "air mobility" OR "UAM" OR "U-SPACE" OR "UTM" OR "AAM" OR "urban airspace" OR "air taxi" OR "flying car*") AND ("corridor*" OR "separation distance*" OR "separation*" OR "airspace design" OR "traffic management" OR "flight path*" OR "route*" OR "trajectory*" OR "flight corridor" OR "air traffic management" OR "conflict" OR "collision*" OR "navigation") ) AND (PY=(2010-2024))
	\end{tcolorbox}
	The search string in IEEE Xplore:
	\begin{tcolorbox}[width=\linewidth,  colback=white!95!black]
		("All Metadata":"?VTOL?" OR "All Metadata":"Vertical Take-Off and Landing" OR "All Metadata":"air mobility" OR "All Metadata":"UAM" OR "All Metadata":"U-SPACE" OR "All Metadata":"UTM" OR "All Metadata":"AAM" OR "All Metadata":"urban airspace" OR "All Metadata":"air taxi" OR "All Metadata":"flying car?") AND ("All Metadata":"corridor?" OR "All Metadata":"separation distance" OR "All Metadata":"separation" OR "All Metadata":"airspace design" OR "All Metadata":"traffic management" OR "All Metadata":"flight path" OR "All Metadata":"route" OR "All Metadata":"trajectory" OR "All Metadata":"flight corridor" OR "All Metadata":"air traffic management" OR "All Metadata":"conflict" OR "All Metadata":"collision?" OR "All Metadata":"navigation")
	\end{tcolorbox}
	
	\renewcommand{\arraystretch}{1}
	\begin{table}[!htb]
		\caption{Number of papers collected from each database}
		%\vspace{0.5cm}
		\centering
		\begin{tabular}{@{}lc@{}}
			\toprule
			\textbf{Database} & \textbf{Number of Papers Collected} \\ \midrule
			IEEE Xplore        & 1116                                 \\
			Web of Science     & 923                                  \\ \midrule
			\textbf{Total}     & \textbf{2039}                        \\ 
			\bottomrule
		\end{tabular}
		\label{tab:numberofpapers}
	\end{table}
	
	\begin{table*}[!htb]
		\caption{Inclusion and exclusion criteria for paper selection}
		\centering
		\begin{tabular}{|p{8cm}|p{8cm}|}
			\hline
			\textbf{\hspace{2.5cm} Inclusion Criteria} & \textbf{\hspace{2.5cm} Exclusion Criteria} \\
			\hline
			\begin{itemize}
				\item Studies within the scope of the research questions.
				\item Publications from January 2010 to June 2024.
				\item Peer-reviewed journal and conference papers.
				\item Articles written in English.
			\end{itemize} & 
			\begin{itemize}
				\item Studies focusing exclusively on rural or non-urban areas.
				\item Studies not addressing passenger or freight eVTOLs, air taxis, or flying cars.
				\item Studies focusing on small Uncrewed Air Vehicles (UAVs)
				\item Studies not covering AAM corridors, separation distances, airspace design, traffic management, flight paths, routes, or trajectories.
				\item Non-peer-reviewed publications (e.g., book chapters, letters to editor), posters, tutorials, magazine articles, abstracts, case reports, comments, reviews, other SLRs.
				\item Articles written in languages other than English.
			\end{itemize} \\
			\hline
		\end{tabular}
		\label{tab:inclusionexclusion}
	\end{table*}

	The initial search (Table \ref{tab:numberofpapers}) identified 2039 papers, 1116 of which were listed in the IEEE Xplore library and 923 in Web of Science.

	\subsubsection{Study Selection}
	The collected papers from the initial search were screened according to the preset inclusion and exclusion criteria (Table \ref{tab:inclusionexclusion}). The paper selection process consisted of two phases. First, based on the inclusion and exclusion criteria, the papers were independently screened by two researchers through title and abstract screening. The publications selected during this phase were then independently assessed by two authors through full-text screening. The authors cross-checked the selection results and resolved any disagreements on the selection decisions. All disagreements in either phase were resolved by consensus.
	
	The inclusion and exclusion criteria were formulated by the authors to effectively select relevant papers as summarized in Table~\ref{tab:inclusionexclusion}. 
	
	\subsubsection{Risk of Bias and Quality Assessment}\label{sec:QualityAssessment}
	
	This systematic literature review adhered to the PRISMA 2020 guidelines to ensure a rigorous and unbiased selection of relevant literature. Recognizing that the selection of keywords and eligibility criteria can introduce bias, the authors implemented several strategies to minimize these risks:
	
	\begin{enumerate}
		\item \textbf{Independent Screening}: Two reviewers independently screened the titles and abstracts of all retrieved studies against the inclusion and exclusion criteria. This process was followed by an independent full-text review.
		\item \textbf{Discrepancy Resolution}: Any disagreements between reviewers during the screening and selection phases were resolved through discussion or by involving a third reviewer to reach consensus.
		\item \textbf{Comprehensive Quality Assessment}: To evaluate the quality and risk of bias in the included studies, the authors employed a quality assessment checklist inspired by the {Critical Appraisal Skills Programme (CASP)}. 
		%\dmh{Should we add a reference to CASP instead of mentioning each?
			The checklist included the following criteria:
			\begin{itemize}
				\item \textbf{Clarity of Objectives}: Are the study objectives clearly stated?
				\item \textbf{Methodological Rigor}: Are the research methods appropriately designed and executed?
				\item \textbf{Data Reporting}: Are the results clearly reported, including measures of accuracy and confidence?
				\item \textbf{Contribution to the Field}: Do the study’s findings offer significant insights, and do they adequately address the research questions?
				\item \textbf{Limitations Acknowledgment}: Are the limitations of the study transparently discussed?
			\end{itemize}
			\item \textbf{Quality Categorization}: Each study was independently assessed by two reviewers, and scores were assigned based on the checklist. %Studies were categorized as high, medium, or low quality, and only those meeting the minimum quality threshold were included in the final synthesis.
		\end{enumerate}
		
		Additionally, the review considered the potential for {publication bias} by including both journal and conference papers and by conducting a comprehensive search across multiple databases. {A complete list of 62 selected studies with their bibliographic data is provided in Table~\ref{tab:slr-comprehensive-summary} in Appendix~\ref{Appendix:overview} along with the CASP scores for each study.}
		
		\subsubsection{Data Extraction}
		
		Relevant data were systematically extracted from the selected publications to facilitate comprehensive analysis and synthesis. The data extraction process was conducted using the \textit{Rayyan} platform independently by two reviewers to minimize bias and ensure accuracy. The following data items were retrieved from each study:
		
		\begin{itemize}
			\item \textbf{Bibliographic Information}: Title, authors, year of publication, journal or conference name.
			\item \textbf{Study Characteristics}: Research objectives, methodology, data sources, type (case-study or theoretical).
			\item \textbf{Technical Aspects}: Specific technical factors addressed, technologies or frameworks used.
			\item \textbf{Environmental and Societal Factors}: Environmental considerations, societal impacts discussed.
			\item \textbf{Findings and Contributions}: Key results, conclusions, and contributions to the field.
			\item \textbf{Limitations}: Identified limitations or gaps in each study.
		\end{itemize}
		
		Data extraction was facilitated using a standardized extraction form developed in Excel, ensuring consistency across all studies.
		
		\subsubsection{Data Synthesis and Analysis}
		
		The data extracted from the included studies were synthesized using a thematic analysis approach. This involved identifying and categorizing recurring themes, patterns, and relationships related to the research questions. The synthesis process included:
		
		\begin{itemize}
			\item \textbf{Thematic Coding}: Assigning codes to key concepts (e.g., distinct corridor design phases) and findings (factors) within each study.
			\item \textbf{Theme Development}: Grouping related codes into broader themes that address the research questions.
			\item \textbf{Integration of Findings}: Combining insights from multiple studies to provide a comprehensive understanding of the factors influencing AAM corridor design and separation distances.
		\end{itemize}
		
		\subsubsection{Limitations of the Methodology}
		
		While this SLR was conducted with rigorous methodological standards, several limitations should be acknowledged:
		
		\begin{itemize}
			\item \textbf{Language Restriction}: Only English-language studies were included, potentially excluding relevant research published in other languages.
			\item \textbf{Database Selection}: Although IEEE Xplore and Web of Science are comprehensive, relevant studies indexed in other databases may have been missed.
			\item \textbf{Publication Bias}: Despite efforts to include grey literature, the review primarily captures peer-reviewed publications, which may overrepresent positive or significant findings.
			\item \textbf{Temporal Limitation}: The search was conducted up to June 2024, and recent publications beyond this date were not included.
			\item \textbf{Subjectivity in Quality Assessment}: Although multiple reviewers were involved, some level of subjectivity in assessing study quality is inherent.
		\end{itemize}

		\section{Results}\label{sec:results}
		This section presents the findings of our systematic review, organized around the three research questions. Section~\ref{sec:results}.\ref{sec:RQ1} covers \textbf{RQ1}, detailing the technical, environmental, and societal factors that guide AAM corridor design. Section~\ref{sec:results}.\ref{sec:RQ2} addresses \textbf{RQ2}, examining which factors influence the definition of separation distances. Finally, Section~\ref{sec:results}.\ref{sec:RQ3} responds to \textbf{RQ3}, discussing the quantitative methodologies used to determine safe separation distances in urban eVTOL operations.
		
		\subsection{RQ1: Corridor Design Factors}\label{sec:RQ1}
		Corridor design for AAM operations involves multiple considerations spanning technical, environmental, and societal domains. The thematic grouping of the AAM corridor design papers revealed three distinct clusters related to the following three phases of AAM corridor design. %We adopt a three-phase approach: 
		Phase~1 addresses vertiport network design; Phase~2 focuses on corridor formulation, emphasizing feasibility, safety, and sustainability; and Phase~3 deals with operational management. This structure aligns with \textbf{RQ1} by revealing the core factors that guide effective corridor design in urban AAM. 
		
\begin{figure}[tbp]
	\centering
	\makebox[\linewidth][c]{%
		\begin{tikzpicture}[
			font=\sffamily\scriptsize,
			thick,
			every node/.style={align=center, text=black, inner sep=0pt, outer sep=0pt} 
			]
			% --- GEOMETRY ---
			\def\Rinner{1.4}  
			\def\Rmid{2.9}    
			\def\Router{4.3} % Set to exactly 8.6cm diameter to perfectly fill standard IEEE columns 
			
			% Text placement radii
			\def\RtextLOne{2.15} 
			\def\RtextLTwo{3.55} 
			
			% --- STRICT BOUNDING BOX ---
			% This trims any empty space on the left/right and locks the width exactly to the outer ring
			\useasboundingbox (-\Router, -\Router) rectangle (\Router, \Router);
			
			% --- COLORS ---
			\definecolor{NavyBlue}{RGB}{0, 41, 142}
			\definecolor{SteelBlue}{RGB}{70, 130, 180}
			
			% ==========================================
			% --- LEVEL 1 (Parent Categories - Phases) ---
			% ==========================================
			
			\filldraw[fill=gray!40, draw=white, line width=1pt] 
			(0:\Rinner) arc (0:85.42:\Rinner) -- (85.42:\Rmid) arc (85.42:0:\Rmid) -- cycle;
			\node[rotate=42.71, font=\sffamily\scriptsize\bfseries] at (42.71:\RtextLOne) {Vertiport\\Network\\(14)};
			
			\filldraw[fill=NavyBlue!50, draw=white, line width=1pt] 
			(85.42:\Rinner) arc (85.42:256.27:\Rinner) -- (256.27:\Rmid) arc (256.27:85.42:\Rmid) -- cycle;
			\node[rotate=170.84-180, font=\sffamily\scriptsize\bfseries] at (170.84:\RtextLOne) {Corridor\\formulation\\(28)};
			
			\filldraw[fill=NavyBlue!20, draw=white, line width=1pt] 
			(256.27:\Rinner) arc (256.27:360:\Rinner) -- (360:\Rmid) arc (360:256.27:\Rmid) -- cycle;
			\node[rotate=308.13-270, font=\sffamily\scriptsize\bfseries] at (308.13:\RtextLOne) {Operations\\(17)};
			
			% ==========================================
			% --- LEVEL 2 ---
			% ==========================================
			
			% --- Vertiport Children ---
			\filldraw[fill=gray!20, draw=white, line width=1pt] (0:\Rmid) arc (0:32.03:\Rmid) -- (32.03:\Router) arc (32.03:0:\Router) -- cycle;
			\node[rotate=16.01] at (16.01:\RtextLTwo) {Demand\\(12)};
			
			\filldraw[fill=gray!10, draw=white, line width=1pt] (32.03:\Rmid) arc (32.03:45.38:\Rmid) -- (45.38:\Router) arc (45.38:32.03:\Router) -- cycle;
			\node[rotate=38.70] at (38.70:\RtextLTwo) {Feasibility\\(5)};
			
			\filldraw[fill=gray!20, draw=white, line width=1pt] (45.38:\Rmid) arc (45.38:61.40:\Rmid) -- (61.40:\Router) arc (61.40:45.38:\Router) -- cycle;
			\node[rotate=53.39] at (53.39:\RtextLTwo) {Regulations\\(6)};
			
			\filldraw[fill=gray!10, draw=white, line width=1pt] (61.40:\Rmid) arc (61.40:85.42:\Rmid) -- (85.42:\Router) arc (85.42:61.40:\Router) -- cycle;
			\node[rotate=73.41] at (73.41:\RtextLTwo) {Economic\\Viability\\(9)};
			
			% --- Corridor Children ---
			\filldraw[fill=NavyBlue!40, draw=white, line width=1pt] (85.42:\Rmid) arc (85.42:138.81:\Rmid) -- (138.81:\Router) arc (138.81:85.42:\Router) -- cycle;
			\node[rotate=112.11-90] at (112.11:\RtextLTwo) {Feasibility\\(15)};
			
			\filldraw[fill=NavyBlue!30, draw=white, line width=1pt] (138.81:\Rmid) arc (138.81:220.67:\Rmid) -- (220.67:\Router) arc (220.67:138.81:\Router) -- cycle;
			\node[rotate=179.74-180] at (179.74:\RtextLTwo) {Safety\\(23)};
			
			\filldraw[fill=NavyBlue!40, draw=white, line width=1pt] (220.67:\Rmid) arc (220.67:256.27:\Rmid) -- (256.27:\Router) arc (256.27:220.67:\Router) -- cycle;
			\node[rotate=238.47-270] at (238.47:\RtextLTwo) {Sustainability\\(10)};
			
			% --- Operations Children ---
			\filldraw[fill=NavyBlue!15, draw=white, line width=1pt] (256.27:\Rmid) arc (256.27:276.35:\Rmid) -- (276.35:\Router) arc (276.35:256.27:\Router) -- cycle;
			\node[rotate=266.31-270] at (266.31:\RtextLTwo) {Flight\\Demand\\(6)};
			
			\filldraw[fill=NavyBlue!10, draw=white, line width=1pt] (276.35:\Rmid) arc (276.35:319.85:\Rmid) -- (319.85:\Router) arc (319.85:276.35:\Router) -- cycle;
			\node[rotate=298.1-270] at (298.1:\RtextLTwo) {Capacity\\(13)}; 
			
			\filldraw[fill=NavyBlue!15, draw=white, line width=1pt] (319.85:\Rmid) arc (319.85:360:\Rmid) -- (360:\Router) arc (360:319.85:\Router) -- cycle;
			\node[rotate=339.92-270] at (339.92:\RtextLTwo) {Operations\\Management\\(12)};
			
			% --- Center Label ---
			\draw[lightgray, line width=0.5pt] (0,0) circle (\Rinner); 
			\node[font=\sffamily\footnotesize\bfseries, inner sep=0pt] at (0,0) {AAM\\Design\\Factors};
			
		\end{tikzpicture}%
	}
	\caption{{Distribution of literature focusing on various AAM design factors, reflecting current research priorities and perceived bottlenecks.}}
	\label{fig:aam-design-sunburst}
\end{figure}
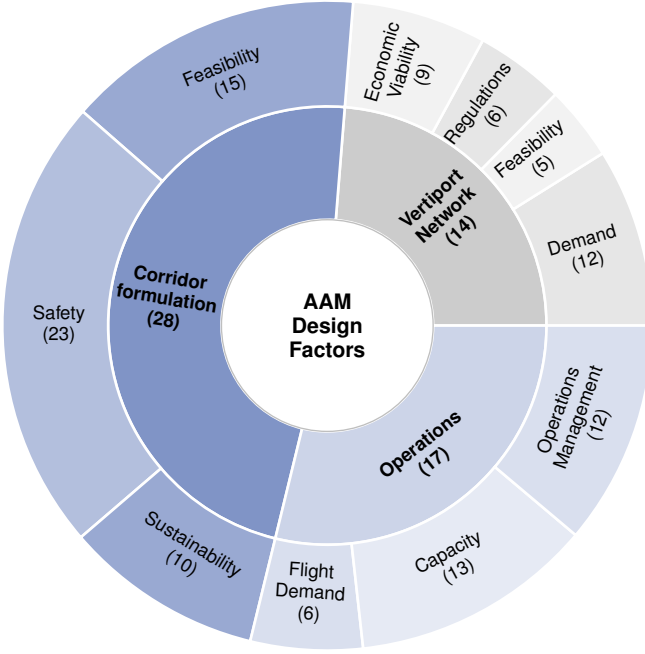

		{To provide a quantitative overview of this specific landscape, Fig.~\ref{fig:aam-design-sunburst} illustrates the distribution of the literature across these three phases. As shown, the majority of the reviewed literature focuses heavily on the Corridor formulation (28 papers), with a specific emphasis on Safety (23 papers), reflecting the immediate perceived bottlenecks for AAM integration.}

		\subsubsection{Phase 1 - Vertiport Network Design} 
		
		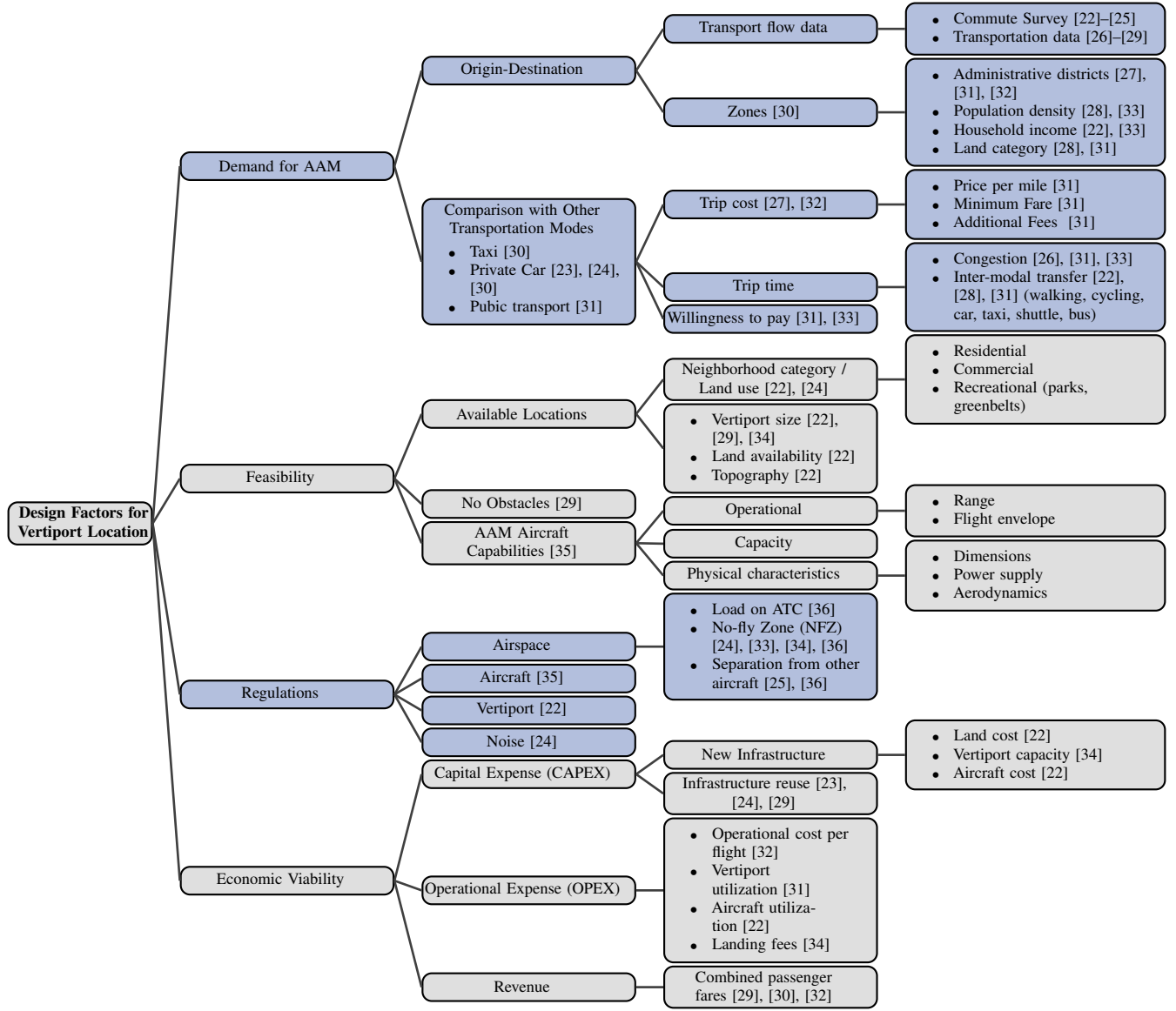
\begin{figure*}[tbph!]
    \centering
    \begin{forest}
        for tree={
            grow=east,
            reversed=true,
            anchor=base west,
            parent anchor=east,
            child anchor=west,
            base=left,
            font=\scriptsize, %
            rectangle,
            draw=black, %
            rounded corners,
            align=left,
            minimum width=2em, %
            edge+={darkgray, line width=1pt},
            s sep=1pt, %
            inner xsep=1pt, %
            inner ysep=2pt, %
            line width=0.8pt,
            text width=9em, %
        },
        [\textbf{\parbox{9em}{\centering Design Factors for\\Vertiport Location}}, text width=6em, fill=lightgray!50
            %% Step 1 factors
            [\parbox{12em}{\centering Demand for AAM}, fill=NavyBlue!30
                [\parbox{12em}{\centering Origin-Destination}, fill=NavyBlue!30
                    [\parbox{12em}{\centering Transport flow data},fill=NavyBlue!30
                                [\parbox{15em}{
                                    \begin{itemize}
                                    \item Commute Survey~\cite{Tarafdar2019nortCalifornia,Lim2019Seoul,Jeong2021kMeansVertiport,Souza2023confManag}
                                    \item Transportation data~\cite{Bulusu2021demand,Peng2022hierarchical,Qu2024demand,Guo2024vtolSite}
                                    \end{itemize}
                                    }, text width=11em, fill=NavyBlue!30, rounded corners
                                ]
                    ]
                    [\parbox{12em}{\centering Zones~\cite{Chae2023vertLocations}},fill=NavyBlue!30 
                                [\parbox{15em}{
                            \begin{itemize}
                            \item Administrative districts~\cite{Rimjha2021factorsVertiport,Peng2022hierarchical,Shin2022Skyport}
                            \item Population density~\cite{Rimjha22AirRestrict,Qu2024demand}
                            \item Household income~\cite{Tarafdar2019nortCalifornia,Rimjha22AirRestrict}
                            \item Land category~\cite{Rimjha2021factorsVertiport,Qu2024demand}
                            \end{itemize}
                            }, text width=11em, fill=NavyBlue!30, rounded corners]
                    ]          
                ]
                [\parbox{12em}{\centering Comparison with Other Transportation Modes 
                                 \begin{itemize}
                                    \item Taxi~\cite{Chae2023vertLocations}
                                    \item Private Car \cite{Lim2019Seoul,Jeong2021kMeansVertiport,Chae2023vertLocations}
                                    \item Pubic transport~\cite{Rimjha2021factorsVertiport}
                                \end{itemize}},fill=NavyBlue!30
                    [\parbox{12em}{\centering Trip cost~\cite{Peng2022hierarchical,Shin2022Skyport}}, fill=NavyBlue!30
                        [\parbox{12em}{
                                \begin{itemize}
                                \item Price per mile~\cite{Rimjha2021factorsVertiport}
                                \item Minimum Fare~\cite{Rimjha2021factorsVertiport}
                                \item Additional Fees ~\cite{Rimjha2021factorsVertiport}
                                \end{itemize}
                                }, text width=11em, fill=NavyBlue!30, rounded corners
                        ]
                    ]
                    [\parbox{12em}{\centering Trip time}, fill=NavyBlue!30
                        [\parbox{15em}{
                                \begin{itemize}
                                %\item Trip length %Distance, 
                                %\item Aircraft speed
                                \item Congestion \cite{Rimjha2021factorsVertiport,Bulusu2021demand,Rimjha22AirRestrict}
                                \item Inter-modal transfer~\cite{Tarafdar2019nortCalifornia,Rimjha2021factorsVertiport,Qu2024demand} (walking, cycling, car, taxi, shuttle, bus)
                                \end{itemize}
                                }, text width=11em, fill=NavyBlue!30, rounded corners
                        ]
                    ]
                    [
                        \parbox{12em}{\centering Willingness to pay~\cite{Rimjha2021factorsVertiport,Rimjha22AirRestrict}}, fill=NavyBlue!30
                    ]
                ]
            ]
            %% Step 2 factors
            [\parbox{12em}{\centering Feasibility}, fill=lightgray!50
                [\parbox{12em}{\centering Available Locations}, fill=lightgray!50
                    [\parbox{12em}{\centering Neighborhood category / Land use~\cite{Tarafdar2019nortCalifornia, Jeong2021kMeansVertiport}},fill=lightgray!50, 
                            [\parbox{12em}{
                                \begin{itemize}
                                \item Residential
                                \item Commercial
                                \item Recreational (parks, greenbelts)                            
                                \end{itemize}
                                }, text width=11em, fill=lightgray!50, rounded corners
                            ]
                    ]
                    [\parbox{12em}{
                            \begin{itemize}
                            \item Vertiport size~\cite{Tarafdar2019nortCalifornia, Rimjha2021LA,Guo2024vtolSite}
                            \item Land availability~\cite{Tarafdar2019nortCalifornia}
                            \item Topography~\cite{Tarafdar2019nortCalifornia}
                            \end{itemize}
                            },fill=lightgray!50, rounded corners
                    ]
                ]
                [\parbox{12em}{\centering No Obstacles~\cite{Guo2024vtolSite}}, fill=lightgray!50
                ]
                [\parbox{12em}{\centering AAM Aircraft Capabilities~\cite{Rakas2021fleetSelection}}, fill=lightgray!50
                    [\parbox{12em}{\centering Operational},fill=lightgray!50, rounded corners
                        [\parbox{12em}{
                                \begin{itemize}
                                \item Range
                                \item Flight envelope                        
                                \end{itemize}
                                }, text width=11em, fill=lightgray!50, rounded corners
                        ]
                    ]
                    [\parbox{12em}{\centering Capacity},fill=lightgray!50, rounded corners
                    ]
                    [\parbox{12em}{\centering Physical characteristics},fill=lightgray!50, rounded corners
                        [\parbox{12em}{
                                \begin{itemize}
                                \item Dimensions %(Wingspan, Height, Length)
                                \item Power supply
                                \item Aerodynamics                  
                                \end{itemize}
                                }, text width=11em, fill=lightgray!50, rounded corners
                        ]
                    ]
                ]
            ]
            %% Step 3 factors
             [\parbox{12em}{\centering Regulations}, fill=NavyBlue!30
                [\parbox{12em}{\centering Airspace}, fill=NavyBlue!30
                    [\parbox{12em}{
                            \begin{itemize}
                            \item Load on ATC~\cite{Vascik2021interoperability}
                            \item No-fly Zone (NFZ) 
                            \cite{Jeong2021kMeansVertiport,Rimjha2021LA, Rimjha22AirRestrict, Vascik2021interoperability}
                            %\item \textcolor{red}{Geofence?}
                            \item Separation from other aircraft~\cite{Vascik2021interoperability,Souza2023confManag}
                            \end{itemize}
                            }, fill=NavyBlue!30  
                    ]
                ]
                [\parbox{12em}{\centering Aircraft~\cite{Rakas2021fleetSelection}}, fill=NavyBlue!30]
                [\parbox{12em}{\centering Vertiport~\cite{Tarafdar2019nortCalifornia}}, fill=NavyBlue!30
                ]
                [\parbox{12em}{\centering Noise \cite{Jeong2021kMeansVertiport}}, fill=NavyBlue!30
                ]
            ]
            %% Step 4 factors
           [\parbox{12em}{\centering Economic Viability}, fill=lightgray!50
                [\parbox{12em}{\centering Capital Expense (CAPEX)}, fill=lightgray!50
                    [\parbox{12em}{\centering New Infrastructure}, fill=lightgray!50
                        [\parbox{15em}{
                            \begin{itemize}
                            \item Land cost~\cite{Tarafdar2019nortCalifornia}
                            \item Vertiport capacity~\cite{Rimjha2021LA}
                            \item Aircraft cost~\cite{Tarafdar2019nortCalifornia}
                            \end{itemize}
                            }, text width=11em, fill=lightgray!50, rounded corners]     
                    ]
                    [\parbox{12em}{\centering Infrastructure reuse~\cite{Lim2019Seoul,Jeong2021kMeansVertiport,Guo2024vtolSite}}, fill=lightgray!50
                    ]
                ]
                [\parbox{12em}{\centering Operational Expense (OPEX)}, fill=lightgray!50
                    [\parbox{12em}{
                            \begin{itemize}
                            \item Operational cost per flight~\cite{Shin2022Skyport} 
                            \item Vertiport utilization~\cite{Rimjha2021factorsVertiport}
                            \item Aircraft utilization~\cite{Tarafdar2019nortCalifornia}
                            \item Landing fees~\cite{Rimjha2021LA}
                            \end{itemize}
                            },fill=lightgray!50, rounded corners
                    ]
                ]
                [\parbox{12em}{\centering Revenue}, fill=lightgray!50
                    [\parbox{12em}{\centering
                        Combined passenger fares~\cite{Shin2022Skyport,Chae2023vertLocations,Guo2024vtolSite} 
                        }, fill=lightgray!50, rounded corners
                    ]
                ]
        	]
    	]
    \end{forest}
    \caption{Taxonomy of design factors for selecting a candidate vertiport location}
    \label{fig:vertiport-design-factors}
\end{figure*}
		
		Fig.~\ref{fig:vertiport-design-factors} presents a taxonomy of design factors for selecting optimal vertiport locations. The location selection process is divided into four tasks:
		\begin{enumerate}
			\item \textbf{Assessing Demand for AAM} aims to define potential coarse vertiport locations (e.g., at the neighborhood level) by analyzing origin-destination patterns, transportation flows, and comparing AAM with existing transportation modes. This foundational assessment ensures that vertiport locations are strategically aligned with areas of high AAM demand.
			\item \textbf{Evaluating Feasibility} focuses on fine-tuning potential vertiport positions by examining land availability, absence of physical obstacles, and the technical capabilities of the AAM aircraft. This evaluation guarantees that selected sites are practically viable and can support the operational requirements of AAM services.
			\item \textbf{Considering Regulations} ensures that chosen locations comply with relevant airspace and aircraft regulations, noise restrictions, and other legal requirements, thereby facilitating smooth integration into existing frameworks. 
			\item \textbf{Analyzing Economic Viability} seeks to select the most suitable vertiport options by assessing both capital and operational expenses, ensuring the economic sustainability of the chosen locations.
		\end{enumerate}
		Collectively, these tasks provide a structured approach to identifying and validating vertiport sites that meet diverse criteria essential for successful AAM integration. Since the modifications in one task may change the output of other phases, several iterations may be required to find a stable solution.
		
		\paragraph{Guidelines for interpreting the forest plot} To effectively interpret Fig.~\ref{fig:vertiport-design-factors}, it is essential to understand its hierarchical structure and the relationships between the nodes at different levels. Each parent node represents a broader category or concept identified in the literature, while the child nodes provide more detailed sub-factors derived from specific studies. For instance, under \textbf{Assessing Demand for AAM}, the parent node \textit{``Zones~\cite{Chae2023vertLocations}''} encompasses the general concept of zoning without detailed distinctions. This indicates that the cited paper \cite{Chae2023vertLocations} recognizes and utilizes the concept of zones in vertiport location selection without specifying the underlying criteria. However, when examining the child nodes such as \textit{``Administrative districts~\cite{Rimjha2021factorsVertiport,Peng2022hierarchical,Shin2022Skyport}''} and \textit{``Population density~\cite{Rimjha22AirRestrict,Qu2024demand}''}, it becomes evident that certain studies provide more granular definitions of the term "zone" based on specific factors. For example, the works \cite{Rimjha2021factorsVertiport,Peng2022hierarchical,Shin2022Skyport} define zones based on administrative districts, while \cite{Peng2022hierarchical,Chae2023vertLocations} use population density as a defining criterion. This hierarchical representation highlights both general and specific applications within the vertiport location selection process, allowing readers to trace the origin of each factor and understand the varying levels of detail provided by different studies.
		
		\paragraph{Explanation of factors for vertiport location definition}
		
		Each of the four tasks in the vertiport location selection taxonomy encompasses a range of specific factors that collectively guide the decision-making process.
		
		\textbf{Assessing Demand for AAM:}
		In this task, the primary focus is on understanding the demand for AAM services. First, transportation \textit{Origin-Destination} (OD) pairs must be defined. Transport flow data derived from commute surveys~\cite{Tarafdar2019nortCalifornia, Lim2019Seoul, Jeong2021kMeansVertiport, Souza2023confManag} and transportation datasets~\cite{Bulusu2021demand, Peng2022hierarchical, Qu2024demand, Guo2024vtolSite} provide insights into existing movement patterns. These OD pairs can be grouped into Zones~\cite{Chae2023vertLocations} to identify the most demanded areas. Zoning can be based on various factors, with the most common approaches utilizing administrative boundaries~\cite{Rimjha2021factorsVertiport,Peng2022hierarchical,Shin2022Skyport}, population density~\cite{Rimjha22AirRestrict, Qu2024demand}, household income~\cite{Tarafdar2019nortCalifornia,Rimjha22AirRestrict}, and land category~\cite{Rimjha2021factorsVertiport,Qu2024demand}.
		
		Next, AAM is compared with \textit{Other Transportation Modes} in terms of trip cost and trip time. Most studies focus on private cars~\cite{Lim2019Seoul,Jeong2021kMeansVertiport,Chae2023vertLocations}, though taxi~\cite{Chae2023vertLocations} and public transport~\cite{Rimjha2021factorsVertiport} are also considered. The final trip cost includes price per mile and various fares~\cite{Rimjha2021factorsVertiport}. Trip time is influenced by factors such as time-varying road congestion~\cite{Rimjha2021factorsVertiport,Bulusu2021demand,Rimjha22AirRestrict} and time spent on inter-modal transfers~\cite{Tarafdar2019nortCalifornia,Rimjha2021factorsVertiport,Qu2024demand} (e.g., walking, cycling, car, taxi, shuttle, bus). Based on the projected passenger value of time, willingness to pay~\cite{Rimjha2021factorsVertiport,Rimjha22AirRestrict} for AAM services can be assessed. It was concluded that willingness to pay in a zone is highly correlated with the density of wealthy populations~\cite{Rimjha22AirRestrict}.

		\textbf{Evaluating Feasibility:}
		The primary objective of this task is to ensure that potential vertiport locations are not only situated in areas with high demand, but also in areas that are practical, physically and technically suitable to support AAM operations.
		
		To evaluate \textit{Available Locations}, the required vertiport size~\cite{Tarafdar2019nortCalifornia, Rimjha2021LA, Guo2024vtolSite} must be defined based on the projected AAM demand.
		A detailed analysis of neighborhood categories and land use types should be performed~\cite{Tarafdar2019nortCalifornia, Jeong2021kMeansVertiport} to exclude areas where it is not possible to construct transportation infrastructure. Next, one should identify land available for sale~\cite{Tarafdar2019nortCalifornia} and exclude the areas where topography~\cite{Tarafdar2019nortCalifornia} does not allow for vertiport construction.
		
		\textit{No Obstacles}~\cite{Guo2024vtolSite} are essential to ensure unobstructed flight paths and safe vertiport operations. As indicated by Rakas~et~al.~\cite{Rakas2021fleetSelection}, the AAM flight envelope depends on the \textit{AAM Aircraft Capabilities}. Moreover, the selected aircraft must be able to travel the distance between vertiports and transport the required number of people. Additionally, aircraft dimensions and power requirements directly influence the design and layout of vertiport infrastructure.
		
		\textbf{Considering Regulations:}
		This task aims at ensuring that chosen vertiport locations comply with all relevant regulatory frameworks.
		
		\textit{Airspace} regulations aim to maintain the safety and efficiency of existing aviation systems following the integration of AAM operations. One critical factor is the \textit{Load on Air Traffic Control (ATC)}~\cite{Vascik2021interoperability}, which assesses the capacity of ATC systems to manage coordination between commercial aviation and AAM traffic around airports. Additionally, the presence of \textit{No-fly Zones (NFZ)}~\cite{Jeong2021kMeansVertiport,Rimjha2021LA,Rimjha22AirRestrict,Vascik2021interoperability} must be considered to avoid restricted areas that could impede vertiport operations or conflict with existing airspace uses. Furthermore, maintaining \textit{Separation from Other Aircraft}~\cite{Vascik2021interoperability,Souza2023confManag} is essential to prevent mid-air collisions and ensure safe distances between AAM flights and other aviation traffic.
		
		\textit{Aircraft} regulations involve ensuring that AAM aircraft meet all required safety and operational standards set by aviation authorities~\cite{Rakas2021fleetSelection}. Currently, there is no specified procedure for AAM aircraft certification; therefore, most studies rely on existing frameworks such as Title 14 of the Code of Federal Regulations (14 CFR) part 135 to assess compliance and operational readiness~\cite{FAA2024_14CFR_I_SubG}.
		
		\textit{Vertiport} regulations establish specific operational guidelines that vertiport facilities must adhere to. Existing work~\cite{Tarafdar2019nortCalifornia} references only legacy helipad regulations.
		
		Lastly, \textit{Noise} regulations focus on minimizing the environmental impact of AAM operations, particularly during the take-off and landing phases, which are the noisiest. It is important to note that permitted noise levels vary depending on the land use types of surrounding areas~\cite{Jeong2021kMeansVertiport}, necessitating careful consideration to ensure compliance and community acceptance.
		
		\textbf{Analyzing Economic Viability:}
		In this task, the economical viability of the selected vertiport locations is inspected.
		
		\textit{Capital Expense (CAPEX)} involves the initial investments required to establish vertiport infrastructure and acquire necessary assets. Key components of CAPEX include New Infrastructure and Infrastructure Reuse. Investments in New Infrastructure depends on Land Cost~\cite{Tarafdar2019nortCalifornia}, which includes the expenses related to acquiring land for vertiport construction. Additionally, the required Vertiport Capacity~\cite{Rimjha2021LA} defines the necessary land size and facilities needed to accommodate projected AAM demand. Finally, Aircraft Cost~\cite{Tarafdar2019nortCalifornia} represents the investment in acquiring AAM vehicles. On the other hand, it is possible to reduce CAPEX through Infrastructure Reuse~\cite{Lim2019Seoul,Jeong2021kMeansVertiport,Guo2024vtolSite}. This involves repurposing existing helipads, rooftop parking and other transportation facilities. Infrastructure reuse not only lowers CAPEX but also accelerates the deployment of vertiport services by utilizing pre-existing structures.
		
		\textit{Operational Expenses (OPEX)} include the ongoing costs associated with the daily functioning of AAM facilities. The first key component of OPEX is Operational Cost per Flight~\cite{Shin2022Skyport}, which encompasses expenses related to fuel, maintenance, and staffing for each individual flight. Additionally, Vertiport Utilization~\cite{Rimjha2021factorsVertiport} and Aircraft Utilization~\cite{Tarafdar2019nortCalifornia} are critical factors that influence OPEX. Efficient utilization of AAM fleet and vertiport infrastructure minimizes idle times and reduces operational costs. 
		Finally, charges for each landing at the vertiport (so-called  Landing Fees~\cite{Rimjha2021LA}) are another significant operational cost. Note that depending on the business model, these fees can also be seen as revenue generators rather than OPEX.

		\textit{Revenue} generation is a vital aspect of economic viability. Multiple papers ~\cite{Shin2022Skyport,Chae2023vertLocations,Guo2024vtolSite} predict revenue based on Combined Passenger Fares collected from users for their AAM journeys. 
		
		\subsubsection{Phase 2 - Corridor Formulation}\label{sec:corridor_factors}
		
		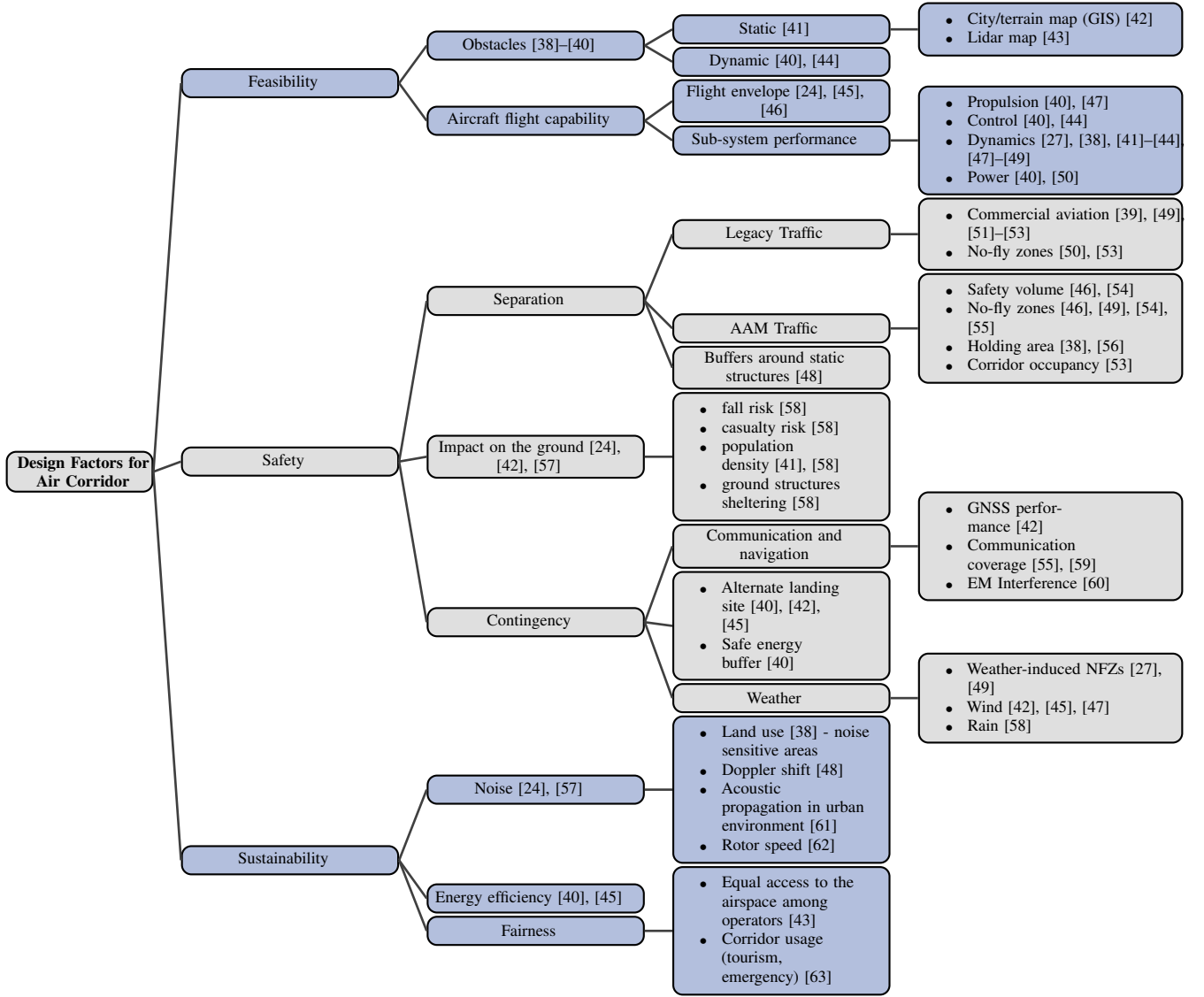
\begin{figure*}[tbph!]
    \centering
    \begin{forest}
        for tree={
            grow=east,
            reversed=true,
            anchor=base west,
            parent anchor=east,
            child anchor=west,
            base=left,
            font=\scriptsize, %
            rectangle,
            draw=black, %
            rounded corners,
            align=left,
            minimum width=2em, %
            edge+={darkgray, line width=1pt},
            s sep=1pt, %
            inner xsep=1pt, %
            inner ysep=2pt, %
            line width=0.8pt,
            text width=9em, %
        },
        [\textbf{\parbox{9em}{\centering Design Factors for\\Air Corridor}}, text width=6em, fill=lightgray!50
            %% Factor Category 1
            [\parbox{12em}{\centering Feasibility}, fill=NavyBlue!30
                [\parbox{12em}{\centering Obstacles\cite{Lee2021holding,Gray2023airport,Hagag2024energy}}, fill=NavyBlue!30
                    [\parbox{12em}{\centering Static\cite{Slama2022corrRoadmap}}, fill=NavyBlue!30
                        [\parbox{16em}{
                            \begin{itemize}
                            \item City/terrain map (GIS)~\cite{Causa2023path}                            
                            \item Lidar map~\cite{Tang2021autoHDflight}
                            \end{itemize}
                            }, text width=11em, fill=NavyBlue!30    
                        ]
                    ]
                    [\parbox{12em}{\centering Dynamic~\cite{Ge2019hierarchical,Hagag2024energy}}, fill=NavyBlue!30
                    ]
               ]
               [\parbox{12em}{\centering Aircraft flight capability}, fill=NavyBlue!30
                    [\parbox{12em}{\centering Flight envelope \cite{Lou2021RTT,Jeong2021kMeansVertiport,Neto2022trajEvaluation}}, fill=NavyBlue!30
                    ]
                    [\parbox{12em}{\centering Sub-system performance}, fill=NavyBlue!30
                        [\parbox{16em}{
                            \begin{itemize}
                            \item Propulsion~\cite{Hagag2024energy,Lu2024vtolTraj}
                            \item Control \cite{Ge2019hierarchical,Hagag2024energy}
                            \item Dynamics \cite{Falck2018acoustic, Ge2019hierarchical,Lee2021holding,Tang2021autoHDflight,Ribeiro2022Geofence,Slama2022corrRoadmap,Peng2022hierarchical,Causa2023path,Lu2024vtolTraj}
                            \item Power \cite{Hagag2024energy,Kotwicz2022restrictions}
                            \end{itemize}
                            }, text width=11em, fill=NavyBlue!30  
                        ]
                    ]
               ]
            ]    
            %% Factor Category 2
            [\parbox{12em}{\centering Safety}, fill=lightgray!50
                [\parbox{12em}{\centering Separation}, fill=lightgray!50
                    [\parbox{12em}{\centering Legacy Traffic}, fill=lightgray!50
                        [\parbox{16em}{
                            \begin{itemize}
                            \item Commercial aviation~\cite{Jiang2022metric,Ribeiro2022Geofence,Verma2022corridors,Kallies2023Frankfurt,Gray2023airport}
                            \item No-fly zones~\cite{Kallies2023Frankfurt,Kotwicz2022restrictions}
                            \end{itemize}
                            }, text width=11em, fill=lightgray!50   
                        ]
                    ]
                    [\parbox{12em}{\centering AAM Traffic}, fill=lightgray!50
                        [\parbox{16em}{
                            \begin{itemize}
                            \item Safety volume~\cite{Thompson2023OpVolume,Neto2022trajEvaluation}
                            \item No-fly zones~\cite{Garcia2022channel,Thompson2023OpVolume,Ribeiro2022Geofence,Neto2022trajEvaluation}
                            \item Holding area~\cite{Lee2021holding,Song2021optSched}
                            \item Corridor occupancy~\cite{Kallies2023Frankfurt}
                            \end{itemize}
                            }, text width=11em, fill=lightgray!50    
                        ]
                    ]
                    [\parbox{12em}{\centering Buffers around static structures~\cite{Falck2018acoustic}}, fill=lightgray!50
                    ]
                ]
                [\parbox{12em}{\centering Impact on the ground~\cite{Glaab2019noise,Jeong2021kMeansVertiport,Causa2023path}}, fill=lightgray!50   
                        [\parbox{10em}
                            {
                            \begin{itemize}
                            \item fall risk~\cite{Ye2024airRoute}
                            \item casualty risk~\cite{Ye2024airRoute}
                            \item population density~\cite{Slama2022corrRoadmap,Ye2024airRoute}
                            \item ground structures sheltering~\cite{Ye2024airRoute}
                            \end{itemize}
                            }, text width=9em, fill=lightgray!50   
                        ]
                ]
                [\parbox{12em}{\centering Contingency}, fill=lightgray!50  
                    [\parbox{12em}{\centering Communication and navigation}, fill=lightgray!50
                        [\parbox{12em}{
                            \begin{itemize}
                                \item GNSS performance~\cite{Causa2023path}
                                \item Communication coverage~\cite{Garcia2022channel,Park2023comm}
                                \item EM Interference~\cite{Nguyen2023radiation}
                            \end{itemize}
                            }, text width=11em, fill=lightgray!50  
                        ]
                    ]
                    [\parbox{10em}
                            {
                            \begin{itemize}
                            \item Alternate landing site~\cite{Lou2021RTT,Causa2023path,Hagag2024energy}
                            \item Safe energy buffer~\cite{Hagag2024energy}
                            \end{itemize}
                            }, text width=9em, fill=lightgray!50  
                    ]
                    [\parbox{12em}{\centering Weather}, fill=lightgray!50
                        [\parbox{16em}
                            {
                            \begin{itemize}
                            \item Weather-induced NFZs~\cite{Peng2022hierarchical,Ribeiro2022Geofence}
                            \item Wind~\cite{Lou2021RTT,Causa2023path,Lu2024vtolTraj}
                            \item Rain~\cite{Ye2024airRoute}
                            \end{itemize}
                            }, text width=11em, fill=lightgray!50    
                        ]
                ]
            ]
            ]
            %% Factor Category 3
            [\parbox{12em}{\centering Sustainability}, fill=NavyBlue!30
                [\parbox{12em}{\centering Noise~\cite{Glaab2019noise,Jeong2021kMeansVertiport}}, fill=NavyBlue!30
                    [\parbox{12em}{
                        \begin{itemize}
                        \item Land use~\cite{Lee2021holding} - noise sensitive areas
                        \item Doppler shift~\cite{Falck2018acoustic}
                        \item Acoustic propagation in urban environment~\cite{Gao2023acoustic}
                        \item Rotor speed~\cite{Cho2024lowNoise}
                        \end{itemize}
                        }, text width=9em, fill=NavyBlue!30   
                    ]
                ]
                [\parbox{12em}{\centering Energy efficiency~\cite{Lou2021RTT,Hagag2024energy}}, fill=NavyBlue!30
                ]
                %[\parbox{12em}{\centering Land use}, fill=lime!30
                %]
                [\parbox{12em}{\centering Fairness}, fill=NavyBlue!30
                    [\parbox{12em}
                            {
                            \begin{itemize}
                            \item Equal access to the airspace among operators~\cite{Tang2021autoHDflight}
                            \item Corridor usage (tourism, emergency)~\cite{Paradis2022visualizing}
                            \end{itemize}
                            }, text width=9em, fill=NavyBlue!30 
                    ]
                ]
            ]
        ]
    \end{forest}
    \caption{\centering Taxonomy of design factors for air corridor formulation}
    \label{fig:corridor-design-factors}
\end{figure*}

		Early papers considered the shortest path between the vertiports as the simplest approach to represent corridors with obvious limited applicability. Next, it was suggested to analyze historical data of helicopter flights to define de-facto existing corridors~\cite{Paradis2022visualizing}. However, the expected scale of AAM operations called for a more holistic approach to defining potential corridors.  
		
		Fig.~\ref{fig:corridor-design-factors} illustrates a comprehensive taxonomy of design factors for formulating effective corridors in AAM systems. Given the vastness of airspace, the corridor formulation process employs a systematic approach to narrow down viable options by gradually removing unsuitable segments. This process is divided into three tasks: 
		\begin{enumerate}
			\item \textbf{Feasibility} check aims at eliminating infeasible flight trajectories.
			\item \textbf{Safety} check focuses on removing areas where flying is unsafe.
			\item \textbf{Sustainability} assessment seeks to indicate corridor candidates that satisfy a broad range of sustainability requirements.
		\end{enumerate}
		Each task encompasses specific factors that collectively ensure the selection of optimal corridors capable of supporting safe, efficient, and environmentally responsible AAM operations.
		
		\textbf{Feasibility:}
		The first task focuses on identifying and eliminating flight paths that are impractical or impossible for AAM operations. 
		
		Multiple contributions~\cite{Lee2021holding,Gray2023airport,Hagag2024energy} indicate \textit{Obstacles} as the primary factor to identify infeasible trajectories. Furthermore, obstacles can be categorized into static and dynamic obstacles. The static obstacles~\cite{Slama2022corrRoadmap} include permanent physical barriers such as city and terrain that can be obtained from a Geographic Information System (GIS)~\cite{Causa2023path} or reconstructed with on-board sensory equipment such as LiDAR~\cite{Tang2021autoHDflight}. On the other hand, dynamic obstacles~\cite{Ge2019hierarchical,Hagag2024energy} refer to temporary or moving objects that can affect flight paths, necessitating (i) a strategic avoidance of these areas or (ii) real-time adjustments and continuous monitoring to maintain operational integrity.
		
		\textit{Aircraft Flight Capability} evaluates the technical performance of AAM aircraft to ensure they can operate effectively within the designated corridors. This can be done based on assessing the Flight Envelope~\cite{Lou2021RTT,Jeong2021kMeansVertiport,Neto2022trajEvaluation}, which describes the operational limits of the aircraft in terms of altitude, speed, and maneuverability. Alternatively, a more thorough approach can include the performance assessment for the aircraft sub-systems: propulsion~\cite{Hagag2024energy,Lu2024vtolTraj}, control systems~\cite{Ge2019hierarchical,Hagag2024energy}, dynamics~\cite{Falck2018acoustic,Ge2019hierarchical,Lee2021holding,Tang2021autoHDflight,Ribeiro2022Geofence,Slama2022corrRoadmap,Peng2022hierarchical,Causa2023path,Lu2024vtolTraj}, and power systems~\cite{Hagag2024energy,Kotwicz2022restrictions}. These sub-systems are vital for ensuring that AAM aircraft can reliably operate within the established flight corridors, maintaining performance standards and operational readiness.
		
		\textbf{Safety:}
		The second task emphasizes the importance of safety by removing segments of airspace that pose risks to (i) other aircraft, (ii) ground, and (iii) passengers on board. This step is aligned with the \textit{Safety} category and encompasses \textit{Separation}, \textit{Impact on the Ground}, and \textit{Contingency} factors, respectively.
		
		\textit{Separation} ensures that AAM traffic maintains adequate distances from legacy aviation traffic, other AAM flights, and adds an extra safety margin to the static structures ensuring that the corridors are not only feasible but also pose no risk of collision. Separation from legacy traffic, such as commercial or general aviation, is a popular topic among researchers~\cite{Jiang2022metric,Ribeiro2022Geofence,Verma2022corridors,Kallies2023Frankfurt,Gray2023airport}. Sometimes it takes a form of adhering to established \textit{No-fly Zones (NFZ)}~\cite{Kallies2023Frankfurt,Kotwicz2022restrictions}. Separation from AAM Traffic involves establishing safety volumes around aircraft~\cite{Thompson2023OpVolume,Neto2022trajEvaluation} and managing corridor occupancy~\cite{Kallies2023Frankfurt} that can be done via establishing no-fly zones~\cite{Garcia2022channel,Thompson2023OpVolume,Ribeiro2022Geofence,Neto2022trajEvaluation}. As it is expected to have a higher density of operating AAM aircraft around vertiports, several papers suggest provisioning of holding areas~\cite{Lee2021holding,Song2021optSched} to prevent mid-air collisions and ensure safe operational distances.
		
		\textit{Impact on the Ground} addresses the potential risks associated with AAM operations affecting populated areas~\cite{Causa2023path}. Early contributions~\cite{Glaab2019noise,Jeong2021kMeansVertiport} favored flying over water (e.g., rivers) to minimize flying overhead people and, consequently, lower the ground risk. A more elaborate approach includes an assessment of casualty risk~\cite{Ye2024airRoute} that is defined by factors such as fall risk~\cite{Ye2024airRoute} and population density~\cite{Slama2022corrRoadmap,Ye2024airRoute}. Additionally, the casualty risk assessment can consider environmental factors such as sheltering provided by ground structures~\cite{Ye2024airRoute}.
		
		\textit{Contingency} planning involves preparing for unforeseen events to ensure operational resilience. This includes assessing Communication and Navigation System (CNS) performance: Global Navigation Satellite System (GNSS) accuracy and availability~\cite{Causa2023path}, cellular communication coverage~\cite{Garcia2022channel,Park2023comm}, and the ability to deal with electromagnetic (EM) interference~\cite{Nguyen2023radiation}. Following the EASA requirement, several papers consider the assignment of alternate landing sites~\cite{Lou2021RTT,Causa2023path,Hagag2024energy}. Moreover, it is important to ensure maintaining a safe battery charge buffer~\cite{Hagag2024energy} allowing for reaching the alternate landing site in case of emergency. Finally, weather must be considered. This can be done via introducing weather-induced no-fly zones~\cite{Peng2022hierarchical,Ribeiro2022Geofence} or via considering effects of wind~\cite{Lou2021RTT,Causa2023path,Lu2024vtolTraj} and rain~\cite{Ye2024airRoute} on the AAM aircraft.
		
		\textbf{Sustainability:}
		This task focuses on evaluating the long-term viability and environmental impact of the proposed AAM corridors. It deals with \textit{Noise}, \textit{Energy Efficiency}, and \textit{Fairness} factors.
		
		\textit{Noise} management and the acoustic footprint of AAM operations is a well-studied topic. Early contributions~\cite{Glaab2019noise,Jeong2021kMeansVertiport} favored flying over water (e.g., rivers) as this minimizes the number of people affected by noise. A more detailed approach considers noise-sensitive areas~\cite{Lee2021holding} (mostly defined by the area land use category) and precise acoustic propagation modeling taking into account Doppler shifts~\cite{Falck2018acoustic}, effects introduced by urban environments~\cite{Gao2023acoustic} (multipath propagation, attenuation, diffraction etc.), and dependency on the rotor speed~\cite{Cho2024lowNoise}.
		
		\textit{Energy Efficiency}~\cite{Lou2021RTT,Hagag2024energy} addresses the need for sustainable energy use in AAM operations, focusing on optimizing the aircraft energy consumption. 
		
		\textit{Fairness} ensures equitable access to airspace among different operators~\cite{Tang2021autoHDflight}. This involves implementing policies that prevent monopolistic practices and ensure that all stakeholders have fair opportunities to utilize the corridors. On the other hand, different operations or corridors may have different priorities and requirements~\cite{Paradis2022visualizing}. For example, the general public tends to better tolerate noise created by an AAM ambulance than by commercial or tourist flights~\cite{LINTAN2021acceptance}.
		
		\subsubsection{Phase 3 - AAM Operations}
		
		\begin{figure*}[tbph!]
    \centering
    \begin{forest}
        for tree={
            grow=east,
            reversed=true,
            anchor=base west,
            parent anchor=east,
            child anchor=west,
            base=left,
            font=\scriptsize, %
            rectangle,
            draw=black, %
            rounded corners,
            align=left,
            minimum width=2em, %
            edge+={darkgray, line width=1pt},
            s sep=1pt, %
            inner xsep=1pt, %
            inner ysep=2pt, %
            line width=0.8pt,
            text width=9em, %
        },
        [\textbf{\parbox{9em}{\centering Design Factors for AAM Operations}}, text width=6em, fill=lightgray!50
            %% Factor Category 1
            [\parbox{12em}{\centering Flight demand}, fill=NavyBlue!30
               [\parbox{12em}{\centering Commute demand~\cite{Rimjha2021factorsVertiport,Souza2023confManag}}, fill=NavyBlue!30
                         [\parbox{12em}{
                            \begin{itemize}
                            \item Proximity to vertiports~\cite{Conrad2024vertManagement}
                            \item Passenger behavior~\cite{Conrad2024vertManagement}
                            \end{itemize}
                            }, fill=NavyBlue!30                         
                    ]
               ]
               [\parbox{12em}{\centering Stochastic modeling}, fill=NavyBlue!30
                    [\parbox{12em}{
                            \begin{itemize}
                            \item Uniform~\cite{Song2021optSched}
                            \item Poisson~\cite{Pooladsanj2023VertiSync}
                            \item Considers peak hours~\cite{Lee2022DCB,Pooladsanj2023VertiSync}
                            \end{itemize}
                            }, fill=NavyBlue!30   
                    ]
               ]
            ]  
            %% Factor Category 2
            [\parbox{12em}{\centering Capacity}, fill=lightgray!50
                [\parbox{12em}{\centering Vertiport Capacity}, fill=lightgray!50
                    [\parbox{12em}{\centering Infrastructure}, fill=lightgray!50
                        [\parbox{16em}{
                            \begin{itemize}
                            \item Number of pads~\cite{Lee2022DCB,Kallies2023Frankfurt,Pooladsanj2023VertiSync,Chen2024arrivalManag}
                            \item Parking stalls~\cite{Rimjha2021factorsVertiport,Pooladsanj2023VertiSync,Schweiger2023wind}
                            \item Charging~\cite{Rimjha2021factorsVertiport}
                            \end{itemize}
                            }, text width=11em,fill=lightgray!50    
                        ]
                    ]
                        [\parbox{12em}{\centering Weather-induced vertiport closure~\cite{Schweiger2023wind}}, fill=lightgray!50
                    ]
                    [\parbox{12em}{\centering Operations}, fill=lightgray!50
                        [\parbox{16em}{
                            \begin{itemize}
                            \item Pads use (landing and take-off)~\cite{Kleinbekman2020rolling,Kallies2023Frankfurt,Pooladsanj2023VertiSync}
                            \item Temporal separation between pad uses~\cite{Kleinbekman2018arrScheduling,Kleinbekman2020rolling,Song2021optSched,Kallies2023Frankfurt,Pooladsanj2023VertiSync,EspejoDiaz2023}
                            \item Service time~\cite{Rimjha2021factorsVertiport,Schweiger2023wind}
                            \item Taxiing~\cite{Rimjha2021factorsVertiport,Schweiger2023wind,EspejoDiaz2023}
                            \item Boarding~\cite{Rimjha2021factorsVertiport,Schweiger2023wind,EspejoDiaz2023}
                            \item Charging~\cite{Rimjha2021factorsVertiport}
                            \item Time on pad~\cite{Rimjha2021factorsVertiport,Kallies2023Frankfurt,Schweiger2023wind,EspejoDiaz2023}
                            \end{itemize}
                            }, text width=11em,fill=lightgray!50  
                        ]
                    ]
                ]
                [\parbox{12em}{\centering Corridor Capacity}, fill=lightgray!50
                   [\parbox{12em}{\centering Configuration}, fill=lightgray!50
                        [\parbox{16em}{
                            \begin{itemize}
                            \item Final approach fix~\cite{Kleinbekman2018arrScheduling,Kleinbekman2020rolling}
                            \item Holding points~\cite{Kleinbekman2020rolling,Chen2024arrivalManag}
                            \item Dynamic no-fly zones~\cite{Lou2021RTT,Ribeiro2022Geofence}
                            \end{itemize}
                            }, text width=11em,fill=lightgray!50    
                        ]
                    ]
                    [\parbox{12em}{\centering Reservation method}, fill=lightgray!50
                        [\parbox{16em}{
                            \begin{itemize}
                            \item Segment (moving keep-out zone)~\cite{Souza2023confManag}
                            \item Full corridor per flight~\cite{Pooladsanj2023VertiSync}
                            \item Time/space separation~\cite{Kleinbekman2018arrScheduling,Kleinbekman2020rolling,Song2021optSched}
                            \end{itemize}
                            }, text width=11em,fill=lightgray!50    
                        ]
                    ]
                ]
            ]       
            %% Factor Category 3
            [\parbox{12em}{\centering Operations Management }, fill=NavyBlue!30
                [\parbox{12em}{\centering Corridor management}, fill=NavyBlue!30
                    [\parbox{12em}{\centering Rerouting~\cite{Lou2021RTT,Souza2023confManag,Suzuki2022flightReplan}}, fill=NavyBlue!30]
                    [\parbox{12em}{\centering Assignment}, fill=NavyBlue!30
                        [\parbox{16em}{
                            \begin{itemize}
                            \item Traffic density/complexity~\cite{Wang2021assignment,Wang2023trafAssign}
                            \item Corridor segment scheduling~\cite{Souza2023confManag}
                            \end{itemize}
                            }, text width=11em,fill=NavyBlue!30 
                        ]
                    ]
                    [\parbox{12em}{\centering Repositioning~\cite{Rimjha2021factorsVertiport,Suzuki2022flightReplan}}, fill=NavyBlue!30]    
                ]
                [\parbox{12em}{\centering Vertiport management}, fill=NavyBlue!30
                    [\parbox{12em}{\centering Arrival management}, fill=NavyBlue!30
                        [\parbox{16em}{
                            \begin{itemize}
                            \item Required time of arrival~\cite{Kleinbekman2018arrScheduling,Kleinbekman2020rolling,EspejoDiaz2023}
                            \item Priority (e.g., low battery)~\cite{Kleinbekman2018arrScheduling,Kleinbekman2020rolling,Conrad2023trafManag}
                            \item Arrivals per time slot~\cite{Lee2022DCB,Pooladsanj2023VertiSync,Chen2024arrivalManag}
                            \item Vertiport infrastructure occupancy~\cite{Rimjha2021factorsVertiport}
                            %\item Operating hours
                            \end{itemize}
                            }, text width=11em,fill=NavyBlue!30   
                        ]
                    ]
                    [\parbox{12em}{\centering Departure management}, fill=NavyBlue!30
                        [\parbox{16em}{
                            \begin{itemize}
                            \item Flight authorization~\cite{Lee2022DCB,Conrad2023trafManag}
                            \item Departure delays~\cite{Lee2022DCB,Chen2024arrivalManag}
                            \end{itemize}
                            }, text width=11em, fill=NavyBlue!30    
                        ]
                    ]
                ]
                ]
            ]        
        ]
    \end{forest}
    \caption{\centering Taxonomy of design factors for AAM operations}
    \label{fig:corridor-ops-design-factors}
\end{figure*}
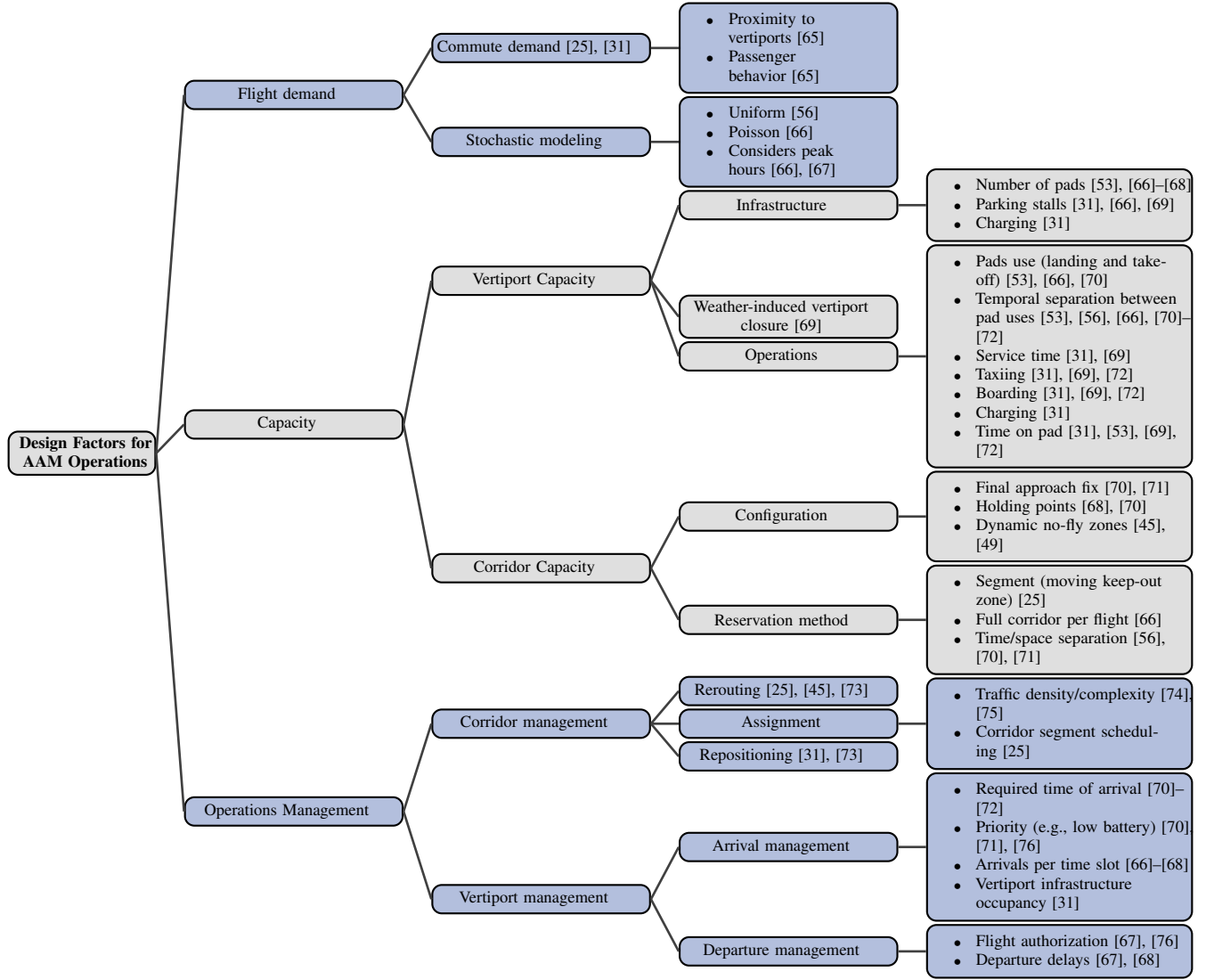

		Following the establishment of a vertiport network and the design of air corridors, the final phase focuses on ensuring the seamless and sustainable operation of AAM services. This phase is critical for maintaining the reliability and efficiency of AAM operations within the established infrastructure. Fig.~\ref{fig:corridor-ops-design-factors} presents a taxonomy of factors essential for operational management in AAM systems. The operational phase is systematically divided into three primary tasks: \textbf{Flight Demand} assessment, \textbf{Capacity} estimation, and \textbf{Operations Management}. Each task depends on specific factors that collectively facilitate the efficient, reliable, and sustainable functioning of AAM operations. The first two tasks can also rely on the outputs of the previous design phases. For instance, the demand analysis should be performed in phase 1 while the projected capacity of vertiports and corridors can re-use results from phases 1 and 2, respectively.
		
		\textbf{Flight Demand:}
		This task involves assessing and forecasting the demand for AAM services to align operational capabilities with passenger needs and ensure optimal resource allocation. This assessment can be done based on \textit{Commute demand}. Several papers~\cite{Rimjha2021factorsVertiport,Souza2023confManag} calculate the  AAM service demand as a percentage of the total transportation demand. The demand can also be predicted based on the proximity of potential users to vertiport locations and their behavioral patterns regarding AAM services~\cite{Conrad2024vertManagement}. Alternatively, \textit{stochastic modeling} can be used toaccount for uncertainty in flight demand. By employing different distribution models such as uniform and Poisson~\cite{Song2021optSched,Pooladsanj2023VertiSync}, and considering peak hours~\cite{Lee2022DCB,Pooladsanj2023VertiSync}, operators can better anticipate demand fluctuations. While this approach is less realistic, it gives more freedom for testing the AAM system behavior in unseen circumstances.
		
		\textbf{Capacity:}
		The second critical task is capacity estimation, focusing on understanding the limits of the vertiports and corridor abilities. 
		
		\textit{Vertiport capacity} encompasses both infrastructure and operational aspects. The infrastructure component involves evaluating the number of pads~\cite{Lee2022DCB,Kallies2023Frankfurt,Pooladsanj2023VertiSync,Chen2024arrivalManag}, parking stalls~\cite{Rimjha2021factorsVertiport,Pooladsanj2023VertiSync,Schweiger2023wind}, and charging facilities~\cite{Rimjha2021factorsVertiport} to determine the vertiport's ability to handle incoming and outgoing traffic. On the operational side, capacity can be influenced by the pad usage approach~\cite{Kleinbekman2020rolling,Kallies2023Frankfurt,Pooladsanj2023VertiSync} in which a given pad can be exclusively used for landing or take-off (a safe choice) or the use can be allocated dynamically. Temporal factors such as time spent for taxiing, boarding, charging, and time spent on the pad~\cite{Rimjha2021factorsVertiport,Kallies2023Frankfurt,Schweiger2023wind,EspejoDiaz2023} also define the vertiport capacity. Another operational constraint is represented by Temporal separation between pad uses~\cite{Kleinbekman2018arrScheduling,Kleinbekman2020rolling,Song2021optSched,Kallies2023Frankfurt,Pooladsanj2023VertiSync,EspejoDiaz2023}. Additionally, it is necessary to assess the probability of weather-induced vertiport closures~\cite{Schweiger2023wind}.
		
		\textit{Corridor capacity} is defined by its configuration (geometry): distances for final approach fix points~\cite{Kleinbekman2018arrScheduling,Kleinbekman2020rolling} and number and locations of holding points~\cite{Kleinbekman2020rolling,Chen2024arrivalManag} will have an effect on capacity. Another factor limiting airspace capacity is dynamic no-fly zones~\cite{Lou2021RTT,Ribeiro2022Geofence} implemented to avoid, for example, areas with adverse weather conditions. Moreover, the airspace capacity is highly dependent on the used reservation method. When a corridor is reserved for the whole duration of the flight~\cite{Pooladsanj2023VertiSync}, it will result in a lower capacity than in the case of a partial reservation~\cite{Souza2023confManag} that can be implemented with a segment-based moving keep-out zone. Alternatively, reserving the corridor can be avoided by relying on time/space separation~\cite{Kleinbekman2018arrScheduling,Kleinbekman2020rolling,Song2021optSched}.
		
		\textbf{Operations Management}
		Due to the dynamic nature of the AAM demand and capacity, the next task focuses on harmonizing them to ensure efficient and smooth AAM operations.  
		Fig.~\ref{fig:corridor-ops-design-factors} depicts the management task consisting of \textit{Corridor Management} and \textit{Vertiport Management}. The primary corridor management goal is to identify (assign) the corridor suitable for the requested flight. As discussed earlier, it may be done for the complete corridor or at the segment level~\cite{Souza2023confManag}. This assignment task can be done based on the traffic density and complexity~\cite{Wang2021assignment,Conrad2023trafManag}. Other corridor management actions include rerouting ~\cite{Lou2021RTT,Souza2023confManag,Suzuki2022flightReplan} (changing the corridor configuration) and repositioning~\cite{Rimjha2021factorsVertiport,Suzuki2022flightReplan} (changing the target vertiport) flights~\cite{Lou2021RTT,Souza2023confManag} in response to appearance of the dynamic no-fly zones and vertiport closures (due to weather or full occupancy), respectively.  
		
		\begin{table*}[h!]
			\centering
			\caption{Notional separation values~\cite{Cotton2019separation}}
			\label{tab:separation}
			\begin{tabular}{|c|c|c|c|c|}
				\hline
				&& AFR & PA$_{\texttt{VFR}}$ & PA$_{\texttt{IFR}}$\\
				\hline
				\multirow{3}{40pt}{AFR}&L:&\cellcolor{NavyBlue!20}  $\omega$& 4000 feet $\approx$ 1200 m& 3 miles $\approx$ 4800 m\\
				&V:&\cellcolor{NavyBlue!20}  300 feet $\approx$ 90 m&450 feet $\approx$ 140 m & 1000 feet $\approx$ 300 m\\
				&S:&\cellcolor{NavyBlue!20}  300 feet $\approx$ 90 m&  1/4 mile $\approx$ 400 m& 2 miles $\approx$ 3200 m\\
				\hline
				\multirow{3}{40pt}{PA$_{\texttt{VFR}}$}&L:& 4000 feet $\approx$ 1200 m&  \multicolumn{2}{c|}{ }\\
				&V:& 450 feet $\approx$ 140 m& \multicolumn{2}{c|}{Well Clear}\\
				&S:&  1/4 mile $\approx$ 400 m&\multicolumn{2}{c|}{ }\\
				\hline\multirow{4}{40pt}{PA$_{\texttt{IFR}}$}&L:&  \multicolumn{3}{c|}{3 miles $\approx$ 4800 m} \\
				&V:&  \multicolumn{3}{c|}{1000 feet $\approx$ 300 m} \\
				&S:&  \multicolumn{3}{c|}{2 miles $\approx$ 3200 m} \\
				\hline
				\multicolumn{5}{l}{Note: wake separation criteria would increase these values when applicable}\\
				\multicolumn{5}{l}{L = Lateral separation}\\
				\multicolumn{5}{l}{V = Vertical separation}\\
				\multicolumn{5}{l}{S = Slow closure or longitudinal separation}\\		
				%		\hline
			\end{tabular}
			\vspace{-0.4cm}
		\end{table*}

		The vertiport management includes arrival and departure management. Effective arrival scheduling must take into account the required time of arrival~\cite{Kleinbekman2018arrScheduling,Kleinbekman2020rolling,EspejoDiaz2023} and current vertiport occupancy~\cite{Rimjha2021factorsVertiport}; prioritize flights with low battery levels~\cite{Kleinbekman2018arrScheduling,Kleinbekman2020rolling,Conrad2023trafManag}; and ensure safety via controlling the number of arrivals per time slot to maintain a safe and steady flow of incoming traffic~\cite{Lee2022DCB,Pooladsanj2023VertiSync,Chen2024arrivalManag}.

		Similarly, departure management focuses on flight authorization and mitigating departure delays~\cite{Lee2022DCB,Conrad2023trafManag,Chen2024arrivalManag}, ensuring that departures are orderly and timely, thereby maintaining the overall efficiency of AAM operations.
		
		\subsection{RQ2: Separation Definition Factors}\label{sec:RQ2}

		\begin{figure*}[]
    \centering
    \begin{forest}
        for tree={
            grow=east,
            reversed=true,
            anchor=base west,
            parent anchor=east,
            child anchor=west,
            base=left,
            font=\scriptsize, %
            rectangle,
            draw=black, %
            rounded corners,
            align=left,
            minimum width=2em, %
            edge+={darkgray, line width=1pt},
            s sep=1pt, %
            inner xsep=1pt, %
            inner ysep=2pt, %
            line width=0.8pt,
            text width=9em, %
        },
        [\textbf{\parbox{9em}{\centering Factors for Definition of Safe Separation}}, text width=6em, fill=lightgray!50
            %% Factor Category 1
            [\parbox{12em}{\centering Vehicle Characteristics}, fill=NavyBlue!30
               [\parbox{12em}{\centering Detect and Avoid (DAA)}, fill=NavyBlue!30
                         [\parbox{16em}{
                            \begin{itemize}
                            \item Detection range~\cite{Geister2018density}
                            \item Tracking~\cite{Cotton2019separation}
                            \end{itemize}
                            },  text width=11em,fill=NavyBlue!30                         
                    ]
               ]
               [\parbox{12em}{\centering Flight dynamics~\cite{Kim2022Geofence}}, fill=NavyBlue!30
                    [\parbox{16em}{
                            \begin{itemize}
                            \item Aircraft velocity~\cite{Geister2018density,Cummings2024airspace}
                            \item Relative velocity~\cite{Cotton2019separation,Gomaa2023windWake}
                            \end{itemize}
                            }, text width=11em,fill=NavyBlue!30
                    ]
               ]
            ]
            %% Factor Category 2
            [\parbox{12em}{\centering Total System Error (TSE)~\cite{Kim2023separation,Gordo2023layered}}, fill=lightgray!50
               [\parbox{12em}{\centering Navigation System Error (NSE)~\cite{Geister2018density}}, fill=lightgray!50
                         [\parbox{16em}{
                            \begin{itemize}
                            \item GNSS
                            \item INS
                            \end{itemize}
                            }, text width=11em,fill=lightgray!50                        
                    ]
               ]
               [\parbox{12em}{\centering Flight Technical Error (FTE)}, fill=lightgray!50
                    [\parbox{16em}{
                            \begin{itemize}
                            \item Communication~\cite{Geister2018density}
                            \item Control~\cite{Kim2022Geofence}
                            \item Weather~\cite{Vascik2021interoperability}
                            \item Pilotage
                            \end{itemize}
                            }, text width=11em,fill=lightgray!50    
                    ]
               ]
            ]
            %% Factor Category 3
            [\parbox{12em}{\centering Environmental }, fill=NavyBlue!30
               [\parbox{12em}{
               	\begin{itemize}
                            \item Wind~\cite{Kim2022Geofence,Gomaa2023windWake}
                            \item Wake vortices~\cite{Geister2018density,Vascik2021interoperability,Verma2022corridors,Gomaa2023windWake}
                            \item Noise~\cite{Gao2023acoustic}
                            \item Obstacles~\cite{Gray2023obstacles}
                            \end{itemize}
                            }, fill=NavyBlue!30
               ]
            ]
        ]
    \end{forest}
    \caption{\centering Taxonomy of factors for definition of safe separation distance}
    \label{fig:safe-sep-factors}
\end{figure*}
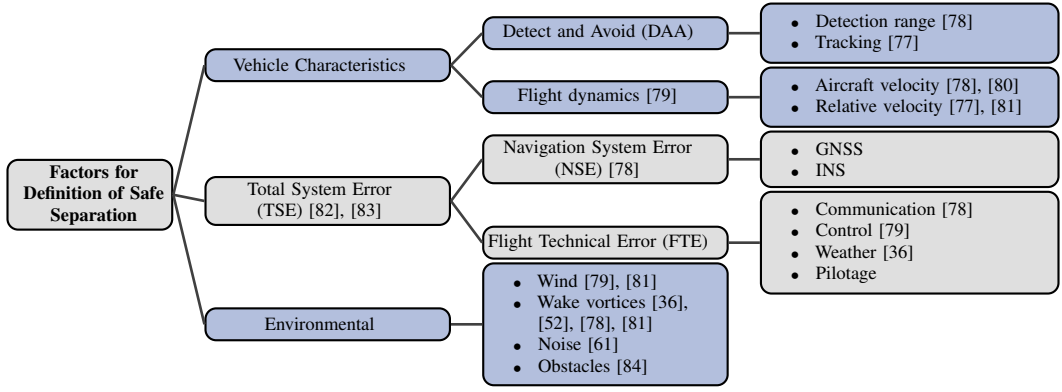

		Proper AAM separation standards are essential to ensure safety, prevent collisions, and optimize the use of airspace. It is widely accepted that the initial AAM operations will be performed by pilots following well-established Visual Flight Rules (VFR) or Instrument Flight Rules (IFR). In these cases, AAM will use the separation distances defined for Piloted Aircraft (PA) separations provided in Table~\ref{tab:separation} adapted from \cite{Cotton2019separation}\footnote{VFR flights operate without participating in any air traffic services and often without the knowledge of Air Traffic controller (ATC). The pilots of VFR flights are responsible for their own separation using "see and be seen" principles (based on subjective "Remain Well-Clear" which is not formally defined) and standard right of way rules. When operation of an aircraft under VFR is not safe, because the visual cues outside the aircraft are obscured by weather, IFR must be used instead. In IFR, ATC ensures appropriate separation of the aircraft from obstacles and other aircraft.}. However, matured AAM systems will require development of a new set of flight rules referred as Autonomous or Digital Flight Rules (AFR/DFR) relying on different separation standards that are still to be defined (highlighted in the table).  Note that in order to minimize the effects on the conventional flights, VFR and IFR traffic is given right of way in every encounter with AFR flights\footnote{For PA$_{\texttt{VFR}}$ encounters with AFR, well-clear should be used, however, we use a more formal definition suggested in the DAA Phase 1 Minimum Operational Performance Standards ~\cite{MOPS} adopted by FAA.}.
		
		{While Table~\ref{tab:separation} defines minimums for distinct vehicle pairings (e.g., AFR vs. Piloted Aircraft), comprehensive studies on the mixed-mode operational phase (where legacy piloted helicopters, semi-automated eVTOLs, and fully autonomous vehicles simultaneously share the same corridor) are  absent from the reviewed literature. We discuss the implications of this critical gap in Section~\ref{sec:discussion}.} 
		In this section, we focus on factors potentially influencing this separation distances.
		
		Fig.~\ref{fig:safe-sep-factors} presents a detailed taxonomy of factors important for defining safe separation distances in mature AAM systems. This taxonomy consists of three main categories: \textbf{Vehicle Characteristics}, \textbf{Total System Error (TSE)}, and \textbf{Environmental Factors}. Each category includes specific elements that must be considered for establishing effective separation distances promoting safe and efficient AAM operations. {From a quantitative perspective, the reviewed literature is relatively evenly distributed among these categories. Environmental factors received the most attention (addressed in 7 papers), closely followed by Vehicle Characteristics (5 papers) and TSE (5 papers).}
		
		\textbf{Vehicle Characteristics} focus on the properties and capabilities of AAM aircraft that influence separation requirements. This category includes \textit{Detect and Avoid (DAA)} systems and \textit{Flight Dynamics}. \textit{Detect and Avoid (DAA)} systems are crucial for identifying potential threats and avoiding collisions. Key components of DAA include Detection Range~\cite{Geister2018density}, which determines how far an aircraft can detect other objects, and {Tracking} capabilities~\cite{Cotton2019separation}, which allow continuous monitoring of nearby aircraft. Effective DAA systems enable AAM vehicles to make timely adjustments to maintain safe distances. \textit{Flight Dynamics} involves accounting for the {Aircraft Velocity}~\cite{Geister2018density,Cummings2024airspace} and {Relative Velocity}~\cite{Cotton2019separation,Gomaa2023windWake} between AAM vehicles. These factors are important for predicting future positions and ensuring that separation maneuvers can be executed smoothly.
		
		\textbf{Total System Error (TSE)} addresses the inaccuracies and uncertainties present in the overall AAM system. This category is divided into \textit{Navigation System Error (NSE)} and \textit{Flight Technical Error (FTE)}. \textit{Navigation System Error (NSE)}~\cite{Geister2018density} includes errors from Global Navigation Satellite Systems (GNSS)~\cite{Geister2018density} and Inertial Navigation Systems (INS)~\cite{Geister2018density}, which can affect the accuracy of an aircraft's position. Accurate navigation is essential for maintaining precise separation distances. \textit{Flight Technical Error (FTE)} encompasses errors related to {Communication} systems~\cite{Geister2018density}, {Control} systems~\cite{Kim2022Geofence}, {Weather} conditions~\cite{Vascik2021interoperability}, and {Pilotage}. Reliable communication ensures that positional and intent information is accurately shared between aircraft, while effective control systems enable precise maneuvers to maintain separation. Additionally, external factors like weather and pilot performance can introduce variability, making robust error management strategies necessary.
		
		\textbf{Environmental Factors} include external conditions that impact the effectiveness of separation measures. This category covers {Wind}, {Wake Vortices}, {Noise}, and {Obstacles}. \textit{Wind}~\cite{Kim2022Geofence,Gomaa2023windWake} can influence aircraft stability and maneuverability, requiring adjustments to separation distances based on wind speed and direction. {Wake Vortices}~\cite{Geister2018density,Vascik2021interoperability,Verma2022corridors,Gomaa2023windWake} generated by one aircraft can pose risks to the following aircraft, especially during takeoff and landing. Managing wake vortices involves maintaining adequate separation and implementing staggered flight schedules. {Noise}~\cite{Gao2023acoustic} considerations are important for minimizing the impact of AAM operations on surrounding communities, influencing how far an aircraft can operate from buildings in noise-sensitive areas. Lastly, {Obstacles}~\cite{Gray2023obstacles}, both permanent and temporary, must be accounted for to ensure that AAM vehicles maintain safe distances from physical barriers within the airspace.
		
		Together, these three categories \textbf{Vehicle Characteristics}, \textbf{Total System Error (TSE)}, and \textbf{Environmental Factors} provide a comprehensive framework for defining safe separation distances in AAM operations. By considering the factors depicted in the taxonomy presented in Fig.~\ref{fig:safe-sep-factors}, researchers aim to ensure that separation distances are effective, adaptable, and capable of maintaining high safety standards. %This approach is vital for the successful integration of AAM services into the existing airspace, ensuring that operations are both safe and efficient.

		\subsection{RQ3: Separation Definition Methodology}\label{sec:RQ3}		
		
		\begin{figure*}[tbph!]
	\centering
	\begin{forest}
		for tree={
			grow=east,
			reversed=true,
			anchor=base west,
			parent anchor=east,
			child anchor=west,
			base=left,
			font=\scriptsize, %
			rectangle,
			draw=black, %
			rounded corners,
			align=left,
			minimum width=2em, %
			edge+={darkgray, line width=1pt},
			s sep=1pt, %
			inner xsep=1pt, %
			inner ysep=2pt, %
			line width=0.8pt,
			text width=9em, %
		},
		[\textbf{\parbox{9em}{\centering Methodologies for Safe Separation Distance}}, text width=6em, fill=lightgray!50
		%% Factor Category 1
		[\parbox{12em}{\centering Model based}, fill=NavyBlue!30
		[\parbox{12em}{\centering Stochastic models}, fill=NavyBlue!30
		[\parbox{16em}{
			\begin{itemize}
				\item Collision risk~\cite{Gray2023obstacles,Tang2022confPlane}
				\item Geofence volumes~\cite{Gomaa2023windWake,Kim2022Geofence,Ribeiro2022Geofence}
			\end{itemize}
		},  text width=11em,fill=NavyBlue!30                         
		]
		]
		[\parbox{12em}{\centering Computational models}, fill=NavyBlue!30
		[\parbox{16em}{
			\begin{itemize}
				\item Fluid dynamics~\cite{Kim2022Geofence}
				\item Gas law~\cite{Cummings2024airspace}
			\end{itemize}
		}, text width=11em,fill=NavyBlue!30
		]
		]
		]
		%% Factor Category 2
		[\parbox{12em}{\centering Empirical}, fill=lightgray!50
		[\parbox{12em}{\centering Simulation}, fill=lightgray!50
		[\parbox{16em}{
			\begin{itemize}
				\item Fast-time simulation~\cite{Tang2022confPlane}
				\item Monte-carlo simulation~\cite{Gordo2023layered}
			\end{itemize}
		}, text width=11em,fill=lightgray!50                        
		]
		]
		[\parbox{12em}{\centering Flight data}, fill=lightgray!50
		[\parbox{16em}{
			\begin{itemize}
				\item Existing studies on VFR/IFR~\cite{Vascik2021interoperability}
				\item Rotorcraft flight data~\cite{Kim2023separation}
			\end{itemize}
		}, text width=11em,fill=lightgray!50    
		]
		]
		]
		]
	\end{forest}
	\caption{\centering Taxonomy of methodologies for safe separation distance %\zh{1) Real time? maybe smth like "flight data" or "measured"? 2) I changed the text a bit to highlight the branches (\textit{Simulation} etc.) as in other sections. but i could not do it for this branch}
	}
	\label{fig:safe-sep-methodologies}
\end{figure*}
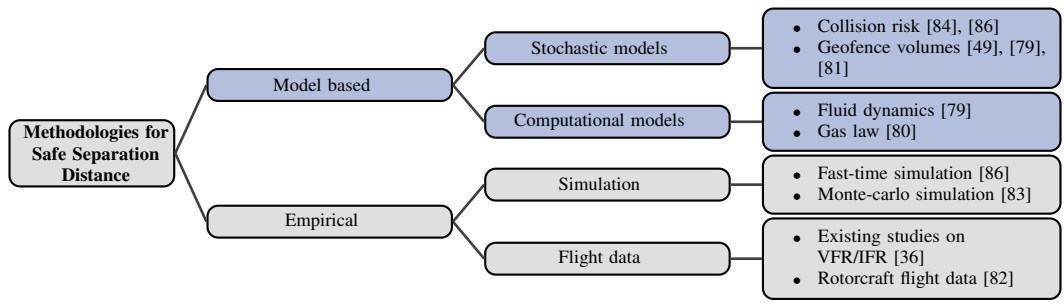

		This section focuses on the adopted methodologies for determining safe separation distances for eVTOL vehicles. Fig.~\ref{fig:safe-sep-methodologies} presents the taxonomy of different methodologies for generation of AAM safe separation distance. According to the literature review, we categorized the existing methodologies into two primary groups: \textbf{model-based} and \textbf{empirical}. Numerous research works utilize model-based methodologies because of their simplicity and ability to represent complex aircraft behavior through simplified abstraction. Within model-based group, \textit{stochastic models} utilize probabilistic estimates to assess collision risks or to delineate the protective geofence surrounding eVTOL in various flight phases of AAM mission. The authors in~\cite{Gray2023obstacles} evaluated a collision risk model of AAM aircraft with obstacles providing means to determine appropriate separation standards to be set for aircraft. The geofence is a virtual airspace boundary that prohibits (keep-out) and restricts (keep-in) access to some or all aircraft to a specific volume of airspace~\cite{Ribeiro2022Geofence}. The keep-in geofence assumes a safe volume (safety bubble) around the aircraft that encapsulates uncertainties in the flight state. The authors in~\cite{Gomaa2023windWake} considered ellipsoidal avoidance volumes around the aircraft accounting different atmospheric phenomena such as wake and wind turbulence. The authors in~\cite{Ribeiro2022Geofence} presents a dynamic keep-out geofencing framework in metroplex environment to ensure safety and performance of AAM traffic. Besides stochastic models, the \textit{computational models} (fluid dynamics, gas laws) form another category of model-based methodologies. The authors in~\cite{Cummings2024airspace} developed macroscopic air-traffic flow models that correlate the vehicle density and relative separation in the airspace to the frequency of conflict occurrence using gas-kinetic analogy (gas-law conflict prediction). The authors in~\cite{Kim2022Geofence} used a Computational Fluid Dynamics (CFD) model to derive safety geofence buffer sizing considering the vehicle dynamics, Guidance, Navigation and Control (GNC) uncertainties and wind effects.    
		
		The second major category of methodologies is \textbf{empirical}, supported by experimental observations and flight data. The safe separation distances established through comprehensive \textit{simulation} results constitute the initial category of empirical group, corroborated by synthetic or actual flight data. The authors in~\cite{Tang2022confPlane} ran fast-time simulations on 20 real-world flight scenarios to estimate the flight collision risks. The minimum horizontal, vertical and lateral separation distances were input to the simulation as per VFR/IFR conditions. The results provide useful insights on identifying, classifying and analyzing the loss of minimum separation. The authors in~\cite{Gordo2023layered} performed comprehensive Monte-Carlo simulations to estimate collision risks of a layered airspace in which Unmanned Aerial System (UAS) operates in the lower part of the layered airspace and AAM aircraft flies on the higher altitudes. The safe distance buffer among this two layers is concluded to be at least 10 meters through the simulation results.   
		
		Another aspect of empirical methodology is attributed to the extensive history of flight operations and \textit{rotorcraft flight data} that inform real-world separation criteria. The early adoption of AAM operations will fundamentally depend on established flight rules (VFR, IFR) and ATC directives~\cite{Vascik2021interoperability}. Hence, the safe separation studies performed already by federal civil aviation agencies (e.g., FAA, EASA and others) on rotorcraft vehicles will be instrumental for initial roll out of AAM operations. The authors in~\cite{Kim2023separation} determined the safety-assured minimum separation boundary for AAM operations by conducting safety risk assessment and analysis through simulations over the data gathered from GNSS/INS integrated navigation system. However, growth of AAM operations may easily overwhelm the ability of ATC to manage all of them efficiently. To ease the ATC load in managing increasing AAM eVTOLs, NASA proposes self-separation using AFR to deal with a high number of AAM flights in mixed airspace environment~\cite{wing2011autonomous, Cotton2019separation}.\\
		
		\section{Discussion and Future trends}\label{sec:discussion}
\begin{figure}[tbph!]
	\centering
	\begin{forest}
		for tree={
			grow=east,
			reversed=true,
			anchor=base west,
			parent anchor=east,
			child anchor=west,
			base=left,
			font=\scriptsize,
			rectangle,
			draw=black,
			rounded corners,
			align=left,
			minimum width=2em,
			edge+={darkgray, line width=1pt},
			s sep=1pt,
			inner xsep=1pt,
			inner ysep=2pt,
			line width=0.8pt,
			text width=8em,
		},
		[\textbf{\parbox{9em}{\centering Design Factors for\\Vertiport Location}}, text width=6em, fill=lightgray!50
		%%% Step 1 factors
		[\parbox{10em}{\centering Demand for AAM~\cite{Goyal2021demand}}, text width=7em, fill=NavyBlue!30
		[\parbox{12em}{\centering Origin-Destination}, fill=NavyBlue!30
		[\parbox{14em}{\centering Transport flow data}, text width=10em, fill=NavyBlue!30 
		[\parbox{20em}{
			\begin{itemize}
				\item Commute Survey~\cite{Tarafdar2019nortCalifornia,Lim2019Seoul,Jeong2021kMeansVertiport,Souza2023confManag}
				\item Transportation data~\cite{Bulusu2021demand,Peng2022hierarchical,Qu2024demand,Guo2024vtolSite}
			\end{itemize}
		}, text width=14em, fill=NavyBlue!30, rounded corners
		]
		]
		[\parbox{14em}{\centering Zones~\cite{Peng2022hierarchical,Chae2023vertLocations}}, text width=10em, fill=NavyBlue!30 
		[\parbox{20em}{
			\begin{itemize}
				\item Administrative districts~\cite{Rimjha2021factorsVertiport,Peng2022hierarchical,Shin2022Skyport}
				\item Population density~\cite{Rimjha22AirRestrict,Qu2024demand}
				\item Household income~\cite{Tarafdar2019nortCalifornia,Rimjha22AirRestrict}
				\item Land category~\cite{Rimjha2021factorsVertiport,Qu2024demand}
			\end{itemize}
		}, text width=14em, fill=NavyBlue!30, rounded corners]
		] 
		[\parbox{14em}{\centering \textcolor{red}{Population Distribution}}, text width=10em, fill=NavyBlue!30 
		[\parbox{20em}{
			\begin{itemize}
				\item \textcolor{red}{Cellular-based heatmap~\cite{Deville2014cellular,Bergroth2022helsinki}}
				%\item \textcolor{red}{Travel purpose} 
				%\item \textcolor{red}{Travel frequency} 
			\end{itemize}
		}, text width=14em, fill=NavyBlue!30, rounded corners]
		]           
		]
		[\parbox{12em}{\centering Comparison with Other Transportation Modes 
			\begin{itemize}
				\item Taxi~\cite{Chae2023vertLocations}
				\item Private Car~\cite{Lim2019Seoul,Jeong2021kMeansVertiport,Chae2023vertLocations}
				\item Public transport~\cite{Rimjha2021factorsVertiport}
			\end{itemize}
		}, fill=NavyBlue!30
		[\parbox{14em}{\centering Trip cost~\cite{Peng2022hierarchical,Shin2022Skyport}}, text width=10em, fill=NavyBlue!30
		[\parbox{20em}{
			\begin{itemize}
				\item Price per mile~\cite{Rimjha2021factorsVertiport}
				\item Minimum Fare~\cite{Rimjha2021factorsVertiport}
				\item Additional Fees~\cite{Rimjha2021factorsVertiport}
			\end{itemize}
		}, text width=14em, fill=NavyBlue!30, rounded corners
		]
		]
		[\parbox{14em}{\centering Trip time}, text width=10em, fill=NavyBlue!30
		[\parbox{20em}{
			\begin{itemize}
				\item Congestion~\cite{Rimjha2021factorsVertiport,Bulusu2021demand,Rimjha22AirRestrict}
				\item Inter-modal transfer~\cite{Tarafdar2019nortCalifornia,Rimjha2021factorsVertiport,Qu2024demand}\\(walking, cycling, car, taxi, bus)
			\end{itemize}
		}, text width=14em, fill=NavyBlue!30, rounded corners
		]
		]
		[\parbox{14em}{\centering Willingness to pay~\cite{Rimjha2021factorsVertiport,Rimjha22AirRestrict}}, text width=10em, fill=NavyBlue!30
		]                   
		[\parbox{14em}{\centering \textcolor{red}{Comfort}}, text width=10em,fill=NavyBlue!30
		[\parbox{20em}{
			\begin{itemize}
				\item \textcolor{red}{Quality of experience (QoE)} 
			\end{itemize}
		}, text width=14em, fill=NavyBlue!30, rounded corners
		]
		]
		[\parbox{14em}{\centering \textcolor{red}{Safety}}, text width=10em,fill=NavyBlue!30
		]
		]
		]
		%%% Step 2 factors
		[\parbox{10em}{\centering Feasibility}, text width=7em, fill=LightGray!50
		[\parbox{12em}{\centering Available Locations}, fill=LightGray!50	
		[\parbox{13em}{
			\begin{itemize}
				\item Vertiport size~\cite{Tarafdar2019nortCalifornia, Rimjha2021LA,Guo2024vtolSite}
				\item Land availability~\cite{Tarafdar2019nortCalifornia}
				\item Topography~\cite{Tarafdar2019nortCalifornia}
			\end{itemize}
		}, text width=10em, fill=LightGray!50, rounded corners
		]
		[\parbox{14em}{\centering Neighborhood category / Land use~\cite{Tarafdar2019nortCalifornia, Jeong2021kMeansVertiport}},text width=10em, fill=LightGray!50, 
		[\parbox{20em}{
			\begin{itemize}
				\item Residential
				\item Commercial
				\item Recreational (parks, greenbelts)                            
			\end{itemize}
		}, text width=14em, fill=LightGray!50, rounded corners
		]
		]
		]
		[\parbox{12em}{\centering No Obstacles~\cite{Guo2024vtolSite}}, fill=LightGray!50
		]
		[\parbox{12em}{\centering \textcolor{red}{Access infrastructure}}, fill=LightGray!50
		[\parbox{12em}{
			\begin{itemize}
				\item \textcolor{red}{Power grid~\cite{Kohlman2018grid}}
				\item \textcolor{red}{Transport proximity}               
			\end{itemize}
		}, text width=10em,fill=LightGray!50, rounded corners
		]
		]
		[\parbox{12em}{\centering AAM Aircraft Capabilities~\cite{Rakas2021fleetSelection}}, fill=LightGray!50
		[\parbox{14em}{\centering Operational}, text width=10em,fill=LightGray!50, rounded corners
		[\parbox{20em}{
			\begin{itemize}
				\item Range
				\item Flight envelope                        
			\end{itemize}
		}, text width=14em, fill=LightGray!50, rounded corners
		]
		]
		[\parbox{14em}{\centering Capacity}, text width=10em,fill=LightGray!50, rounded corners
		]
		[\parbox{14em}{\centering Physical characteristics}, text width=10em,fill=LightGray!50, rounded corners
		[\parbox{20em}{
			\begin{itemize}
				\item Dimensions %(Wingspan, Height, Length)
				\item Power supply
				\item Aerodynamics                  
			\end{itemize}
		}, text width=14em, fill=LightGray!50, rounded corners
		]
		]
		]
		[\parbox{12em}{\centering \textcolor{red}{Climate}}, fill=LightGray!50
		[\parbox{12em}{
			\begin{itemize}
				\item \textcolor{red}{Wind}
				\item \textcolor{red}{Seasonal rain, fog}
				\item \textcolor{red}{Precipitation}
				%\item \textcolor{red}{Weather-dependent noise terrain}
			\end{itemize}
		}, text width=10em,fill=LightGray!50, rounded corners
		]
		]
		]            
		%%% Step 3 factors
		[\parbox{10em}{\centering Regulations}, text width=7em, fill=NavyBlue!30
		[\parbox{12em}{\centering \textcolor{red}{Safety~(Table~\ref{tab:regulations})}}, fill=NavyBlue!30                
		[\parbox{14em}{\centering Airspace}, text width=10em,fill=NavyBlue!30
		[\parbox{20em}{
			\begin{itemize}
				\item Load on ATC~\cite{Vascik2021interoperability}
				\item No-fly Zone (NFZ) \cite{Jeong2021kMeansVertiport,Rimjha2021LA, Rimjha22AirRestrict, Vascik2021interoperability}
				%\item \textcolor{red}{Geofence?}
				\item Separation from other aircraft~\cite{Vascik2021interoperability,Souza2023confManag}
			\end{itemize}
		}, text width=14em, fill=NavyBlue!30    
		]
		]
		[\parbox{14em}{\centering Aircraft~\cite{Rakas2021fleetSelection}},text width=10em, fill=NavyBlue!30
		]
		[\parbox{14em}{\centering \textcolor{red}{Vertiport}}, text width=10em,fill=NavyBlue!30
		[\parbox{20em}{
			\begin{itemize}
				\item \textcolor{red}{Safeguarding}
			\end{itemize}
		}, text width=14em, fill=NavyBlue!30    
		]
		]
		]
		[\parbox{12em}{\centering \textcolor{red}{Sustainability}}, fill=NavyBlue!30             
		[\parbox{14em}{\centering Noise~\cite{Jeong2021kMeansVertiport}},text width=10em, fill=NavyBlue!30
		]
		[\parbox{14em}{\centering \textcolor{red}{Urban development plan}},text width=10em, 
		fill=NavyBlue!30
		]
		[\parbox{14em}{\centering \textcolor{red}{Economic growth}},text width=10em, 
		fill=NavyBlue!30
		]
		[\parbox{14em}{\centering \textcolor{red}{Emission/Pollution}},text width=10em, 
		fill=NavyBlue!30
		]
		[\parbox{14em}{\centering \textcolor{red}{Ground transport norms}},text width=10em, 
		fill=NavyBlue!30
		]
		[\parbox{14em}{\centering \textcolor{red}{Land use}}, text width=10em,fill=NavyBlue!30
		[\parbox{20em}{
			\begin{itemize}
				\item \textcolor{red}{Wildlife}
			\end{itemize}
		}, text width=14em, fill=NavyBlue!30    
		]
		]
		]
		]
		%%% Step 4 factors
		[\parbox{10em}{\centering Economic Viability~\cite{Pertz2023economic}}, text width=7em,fill=LightGray!50
		[\parbox{12em}{\centering Capital Expense (CAPEX)}, fill=LightGray!50
		[\parbox{14em}{\centering Infrastructure reuse~\cite{Lim2019Seoul,Jeong2021kMeansVertiport,Guo2024vtolSite}},text width=10em, fill=LightGray!50
		]
		[\parbox{14em}{\centering New Infrastructure}, text width=10em,fill=LightGray!50
		[\parbox{20em}{
			\begin{itemize}
				\item Land cost~\cite{Tarafdar2019nortCalifornia}
				\item Vertiport capacity~\cite{Rimjha2021LA}
				\item Aircraft cost~\cite{Tarafdar2019nortCalifornia}
			\end{itemize}
		}, text width=14em, fill=LightGray!50, rounded corners]     
		]
		[\parbox{14em}{\centering \textcolor{red}{Interconnection to public transportation}}, text width=10em, fill=LightGray!50
		]
		[\parbox{14em}{\centering \textcolor{red}{Interconnection to grid}},text width=10em, fill=LightGray!50
		]
		]
		[\parbox{12em}{\centering Operational Expense (OPEX)}, fill=LightGray!50
		[\parbox{14em}{
			\begin{itemize}
				\item Operational cost per flight~\cite{Shin2022Skyport} 
				\item Vertiport utilization~\cite{Rimjha2021factorsVertiport}
				\item Aircraft utilization~\cite{Tarafdar2019nortCalifornia}
				\item Landing fees~\cite{Rimjha2021LA}
				\item \textcolor{red}{Ground staffing}
				\item \textcolor{red}{Energy charges}
			\end{itemize}
		}, text width=10em, fill=LightGray!50, rounded corners
		]
		]
		[\parbox{12em}{\centering Revenue}, fill=LightGray!50
		[\parbox{14em}{\centering
			Combined passenger fares~\cite{Shin2022Skyport,Chae2023vertLocations,Guo2024vtolSite} 
		}, text width=10em,fill=LightGray!50, rounded corners
		]
		]
		]        
		]
	\end{forest}
    \caption{Completed taxonomy of design factors for selecting a candidate vertiport location}
	\label{fig:new-vertiport-design-factors}
\end{figure}
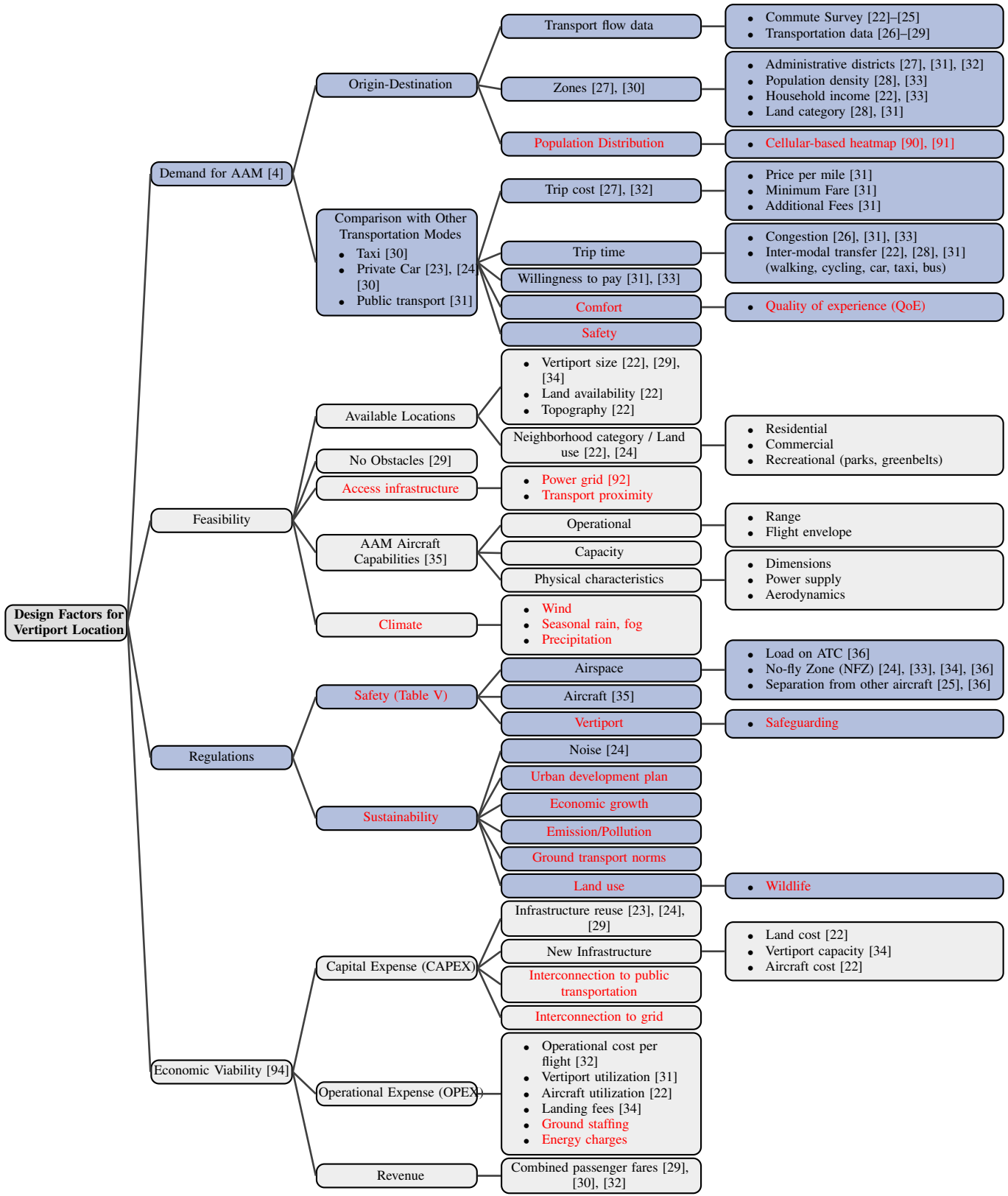

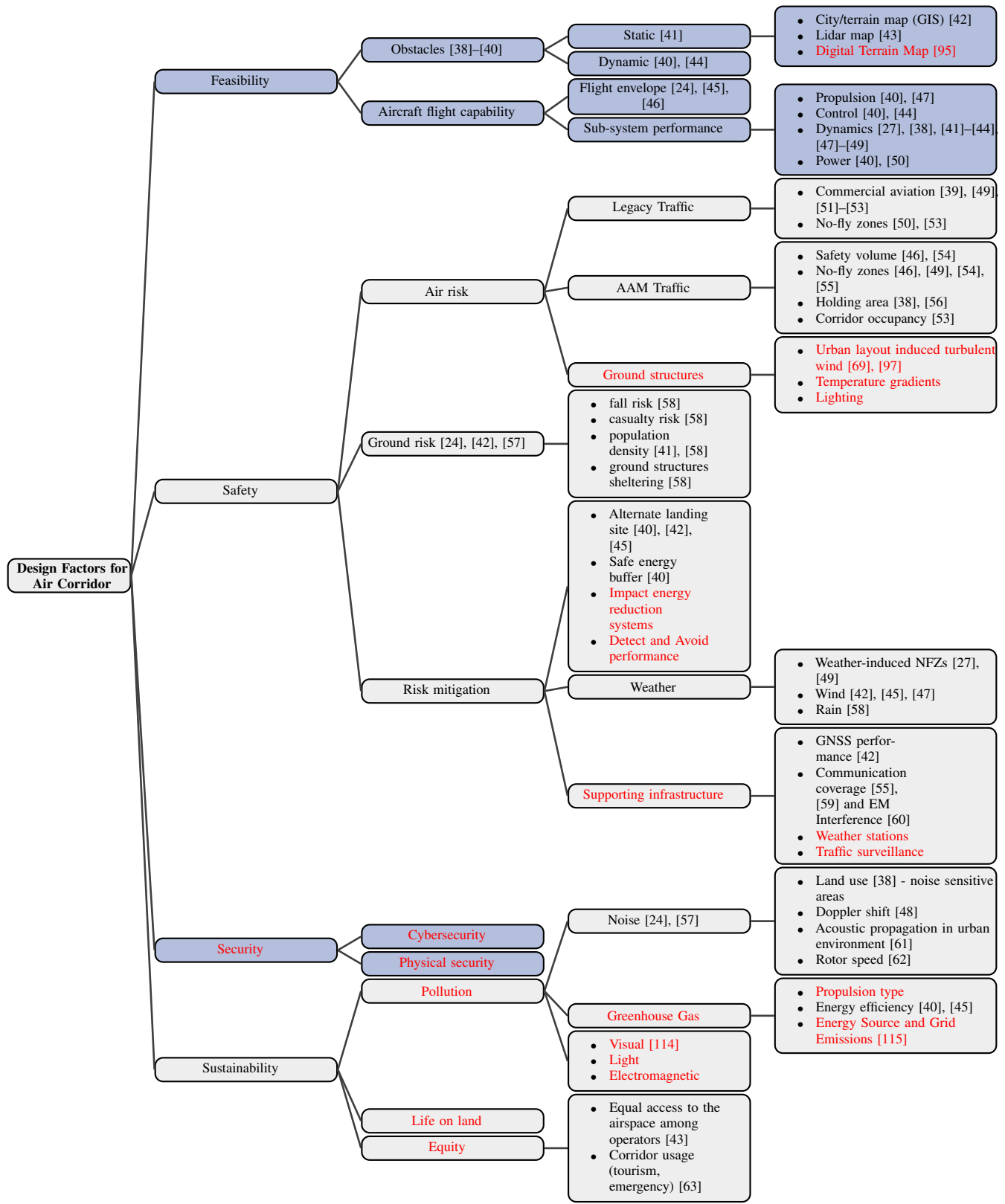
\begin{figure*}[tbph!]
	\centering
	\begin{forest}
		for tree={
			grow=east,
			reversed=true,
			anchor=base west,
			parent anchor=east,
			child anchor=west,
			base=left,
			font=\scriptsize, %
			rectangle,
			draw=black, %
			rounded corners,
			align=left,
			minimum width=2em, %
			edge+={darkgray, line width=1pt},
			s sep=1pt, %
			inner xsep=1pt, %
			inner ysep=2pt, %
			line width=0.8pt,
			text width=9em, %
		},
		[\textbf{\parbox{9em}{\centering Design Factors for\\Air Corridor}}, text width=6em, fill=LightGray!50
		%% faibility
		%% Factor Category 1
		[\parbox{12em}{\centering Feasibility}, fill=NavyBlue!30
		[\parbox{12em}{\centering Obstacles\cite{Lee2021holding,Gray2023airport,Hagag2024energy}}, fill=NavyBlue!30
		[\parbox{12em}{\centering Static\cite{Slama2022corrRoadmap}}, fill=NavyBlue!30
		[\parbox{16em}{
			\begin{itemize}
				\item City/terrain map (GIS)~\cite{Causa2023path}                            
				\item Lidar map~\cite{Tang2021autoHDflight}
				\item \textcolor{red}{Digital Terrain Map~\cite{DTM_nasa}}
			\end{itemize}
		}, text width=11em, fill=NavyBlue!30    
		]
		]
		[\parbox{12em}{\centering Dynamic~\cite{Ge2019hierarchical,Hagag2024energy}}, fill=NavyBlue!30
		]
		]
		[\parbox{12em}{\centering Aircraft flight capability}, fill=NavyBlue!30
		[\parbox{12em}{\centering Flight envelope \cite{Lou2021RTT,Jeong2021kMeansVertiport,Neto2022trajEvaluation}}, fill=NavyBlue!30
		]
		[\parbox{12em}{\centering Sub-system performance}, fill=NavyBlue!30
		[\parbox{16em}{
			\begin{itemize}
				\item Propulsion~\cite{Hagag2024energy,Lu2024vtolTraj}
				\item Control \cite{Ge2019hierarchical,Hagag2024energy}
				\item Dynamics \cite{Falck2018acoustic, Ge2019hierarchical,Lee2021holding,Tang2021autoHDflight,Ribeiro2022Geofence,Slama2022corrRoadmap,Peng2022hierarchical,Causa2023path,Lu2024vtolTraj}
				\item Power \cite{Hagag2024energy,Kotwicz2022restrictions}
			\end{itemize}
		}, text width=11em, fill=NavyBlue!30   
		]
		]
		]
		%[\parbox{12em}{\centering \textcolor{red}{Airspace Capacity}}, fill=NavyBlue!30]
		]
		%% Factor Category 2 - Nothing to add.   
		[\parbox{12em}{\centering Safety}, fill=LightGray!50
		[\parbox{12em}{\centering Air risk}, fill=LightGray!50
		[\parbox{12em}{\centering Legacy Traffic}, fill=LightGray!50
		[\parbox{16em}
		{
			\begin{itemize}
				\item Commercial aviation~\cite{Jiang2022metric,Ribeiro2022Geofence,Verma2022corridors,Kallies2023Frankfurt,Gray2023airport}
				\item No-fly zones~\cite{Kallies2023Frankfurt,Kotwicz2022restrictions}
			\end{itemize}
		}, text width=11em, fill=LightGray!50    
		]
		]
		[\parbox{12em}{\centering AAM Traffic}, fill=LightGray!50
		[\parbox{16em}
		{
			\begin{itemize}
				\item Safety volume~\cite{Thompson2023OpVolume,Neto2022trajEvaluation}
				\item No-fly zones~\cite{Garcia2022channel,Thompson2023OpVolume,Ribeiro2022Geofence,Neto2022trajEvaluation}
				\item Holding area~\cite{Lee2021holding,Song2021optSched}
				\item Corridor occupancy~\cite{Kallies2023Frankfurt}
			\end{itemize}
		}, text width=11em, fill=LightGray!50    
		]
		]
		[\parbox{12em}{\centering \textcolor{red}{Ground structures}}, fill=LightGray!50
		[\parbox{16em}
		{
			\begin{itemize}
				\item \textcolor{red}{Urban layout induced turbulent wind \cite{Schweiger2023wind,Giersch2022athmFlow}}
				\item \textcolor{red}{Temperature gradients}
				\item \textcolor{red}{Lighting}
				%\item \textcolor{red}{Additional Item} % Uncomment and add content if needed
			\end{itemize}
		}, text width=11em, fill=LightGray!50    
		]
		]
		]
		[\parbox{12em}{\centering Ground risk~\cite{Glaab2019noise,Jeong2021kMeansVertiport,Causa2023path}}, fill=LightGray!50    
		[\parbox{10em}
		{
			\begin{itemize}
				\item fall risk~\cite{Ye2024airRoute}
				\item casualty risk~\cite{Ye2024airRoute}
				\item population density~\cite{Slama2022corrRoadmap,Ye2024airRoute}
				\item ground structures sheltering~\cite{Ye2024airRoute}
			\end{itemize}
		}, text width=9em, fill=LightGray!50    
		]
		]
		[\parbox{12em}{\centering Risk mitigation}, fill=LightGray!50    
		[\parbox{10em}
		{
			\begin{itemize}
				\item Alternate landing site~\cite{Lou2021RTT,Causa2023path,Hagag2024energy}
				\item Safe energy buffer~\cite{Hagag2024energy}
				\item \textcolor{red}{Impact energy reduction systems}                   
				\item \textcolor{red}{Detect and Avoid performance}
			\end{itemize}
		}, text width=9em, fill=LightGray!50    
		]
		[\parbox{12em}{\centering Weather}, fill=LightGray!50
		[\parbox{16em}
		{
			\begin{itemize}
				\item Weather-induced NFZs~\cite{Peng2022hierarchical,Ribeiro2022Geofence}
				\item Wind~\cite{Lou2021RTT,Causa2023path,Lu2024vtolTraj}
				\item Rain~\cite{Ye2024airRoute}
			\end{itemize}
		}, text width=11em, fill=LightGray!50    
		]
		]
		[\parbox{12em}{\centering \textcolor{red}{Supporting infrastructure}}, fill=LightGray!50
		[\parbox{12em}
		{
			\begin{itemize}
				\item GNSS performance~\cite{Causa2023path}
				\item Communication coverage~\cite{Garcia2022channel,Park2023comm} and EM Interference~\cite{Nguyen2023radiation}
				\item \textcolor{red}{Weather stations}
				\item \textcolor{red}{Traffic surveillance}
			\end{itemize}
		}, text width=11em, fill=LightGray!50    
		]
		]
		]
		]
		%% Factor Category 3
		[\parbox{12em}{\centering \textcolor{red}{Security}}, fill=NavyBlue!30
		[\parbox{12em}{\centering \textcolor{red}{Cybersecurity}}, fill=NavyBlue!30
		%[\parbox{12em}{
		%	\begin{itemize}
		%		\item \textcolor{red}{aaaa?} 
		%		\item \textcolor{red}{bbbb?} 
		%	\end{itemize}
		%}, text width=9em, fill=NavyBlue!30    
		%]
		]
		[\parbox{12em}{\centering \textcolor{red}{Physical security}}, fill=NavyBlue!30
		%[\parbox{12em}{
		%	\begin{itemize}
		%		\item \textcolor{red}{Cyber threat?}
		%		\item \textcolor{red}{Physical threat?}
		%	\end{itemize}
		%}, text width=9em, fill=NavyBlue!30    
		%]
		]
		]
		%% Factor Category 4
		[\parbox{12em}{\centering Sustainability}, fill=LightGray!50
		[\parbox{12em}{\centering \textcolor{red}{Pollution}}, fill=LightGray!50
		[\parbox{12em}{\centering Noise~\cite{Glaab2019noise,Jeong2021kMeansVertiport}}, fill=LightGray!50
		[\parbox{16em}{
			\begin{itemize}
				\item Land use~\cite{Lee2021holding} - noise sensitive areas
				\item Doppler shift~\cite{Falck2018acoustic}
				\item Acoustic propagation in urban environment~\cite{Gao2023acoustic}
				\item Rotor speed~\cite{Cho2024lowNoise}
			\end{itemize}
		}, text width=11em, fill=LightGray!50    
		]
		]
		[\parbox{12em}{\centering \textcolor{red}{Greenhouse Gas}}, fill=LightGray!50
		[\parbox{16em}{
			\begin{itemize}
				\item \textcolor{red}{Propulsion type}
				\item Energy efficiency~\cite{Lou2021RTT,Hagag2024energy}
				\item \textcolor{red}{Energy Source and Grid Emissions~\cite{Vashi2024co2}}
			\end{itemize}
		}, text width=11em, fill=LightGray!50    
		]
		]
		%[\parbox{12em}{\centering Land use}, fill=NavyBlue!30
		%]
		[\parbox{12em}{
			\begin{itemize}
				\item \textcolor{red}{Visual~\cite{Kilian2023visPollution}}
				\item \textcolor{red}{Light}
				\item \textcolor{red}{Electromagnetic}
			\end{itemize}
		}, text width=9em, fill=LightGray!50   
		]
		]
		[\parbox{12em}{\centering \textcolor{red}{Life on land}}, fill=LightGray!50]
		[\parbox{12em}{\centering \textcolor{red}{Equity}}, fill=LightGray!50
		[\parbox{12em}
		{
			\begin{itemize}
				\item Equal access to the airspace among operators~\cite{Tang2021autoHDflight}
				\item Corridor usage (tourism, emergency)~\cite{Paradis2022visualizing}
			\end{itemize}
		}, text width=9em, fill=LightGray!50    
		]
		]
		]
		]
	\end{forest}
	\caption{\centering Completed taxonomy of design factors for air corridor formulation}
	\label{fig:new-corridor-design-factors}
\end{figure*}

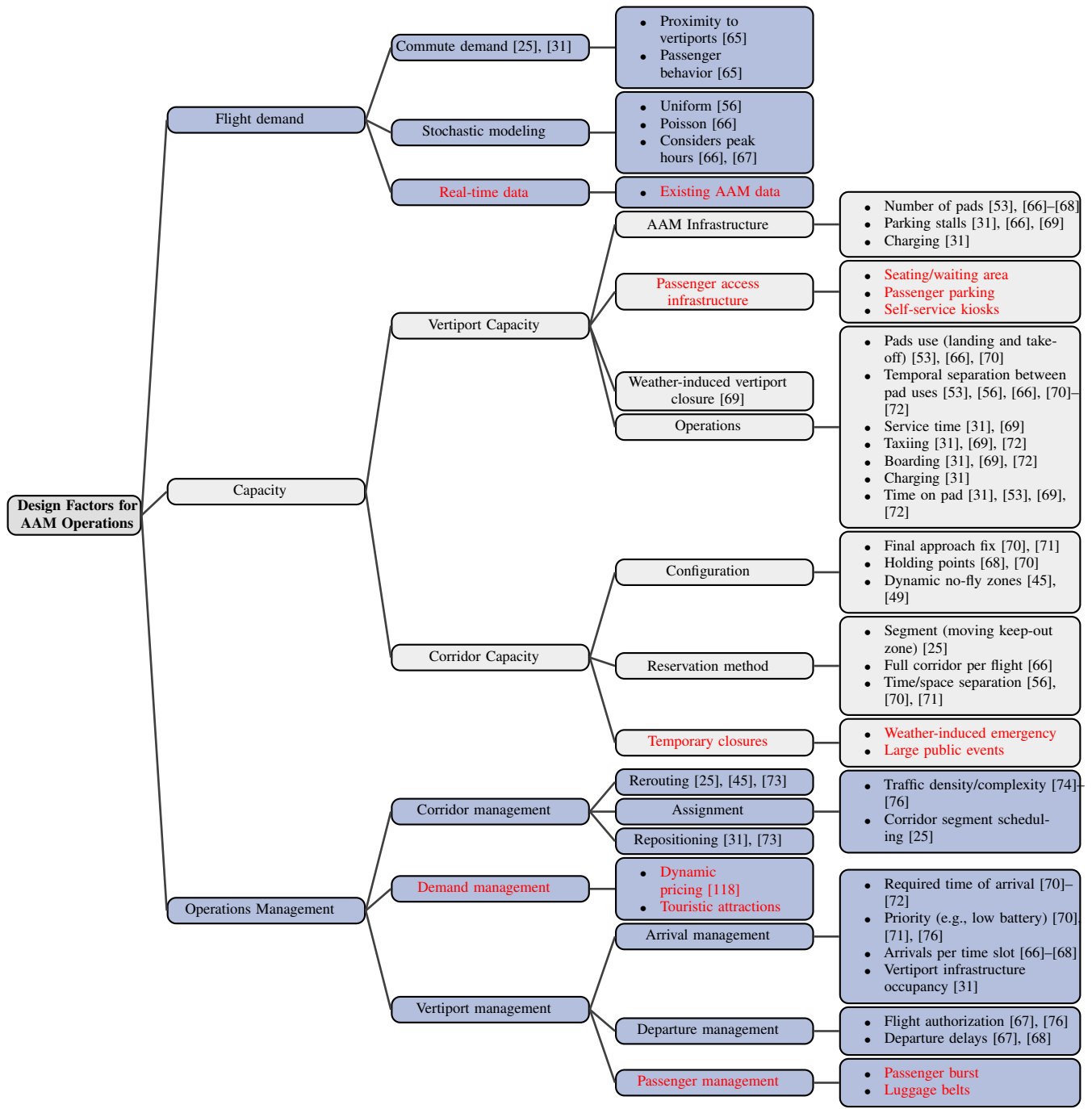
\begin{figure*}[tbph!]
	\centering
	\begin{forest}
		for tree={
			grow=east,
			reversed=true,
			anchor=base west,
			parent anchor=east,
			child anchor=west,
			base=left,
			font=\scriptsize, %
			rectangle,
			draw=black, %
			rounded corners,
			align=left,
			minimum width=2em, %
			edge+={darkgray, line width=1pt},
			s sep=1pt, %
			inner xsep=1pt, %
			inner ysep=2pt, %
			line width=0.8pt,
			text width=9em, %
		},
		[\textbf{\parbox{9em}{\centering Design Factors for AAM Operations}}, text width=6em, fill=lightgray!50
		%% Factor Category 1
		[\parbox{12em}{\centering Flight demand}, fill=NavyBlue!30
		[\parbox{12em}{\centering Commute demand~\cite{Rimjha2021factorsVertiport,Souza2023confManag}}, fill=NavyBlue!30
		[\parbox{12em}{
			\begin{itemize}
				\item Proximity to vertiports~\cite{Conrad2024vertManagement}
				\item Passenger behavior~\cite{Conrad2024vertManagement}
			\end{itemize}
		}, fill=NavyBlue!30                         
		]
		]
		[\parbox{12em}{\centering Stochastic modeling}, fill=NavyBlue!30
		[\parbox{12em}{
			\begin{itemize}
				\item Uniform~\cite{Song2021optSched}
				\item Poisson~\cite{Pooladsanj2023VertiSync}
				\item Considers peak hours~\cite{Lee2022DCB,Pooladsanj2023VertiSync}
			\end{itemize}
		}, fill=NavyBlue!30    
		]
		]
		[\parbox{12em}{\centering \textcolor{red}{Real-time data}}, fill=NavyBlue!30
		[\parbox{12em}{
			\begin{itemize}
				\item \textcolor{red}{Existing AAM data}
			\end{itemize}
		}, fill=NavyBlue!30                         
		]
		]
		]  
		%% Factor Category 2
		[\parbox{12em}{\centering Capacity}, fill=LightGray!50
		[\parbox{12em}{\centering Vertiport Capacity}, fill=LightGray!50
		[\parbox{12em}{\centering AAM Infrastructure}, fill=LightGray!50
		[\parbox{16em}{
			\begin{itemize}
				\item Number of pads~\cite{Lee2022DCB,Kallies2023Frankfurt,Pooladsanj2023VertiSync,Chen2024arrivalManag}
				\item Parking stalls~\cite{Rimjha2021factorsVertiport,Pooladsanj2023VertiSync,Schweiger2023wind}
				\item Charging~\cite{Rimjha2021factorsVertiport}
			\end{itemize}
		},text width=11em, fill=LightGray!50    
		]
		]
		[\parbox{12em}{\centering \textcolor{red}{Passenger access infrastructure}}, fill=LightGray!50
		[\parbox{16em}{
			\begin{itemize}
				\item \textcolor{red}{Seating/waiting area}
				\item \textcolor{red}{Passenger parking}
				\item \textcolor{red}{Self-service kiosks}
			\end{itemize}
		}, text width=11em,fill=LightGray!50    
		]
		]
		[\parbox{12em}{\centering Weather-induced vertiport closure~\cite{Schweiger2023wind}}, fill=LightGray!50
		]
		[\parbox{12em}{\centering Operations}, fill=LightGray!50
		[\parbox{16em}{
			\begin{itemize}
				\item Pads use (landing and take-off)~\cite{Kleinbekman2020rolling,Kallies2023Frankfurt,Pooladsanj2023VertiSync}
				\item Temporal separation between pad uses~\cite{Kleinbekman2018arrScheduling,Kleinbekman2020rolling,Song2021optSched,Kallies2023Frankfurt,Pooladsanj2023VertiSync,EspejoDiaz2023}
				\item Service time~\cite{Rimjha2021factorsVertiport,Schweiger2023wind}
				\item Taxiing~\cite{Rimjha2021factorsVertiport,Schweiger2023wind,EspejoDiaz2023}
				\item Boarding~\cite{Rimjha2021factorsVertiport,Schweiger2023wind,EspejoDiaz2023}
				\item Charging~\cite{Rimjha2021factorsVertiport}
				\item Time on pad~\cite{Rimjha2021factorsVertiport,Kallies2023Frankfurt,Schweiger2023wind,EspejoDiaz2023}
			\end{itemize}
		}, text width=11em,fill=LightGray!50    
		]
		]
		]
		[\parbox{12em}{\centering Corridor Capacity}, fill=LightGray!50
		[\parbox{12em}{\centering Configuration}, fill=LightGray!50
		[\parbox{16em}{
			\begin{itemize}
				\item Final approach fix~\cite{Kleinbekman2018arrScheduling,Kleinbekman2020rolling}
				\item Holding points~\cite{Kleinbekman2020rolling,Chen2024arrivalManag}
				\item Dynamic no-fly zones~\cite{Lou2021RTT,Ribeiro2022Geofence}
			\end{itemize}
		}, text width=11em,fill=LightGray!50    
		]
		]
		[\parbox{12em}{\centering Reservation method}, fill=LightGray!50
		[\parbox{16em}{
			\begin{itemize}
				\item Segment (moving keep-out zone)~\cite{Souza2023confManag}
				\item Full corridor per flight~\cite{Pooladsanj2023VertiSync}
				\item Time/space separation~\cite{Kleinbekman2018arrScheduling,Kleinbekman2020rolling,Song2021optSched}
			\end{itemize}
		}, text width=11em,fill=LightGray!50    
		]
		]
		[\parbox{12em}{\centering \textcolor{red}{Temporary closures}}, fill=LightGray!50
		[\parbox{16em}{
			\begin{itemize}
				\item \textcolor{red}{Weather-induced emergency}
				\item \textcolor{red}{Large public events}
			\end{itemize}
		}, text width=11em,fill=LightGray!50    
		]
		]
		]
		]              
		%% Factor Category 3
		[\parbox{12em}{\centering Operations Management }, fill=NavyBlue!30
		[\parbox{12em}{\centering Corridor management}, fill=NavyBlue!30
		[\parbox{12em}{\centering Rerouting~\cite{Lou2021RTT,Souza2023confManag,Suzuki2022flightReplan}}, fill=NavyBlue!30]
		[\parbox{12em}{\centering Assignment}, fill=NavyBlue!30
		[\parbox{16em}{
			\begin{itemize}
				\item Traffic density/complexity~\cite{Wang2021assignment,Conrad2023trafManag,Wang2023trafAssign}
				\item Corridor segment scheduling~\cite{Souza2023confManag}
			\end{itemize}
		}, text width=11em,fill=NavyBlue!30    
		]
		]
		[\parbox{12em}{\centering Repositioning~\cite{Rimjha2021factorsVertiport,Suzuki2022flightReplan}}, fill=NavyBlue!30]    
		]
		[\parbox{12em}{\centering \textcolor{red}{Demand management}}, fill=NavyBlue!30
		[\parbox{12em}{
			\begin{itemize}
				\item \textcolor{red}{Dynamic pricing~\cite{Kirste2024dynPricing}}
				\item \textcolor{red}{Touristic attractions}
			\end{itemize}
		}, fill=NavyBlue!30    
		]    
		]
		[\parbox{12em}{\centering Vertiport management}, fill=NavyBlue!30
		[\parbox{12em}{\centering Arrival management}, fill=NavyBlue!30
		[\parbox{16em}{
			\begin{itemize}
				\item Required time of arrival~\cite{Kleinbekman2018arrScheduling,Kleinbekman2020rolling,EspejoDiaz2023}
				\item Priority (e.g., low battery)~\cite{Kleinbekman2018arrScheduling,Kleinbekman2020rolling,Conrad2023trafManag}
				\item Arrivals per time slot~\cite{Lee2022DCB,Pooladsanj2023VertiSync,Chen2024arrivalManag}
				\item Vertiport infrastructure occupancy~\cite{Rimjha2021factorsVertiport}
				%\item Operating hours
			\end{itemize}
		}, text width=11em,fill=NavyBlue!30    
		]
		]
		[\parbox{12em}{\centering Departure management}, fill=NavyBlue!30
		[\parbox{16em}{
			\begin{itemize}
				\item Flight authorization~\cite{Lee2022DCB,Conrad2023trafManag}
				\item Departure delays~\cite{Lee2022DCB,Chen2024arrivalManag}
			\end{itemize}
		}, text width=11em,fill=NavyBlue!30   
		]
		]
		[\parbox{12em}{\centering \textcolor{red}{Passenger management}}, fill=NavyBlue!30
		[\parbox{16em}{
			\begin{itemize}
				\item \textcolor{red}{Passenger burst}
				\item \textcolor{red}{Luggage belts}
			\end{itemize}
		}, text width=11em,fill=NavyBlue!30    
		]
		]
		]
		]
		]        
		]
	\end{forest}
	\caption{\centering Completed taxonomy of design factors for AAM operations}
	\label{fig:new-corridor-ops-design-factors}
\end{figure*}

		The previous section outlined a structured taxonomy of AAM design factors and helped to identify critical gaps in current methodologies. To address these gaps, this section expands the analysis with additional insights and factors, as indicated in the forest charts in Figs.~\ref{fig:new-vertiport-design-factors}~-~\ref{fig:new-corridor-ops-design-factors}. 
		
		{To explicitly differentiate the maturity of these factors, we applied a visual coding scheme. Factors that are empirically supported by the current AAM literature are presented in standard black font. Conversely, factors that are currently hypothetical, projected, or represent critical gaps are emphasized in \textbf{red font}. These hypothetical factors are not entirely speculative; rather, they are derived from future research directions identified in the primary literature, expert assessment, and established practices in adjacent disciplines (e.g., small UAVs, intelligent transportation systems, and conventional aviation), which are cited in Figs.~\ref{fig:new-vertiport-design-factors}~-~\ref{fig:new-corridor-ops-design-factors} and throughout this section.}
		
		Moreover, since this systematic review is focused on corridor design, we excluded papers focusing on smaller-scale tasks (e.g., AAM demand estimation~\cite{Goyal2021demand}) from the text presented above. In this section, we enrich the taxonomy with these works.
		\subsection{Identified Gaps: RQ1 - Corridor Design Factors}
		\subsubsection{Phase 1 - Vertiport Network Design} Though this phase has received significant attention from the research community, there are still several factors and data sources that were not considered~\cite{Mendonca2022vertiport}. Let us briefly describe them for each task.
		
		\textbf{AAM Demand:}
		First of all, we would like to mention several works focusing exclusively on the demand analysis and market potential assessment~\cite{Long2023demand,Goyal2021demand}. Even though the presented approaches do not significantly differ from the ones presented in our paper, these works contain an in-depth analysis of the demand. 
		
		The skewed nature of \textit{Origin-Destination} data from surveys may not accurately capture real-world dynamics. This bias can be also observed in other datasets based on self-reported location information~\cite{Zhang2024data}. More reliable alternatives, such as ITS or transportation authority datasets, should be prioritized. When the transportation data is not available, one may reasonably assume that transportation demand reflects population dynamics. In this case, cellular network data presents a viable option for tracking real-time population density and estimating transportation demand~\cite{Deville2014cellular,Bergroth2022helsinki}. 
		
		\textit{Comparison with Other Transportation Modes} currently highlights AAM’s potential as a premium service, leveraging advantages in comfort, safety, and efficiency. With shorter travel times, fewer transfers, and strict aviation safety standards, AAM offers a superior passenger experience compared to taxis or private cars, enhancing public trust and adoption.
		However, unsatisfied transportation demand does not always require an AAM solution. In some cases, adding a new road or improving public transport may be more efficient and sustainable. AAM should be deployed strategically, focusing on areas where it provides the greatest value, such as underserved regions or congested urban centers. Aligning AAM demand assessments with broader urban mobility goals can foster equitable and sustainable growth. 
		
		\textbf{Feasibility:}
		Feasibility assessments for vertiports often overlook critical factors. \textit{Climatic} conditions, such as seasonal winds or rains, can make vertiports inoperable for months and must be accounted for to ensure operational resilience. \textit{Infrastructure} availability, including power grid capacity \cite{Kohlman2018grid} and ground transportation access, is equally essential for seamless operations.
		
		\textbf{Regulations:}
		First of all, the extended list of regulations requires refined categories. Hence, we distinguish two types of regulations by their main focus: \textit{Safety}- and \textit{Sustainability}-related regulatory documents. 
		
		The \textit{Safety}-related regulations are well-covered in the literature. We provide an overview of relevant regulatory documents in Table~\ref{tab:regulations}. However, a comprehensive vertiport Safeguarding strategy is still overlooked in academic literature. Safeguarding should address key operational risks, including ensuring that lighting installations are not obscured or confused with non-aeronautical lighting, mitigating wildlife strike risks such as bird strikes near waste disposal sites, and preventing construction-related interference like dust, smoke, or disruptions to navigational aids. Moreover, forward-looking safeguarding strategies consider urban development plans to ensure that no obstacles or other hazardous objects will appear in the future. For more details on safeguarding as well as required vertiport layouts, we refer to the vertiport regulations provided in Table~\ref{tab:regulations}. 
		
		In the reviewed literature, \textit{Sustainability} in vertiport location definition is often limited to noise-related issues. A more holistic approach is needed to ensure the harmonious development of AAM and the environment surrounding it. First, vertiport placement should align with urban development plans to account for future transportation needs. Additionally, insights from {ground public transportation norms} can improve route optimization and passenger flow management, enhancing operational efficiency. Second, {economic growth} potential must be considered, as well-placed vertiports can boost local business activity and connectivity, particularly in underserved areas. Environmental concerns, such as {emissions and pollution}, remain significant when aircraft are not fully electric. However, AAM has a non-zero greenhouse footprint even for fully-electric vehicles \cite{Stolaroff2018EnergyUA} due to the grid emissions. Finally, location selection must minimize {wildlife impact}, avoiding creating disturbances to sensitive ecosystems.
		
		\textbf{Economic Viability:}
		Similar to the task of AAM demand assessment, some works are focused exclusively on economic viability~\cite{Pertz2023economic} providing a greater level of details spanning from the factors we mentioned in Fig.~\ref{fig:vertiport-design-factors} to accounting for maintenance, overhaul, and depreciation. 
		
		\textbf{Takeaway:} By expanding the scope of demand assessments, feasibility analyses, regulatory considerations, and sustainability criteria, the proposed taxonomy offers a more comprehensive framework for vertiport location definition. Integrating reliable data sources, considering long-term climatic and urban development factors, incorporating lessons from existing transportation modes, and evaluating economic viability more thoroughly will help ensure that AAM infrastructure is both efficient and resilient, ultimately contributing to a balanced and sustainable urban mobility ecosystem.
		
		\subsubsection{Phase 2 - Corridor design}\label{sec:Discussion.RQ1.2}
		Fig.~\ref{fig:new-corridor-design-factors} presents a modified taxonomy of factors to be considered during the corridor design phase. 
		While the \textbf{Feasibility} branch remains unchanged due to its foundational nature\footnote{The only minor change is related to another source of terrain information that can be presented in shape of NASA-baked Digital Terrain Maps~\cite{DTM_nasa}.}, the key improvements focus on refining the \textbf{Safety} branch with Specific Operations Risk Assessment (SORA), incorporating a new \textbf{Security} aspect, and significantly expanding the \textbf{Sustainability} definition. Below, we elaborate on these enhancements and their implications.

		\textbf{Safety:}
		SORA, widely used for assessing complex UAS operations, systematically identifies risks as a combination of occurrence probability and severity. Safety in this context refers to a state where the risk is acceptable. While the current version (2.5)~\cite{jarus2024sora} does not include passenger missions, its core principles remain applicable to AAM. {Considering AAM-adapted SORA will become particularly important when we face the need of reevaluating the current airspace configurations to increase the \textit{airspace capacity} by identifying new airspace segments that has not yet been thoroughly investigated or utilized but are vital for overall operational efficiency of AAM.}
		
		Future AAM-specific SORA frameworks could integrate additional factors identified in this review. For instance, \textit{Air risks} can be enhanced by considering ground structure influences, such as wind patterns~\cite{Giersch2022athmFlow} and temperature gradients caused by urban environments, as well as ground lighting effects on pilots. {Similarly, \textit{ground risks} can be assessed by advanced Machine Learnig techniques prior or during the flight based on the data collected by the aircraft \cite{Andrade2026}.} \textit{Risk mitigation} measures should extend beyond on-board systems such as impact reduction (e.g., parachute) and DAA solutions: the corridor definition additionally should consider coverage and performance of supporting infrastructure providing weather data and traffic surveillance. Examples of the former include the Integrated Terminal Weather System (ITWS) and the Digital Automatic Terminal Traffic surveillance systems should include both cooperative sources (e.g., ADS-B, Mode S transponders, or Remote ID~\cite{Vinogradov2022rid,Tedeschi2023rid,Vinogradov2023rid}) and non-cooperative sources (e.g., primary radar systems). Additionally, as it is pointed out in Section~\ref{sec:results}.\ref{sec:RQ1}.\ref{sec:corridor_factors}, ground communication network performance should be considered. However, the performance does not have to be static as the network can be reconfigured to serve AAM traffic~\cite{Prabhath2023comm4corr_UAV,Maeng2023uptilt,Bernabe2024highway}. Alternatively, multiple independent networks can be used to boost reliability~\cite{Colpaert2022multi}. Moreover, future generations of cellular networks can be used for environmental sensing including cooperative/non-cooperative traffic surveillance as well as weather monitoring~\cite{Ostrometzky2024weather}.

		\textbf{Security:}
		While aircraft-related security issues are {addressed by traditional aviation cybersecurity frameworks (such as the technically identical standards jointly published by RTCA and EUROCAE, e.g., DO-326A/ED-202A, DO-356A/ED-203A, and DO-355/ED-204~\cite{EUROCAE_ED202A,EUROCAE_ED203A,EUROCAE_ED204}) and} explored in~\cite{Ferrao2022_SLR_security}, secure corridor design is an essential yet underexplored area. Threats include cyberattacks targeting navigation and communication systems~\cite{Tedeschi2023rid}, as well as physical risks to infrastructure. Mitigation strategies should include:
		\begin{itemize}
			\item \textit{Cybersecurity}: Encrypted communications, intrusion detection systems, and network hardening for traffic and weather data exchange.
			\item \textit{Physical Security}: Robust protections for vertiport and corridor infrastructure to prevent unauthorized access or sabotage.
		\end{itemize}
		Addressing these gaps will enhance the resilience and integrity of AAM operations, ensuring secure operations in complex urban environments. Consequently, the presence of secure communication and navigation infrastructure will become a corridor design factor. {Because legacy aircraft-centric standards are insufficient for multi-agent network architectures, compliance with emerging Unmanned Aircraft System Traffic Management (UTM) security standards (e.g., ASTM F3548-21~\cite{ASTM2021_f3548}), regional frameworks like EASA U-space~\cite{EASA_Uspace_2021}, and global directives like the ICAO Aviation Cybersecurity Strategy~\cite{ICAO_CyberSecurity_2019} is critical to protect AAM networks.} Additionally, AAM requires new security regulations and protocols aligned with piloted aviation standards, tailored to the unique challenges of AAM corridors.
		
		\textbf{Sustainability:}
		Sustainability in corridor design must extend beyond noise emissions to include other environmental and social impacts~\cite{EASA2024_EFA}. All types of \textit{Pollution} such as visual~\cite{Kilian2023visPollution} {(e.g., the daytime aesthetic impact of physical infrastructure and low-flying aircraft)}, light {(e.g., the nighttime impact of mandatory aviation lighting and beacons)}, electromagnetic, and greenhouse gas emissions~\cite{Vashi2024co2} must be minimized through optimized corridor design.
		
		\textit{Equity} is another crucial factor, going beyond "Fairness" in Section~\ref{sec:corridor_factors}. Communities negatively affected by AAM operations, such as those exposed to noise, should also receive tangible benefits from the service. {It was demonstrated that a wider set of social and societal factors can be embedded into corridor design \cite{Hohmann2024social}}. Moreover, wildlife protection must be a priority, ensuring that corridors avoid disrupting ecosystems and align with the UN’s \textit{Life on Land} goal.
		
		\textbf{Takeaway:} The Safety aspect of corridor design will benefit from incorporating AAM-specific risks, including urban environmental factors and enhanced infrastructure-enabled mitigation measures using data from weather services, traffic surveillance, and cellular networks. The Security aspect will address cyber and physical threats, emphasizing encryption, intrusion detection, and alignment with aviation security standards. The broadened {Sustainability} branch will integrate pollution management, equity considerations, and wildlife protection, ensuring AAM aligns with environmental and social goals. These improvements provide an actionable framework for safe, secure, and sustainable deployment of AAM corridors.
		
		\subsubsection{Phase 3 - AAM operations}\label{sec:Discussion.RQ1.3}
		Fig.~\ref{fig:new-corridor-ops-design-factors} presents an enhanced taxonomy for AAM operations, focusing on critical aspects of demand, capacity, and their dynamic balancing with management of operations. Below, we provide further insights into these components.
		
		\textbf{Flight Demand:}
		Accurately assessing and forecasting demand for AAM services remains a challenge due to the lack of real-world operational data. Current models serve as proxies, but once \textit{real AAM data} becomes available, it must be integrated to refine demand estimation and forecasting.
		
		\textbf{Capacity:}
		Vertiport capacity assessment can be significantly enhanced by accounting for passenger-centered infrastructure, such as car parking, lobby and waiting areas, and self-service kiosks. These elements are critical for efficient passenger flow and overall service quality.
		Additionally, corridor capacity is subject to temporary closures caused by hazards such as severe weather events or large gatherings, motivating the need for adaptive planning to maintain operational efficiency.
		
		\textbf{Operations Management:}
		Managing operations in AAM systems is a dynamic challenge that extends beyond traditional infrastructure management, {as robustness against operational uncertainties depends on a complex interplay between pre-departure strategic decisions and the effects of tactical manoeuvring \cite{Badea2025}.} Moreover, the list of factors should be expanded: alongside corridor and vertiport management actions proposed in the literature, we suggest adding \textit{Demand Management} strategies to the taxonomy. Dynamic pricing~\cite{Kirste2024dynPricing} is a promising approach, allowing for price adjustments based on demand levels. For instance, higher fares during peak hours can lower demand (causing overflow to other transportation modes), while discounts during off-peak periods can encourage usage and improve utilization rates.
		
		Another innovative strategy is enhancing the passenger {Quality of Experience}. Offering value-added options, such as scenic routes or premium services during low-demand periods, can attract passengers while optimizing operational efficiency. These strategies not only improve resource allocation but also contribute to passenger satisfaction and system profitability.
		
		\textbf{Takeaway:} The availability of real-world AAM data will mark a turning point for the industry, enabling more accurate models for dynamic demand and capacity estimation. These models will inform infrastructure planning and operational strategies. Real operations will also allow for the evaluation of proposed strategies, such as dynamic pricing and quality-enhancing measures, providing insights into their efficiency and scalability. By implementing these advancements, AAM systems can achieve the adaptability and resilience needed to meet the demands of modern urban mobility.
		\subsection{Identified Gaps: RQ2 - Separation Distance Definition}

		While the current taxonomy (Fig.~\ref{fig:safe-sep-factors}) provides a robust foundation for defining safe separation distances in AAM systems, several critical factors remain unexplored. Addressing these gaps is essential for enhancing safety, scalability, and adaptability in AAM operations. Below, we present these factors organized within existing and newly proposed groups.
		
		\textbf{Vehicular Characteristics:}
		The energy state of eVTOLs significantly affects their ability to maintain safe separation distances. Low battery levels may limit maneuverability, such as performing evasive actions or sustaining prolonged hover modes. Hence, separation protocols must account for energy state and battery dynamics related operational constraints to ensure safety during extended or emergency operations.
		
		\textbf{Total System Error:} 
		Current TSE subdivision on NSE and FTE does not address vulnerabilities introduced by cyberattacks. Threats like GPS spoofing or communication jamming can disrupt navigation accuracy and coordination. Incorporating the resilience to cyber threats and redundancy measures will be critical for maintaining separation in contested environments.
		
		In cooperative airspace systems, navigation or communication errors in one vehicle can propagate, impacting neighboring aircraft and the broader system. Separation protocols should account for such cascading effects of error propagation in cooperative systems, incorporating strategies like error isolation, real-time error correction, and multi-source validation.
		
		\textbf{Environmental Factors:}
		Current separation protocols do not consider the long-term effects of climate change, such as increased frequency of extreme winds or turbulence. Considering the climate change impacts through climate projections and dynamic weather adaptation into safe separation strategies will enhance the resilience of AAM operations to evolving environmental conditions.
		
		\textbf{System-Level and Technological Factors:}
		{\textit{Mixed-Mode Operations}: As highlighted in Section~\ref{sec:results}.\ref{sec:RQ2}, the transitional phase where piloted, automated, and fully autonomous aircraft share airspace is a major gap in the current literature. This phase is highly debated among regulators and practitioners, with consensus far from being reached. Separation definition becomes highly complex due to discrepancies in trajectory predictability, reaction times, and communication methods (voice vs. digital). Future corridor design during this transitional phase will likely require dynamic separation buffers that default to the most conservative standard of the least equipped aircraft, or physical segregation within the corridor (e.g., dedicated altitude strata for autonomous vs. piloted flight) to maintain safety without severely degrading throughput.}
		
		As AAM operations scale, ensuring safe separation in dense corridors requires multi-agent coordination mechanism i.e., seamless coordination among multiple vehicles. Collaborative flight planning, decentralized decision-making, and shared situational awareness are vital for scalable and efficient AAM.
		
		The variability in the performance of communication network, particularly coverage and latency, directly impacts real-time separation adjustments. Addressing these variabilities, especially in urban environments, will be essential to ensure reliable and adaptive operations.
		
		Separation protocols currently lack validation through Digital Twin (DT) simulations. These virtual replicas can simulate real-world scenarios, such as high-density traffic or extreme weather, providing a safe and cost-effective way to test and refine separation strategies.
		
		\textbf{Takeaway:} The inclusion of these new factors warrants expanding the current taxonomy to introduce a new group, System-Level and Technological Factors, alongside the existing categories of Vehicular Characteristics, Total System Error, and Environmental Factors. This updated taxonomy would provide a more comprehensive framework for defining safe separation distances, addressing both operational challenges and emerging technological opportunities.
		
		By integrating these overlooked factors, AAM systems can adopt safer, more adaptive separation protocols that reflect the complexities of urban airspace, evolving environmental conditions, and advanced technologies. 
		
		\subsection{{Identified Gaps: RQ3 - Separation Distance Methodology}} 
		{The methodologies described in the current taxonomy (Fig.\ref{fig:safe-sep-methodologies}) are exhaustively examined in literature and are recognized as established methodologies in the legacy aviation domain. As AI capabilities and data-driven decision-making continue to develop, we envision the future research on AAM separation distances will incorporate (1) AI/ML models to complement traditional stochastic and computational approaches, and (2) data-driven techniques at the forefront of innovation that will be accumulated from various AAM experimental campaigns.}
		
		{\textbf{Model-based:}
			Alongside computational and stochastic models, AI/ML models will persist in being extensively utilized to evaluate various separation requirements prior to their incorporation into regulation. The current application of such models is still in its infancy and requires further research efforts.}  
		
		{\textbf{Empirical:}
			The lack of operational AAM flights has resulted in insufficient data regarding flight behavior, separation distances, and associated hazards, thereby hindering subsequent data-driven research on AAM safety. Data-driven simulations, combined with the integration of AI/ML models, will be essential for developing safe separation methodologies in the future. This will involve comparing and contrasting various separation standards based on eVTOL capabilities, airspace structure, and air traffic density.}
		
		{\textbf{Aircraft reference model:}
			The reference model for eVTOL vehicles and the avionics equipment significantly affect the design of safe separation minima. Aside from the NASA UAM reference model~\cite{NASA-eVTOL-Model}, there are no publicly accessible eVTOL models for researchers. This impedes additional research and development by scholars, with such studies predominantly conducted in isolation by eVTOL manufacturers. Introducing the vehicle model to a broader research community will facilitate more comprehensive results and accelerate innovation in this domain.}
		
		{\textbf{Takeaway:}
			Derivation safe separation distance relies on data assets such as eVTOL vehicle reference models and AAM operation data. With these models as a guide, AI/ML frameworks and smart methodologies may be applied to AAM flights, making them safer and more efficient overall.}

		\subsection{{Design choice for real-world corridors}}\label{sec:insights}
		Despite the fact that we have provided a comprehensive taxonomy of factors for each phase of AAM corridor design, there is no "one-size-fits-all" formula to determine the optimal factors when the actual AAM corridors are placed in the national airspace. The best-fitting factors are selected on a case-by-case basis, contingent upon the regulatory frameworks and geographical region of a particular country. The "feasibility and safety" factors must be prioritized as a general approach, prior to any consideration of the "sustainability" or "operations management" factors. For instance, the most critical factors in selecting the optimal vertiport location are "demand" and "feasibility," with "national regulation" following in terms of importance. Economic factors are considered at the end. 
		
		{To illustrate the practical utility of these taxonomies, we present a step-by-step application for designing a hypothetical AAM corridor connecting a major international airport to a downtown financial district. Instead of an ad-hoc approach, a planner would systematically traverse the proposed taxonomies to transform high-level AAM goals into actionable design constraints:
			\begin{itemize}
				\item \textbf{Phase 1 - Vertiport Location (Fig.~\ref{fig:new-vertiport-design-factors}):} The process begins by evaluating \textit{Demand for AAM} (using \textit{Commute Survey} data and projected \textit{Trip time} savings) to justify the origin and destination. \textit{Feasibility} dictates the exact physical placement; planners must find sites with adequate \textit{Land availability} that integrate with existing \textit{Access infrastructure} (e.g., railway or subway networks) while ensuring \textit{No Obstacles} block approach paths. \textit{Regulations} (such as local \textit{Land use} constraints) are then applied to filter out non-compliant sites before \textit{Economic viability} is assessed.
				\item \textbf{Phase 2 - Air Corridor Formulation (Fig.~\ref{fig:new-corridor-design-factors}):} With the vertiports fixed, the air route is formulated by prioritizing \textit{Feasibility} and \textit{Safety}. Planners minimize \textit{Ground risk} (specifically \textit{population density}) by routing the corridor over rivers or the sea. To manage \textit{Air risk} at the airport node, planners must safely integrate with high-volume \textit{Legacy Traffic}, establishing strict separation standards to deconflict commercial jets from AAM approach sectors. \textit{Sustainability} factors, specifically \textit{Noise}, further constrain the route away from residential zones, while \textit{Weather} conditions (such as local \textit{Wind} patterns over the water or in urban canyons) dictate the physical width and altitude limits of the corridor.
				\item \textbf{Phase 3 - AAM Operations (Fig.~\ref{fig:new-corridor-ops-design-factors}):} Finally, operational rules are established to manage traffic safely within the newly formulated corridor. Planners utilize \textit{Operations Management} tools, selecting a specific \textit{Reservation method} to balance peak \textit{Flight demand} and ensure the structural \textit{Corridor Capacity} is never exceeded. Operational rules are defined to establish the flight \textit{Priority} (e.g., expediting emergency or low-battery flights) and smoothly integrate the corridor with the surrounding managed airspace.
			\end{itemize}
			This structured approach demonstrates how the taxonomies serve as a sequential decision-making framework for urban airspace integration.}

		\subsection{Future Trends}\label{sec:future_trends}
		Looking ahead, corridor design and separation protocols in AAM are expected to evolve toward more dynamic/adaptive, data-centric, and interconnected frameworks. Below, we summarize several key directions that are likely to shape the next generation of research and practice in this domain:

\subsubsection{Data Management and Standardization}  
As corridor operations become increasingly dense and dynamic, a unified data taxonomy covering flight states, weather updates, and communication health will be essential. Ongoing efforts by industry consortia and regulators seek to establish common data standards and best practices regarding data collection, storage, and ownership.

\subsubsection{Sensing and Communication technologies}
Ongoing research points toward dynamic corridor reconfiguration requiring real-time data on weather, aerial traffic, and local ground conditions. Some of these data is unavailable (e.g., weather for the AAM altitudes over urban environments) and we need efficient ways to collect this information. Additionally, we will need to merge information from cooperative sources (e.g., ADS-B, Remote~ID, Mode~S transponders) with non-cooperative sensing (e.g., radar, LiDAR, cellular-based detection). Exchanging and processing these data requires reliable, low-latency communication links (e.g., 5G/6G cellular, satellite) and robust sensor fusion algorithms that rapidly process data from multiple sources. Potentially, 5G/6G networks can also provide Integrated Communications, Navigations, and Survellance (ICNS) services~\cite{Ullah2025icns}.

\subsubsection{Digital Twins for Simulation, Testing, and Exploitation}\label{sec:Future.DT} 
High-fidelity digital twins can serve as continuously updated mirrors of real-world corridors, injecting live telemetry (e.g., weather, traffic density, aircraft states) into virtual replicas. By stress-testing new layouts, separation rules, and emergency scenarios in a risk-free setting, planners can refine operational strategies and identify vulnerabilities before implementing changes in actual airspace. {Moreover, infrastructure DTs (i.e., aircraft, vertiport DTs) will enable predictive maintenance and efficient resource utilization \cite{Kabir2025DT}}.

\subsubsection{Coordinated ATM--AAM Traffic Management}  
Growing AAM volumes in low-altitude airspace intensify the need to unify AAM-specific procedures with conventional Air Traffic Management (ATM). Standardized interfaces for exchanging flight plans, route clearances, and no-fly advisories will help controllers—or automated decision systems—treat AAM flights alongside regular air traffic. This integration becomes particularly urgent when corridor segments intersect with conventional approach paths or controlled airspace, underscoring the importance of a consistent, real-time information flow. Since human controllers will need to interact with highly automated AAM systems, the danger of overloading the controllers must be overcome with careful design of AAM Human-Machine Interfaces (HMI).

\subsubsection{Autonomous Flight Rules and Onboard Autonomy.}  
Though the AFR concept was suggested almost a decade ago, it is not yet mature. Future concepts envision autonomous or semi-autonomous eVTOLs that handle collision avoidance and minor route adjustments in flight, drawing on local sensor fusion and short-range inter-aircraft communication, although comprehensive fallback procedures and certification frameworks will be essential for practical deployment.
{\subsubsection{Large Language Models and Agent-Based Autonomy}\label{sec:Future.LLM}
As autonomous systems increasingly transition from rigid, rule-based algorithms to reasoning-driven, multimodal intelligence, the integration of Large Language Models (LLMs) and AI agents presents a transformative direction for AAM~\cite{Ferrag2026}. Future AAM corridor management could leverage LLM-powered agents to process complex, unstructured data streams (legacy voice ATC instructions, text-based NOTAMs, and dynamic weather reports) and fuse them with spatial sensor data to make real-time separation and routing decisions. Furthermore, generative AI agents can serve as an intelligent bridge in HMI, alleviating controller workload by summarizing complex traffic scenarios and explaining automated conflict-resolution decisions in natural language. However, adapting these foundational models for safety-critical AAM operations will require rigorous research into their security and reliability, specifically focusing on mitigating AI hallucinations, and defending against adversarial inputs (e.g., prompt injection). Perhaps the most challenging issue will be to establish verification methods suitable for aviation certification frameworks of AI solutions.}
\subsubsection{Integration into Broader Intelligent Transportation Systems (ITS).}  
As AAM becomes part of the urban mobility landscape, interoperability with established ground networks will be vital. Shared data platforms that unify AAM flight schedules with bus, rail, or micro-mobility services allow passengers to plan door-to-door trips seamlessly. Furthermore, integrating AAM routes into citywide congestion-monitoring systems can help alleviate roadway overload, advancing both reliability and sustainability in metropolitan transport ecosystems.

In addition to these corridor- and separation-focused directions, future AAM research will increasingly address broader challenges such as AAM regulatory frameworks,  alternative propulsion methods (e.g., hydrogen), new business models for on-demand AAM services, enhanced security protocols for preventing cyber intrusions, and novel approaches for infrastructure design. Efforts to streamline operational certifications, develop rigorous pilot or pilotless (autonomous) training modules, and foster public acceptance through transparent noise and safety assessments will further round out the AAM landscape. By tackling these cross-cutting issues, the AAM ecosystem stands to achieve higher levels of scalability, environmental responsibility, and societal benefit.

		\section{Conclusions}\label{conclusion}
		This work presents a systematic literature review of the AAM corridor design and the definition of safe separation. The review applies established PRISMA guidelines to reduce bias and ensure the reproducibility of results. The systematic review was structured around three research questions that integrate vertiport network design, corridor design, operational management, and the definition of safe separation. A total of 1949 unique papers were identified during the screening process, from which 175 papers were selected for full-text review. Utilizing inclusion/exclusion criteria, we identified 62 articles that addressed at least one of the research questions. A comprehensive taxonomy of corridor design factors is synthesized, and we have expanded it with additional factors that have been overlooked in existing studies. According to the results of our systematic review, we delineate the effective AAM corridor design approach (RQ1) into three separate phases. 
		
		Phase 1 aims to identify suitable vertiport locations that meet commuting demand, are feasible, comply with national regulations, and are economically viable. The vertiports must be strategically positioned in geographical areas characterized by high AAM service demand that can be estimated based on population density and income levels. The vertiport construction must be feasible considering land availability, absence of obstructions, availability of public access and power infrastructure, and aircraft capabilities. The chosen vertiport sites must adhere to national regulations concerning airspace, aircraft, vertiport standards, and noise limitations. Moreover, the enduring economic viability of the chosen sites relies on evaluating both operational and capital expenditures. 

		In phase 2, AAM corridors are established to connect vertiport locations, guaranteeing that the flight path is feasible, safe, and environmentally sustainable. The feasibility component emphasizes the strategic avoidance of flight paths with static and dynamic obstacles, as well as evaluating the aircraft's ability to maneuver through designated flight corridors. The safety of the AAM corridor depends on air and ground risks (e.g., probability of colliding with another aircraft or damaging people/infrastructure). Furthermore, these risks can be reduced by contingency measures accounting for unplanned or emergency situations caused by aircraft sub-system malfunctioning or meteorological conditions. The sustainability aspect aims to link AAM corridors with long-term viability, societal and environmental impact concerning many aspects including noise footprint, energy efficiency, and equitable access to airspace.
		
		Phase 3 ensures the effective and efficient operation of AAM services across vertiports and corridors established in previous phases. This phase begins with the assessment and projection of flight demand for AAM services. It is followed by an analysis of the capacity constraints of vertiports and corridors. Prior analysis of demand and capacity is then used by dynamic operation management ensuring that vertiports are adequately prepared for efficient and timely arrival and departure management. Additionally, this approach allows the AAM corridors to effectively manage varying air traffic density, flow management, and time-critical rerouting. 
		
		In addition to corridor design, our work provides a comprehensive examination of the definition of safe separation (RQ2), which is essential for preventing collisions and optimizing airspace utilization. The first implementation of AAM will depend on current VFR and IFR, in conjunction with ATC, utilizing a rule-based separation approach; nevertheless, a novel framework of performance-based Digital and Autonomous Flight Rules (DFR/AFR) must be developed to progress towards a complex AAM ecosystem during the next decade. The aircraft performance, particularly regarding DAA capabilities and flight dynamics, are primary factors influencing the safe distance between aircraft. Given that the onboard avionics may experience performance uncertainties and inaccuracies, it is essential to implement robust error management strategies to estimate the total system error related to the aircraft. The total system error consists of the navigation system error resulting from GNSS/INS sensors and the flight technical error stemming from communication, control subsystems of the aircraft.  Furthermore, external environmental conditions such as wind, wake vortices are critical factors that must be considered to uphold the highest safety standards.
		Numerous model-based separation approaches utilize stochastic models to assess collision risks, macroscopic air traffic flow models, gas-law-based conflict prediction models, computational fluid dynamics models, and geometric models to derive safe separation distance of AAM aircraft (RQ3). Many existing works have also illustrated the simulation-based methodology utilizing real or synthetic flight data to get insights regarding the safe separation distances. 
		
		We believe that this work on AAM corridor design and safe separation distance definition will serve as a valuable resource for researchers and practitioners in this field.
		
		%\section*{Conflict of Interest Statement}
		%The authors declare that the research was conducted in the absence of any commercial or financial relationships that could be construed as a potential conflict of interest.
		
		\section*{Author Contributions}
		
		Conceptualization - E.V,  D.M., M.A., J.A., A.A, J.S., E.N.; methodology - E.V; data collection - E.V, D.M.; paper screening - E.V, D.M.; quality assessment  - E.V, D.M.; data extraction - E.V, D.M.;  writing of the manuscript  - E.V, D.M.; review and editing - M.A., J.A., A.A, J.S., E.N..

		\ifCLASSOPTIONcaptionsoff
		\newpage
		\fi

		\bibliographystyle{IEEEtran}
		\bibliography{bibliography}
		\newpage \noindent
		
		\appendices

		\onecolumn
		\section{Overview of the included literature}\label{Appendix:overview}
		
		\begin{footnotesize}
	\begin{longtblr}[
		caption = {{Comprehensive summary and quality assessment of selected papers}},
		label = {tab:slr-comprehensive-summary},
		remark{Note} = {CASP breakdown: (1) Objectives, (2) Rigor, (3) Data Reporting, (4) Contribution, (5) Limitations. Score 1 = Criterion Met; 0 = Not Met/Unclear.}
		]{
			colspec = {|m{0.7cm}|m{3cm}|m{3.2cm}|m{4cm}|m{2.5cm}|m{1.8cm}|},
			rowhead = 1,
			hlines,
			row{even} = {gray9},
			row{1} = {NavyBlue!30},
		}
		Ref\# & Author(s) \& Year & Study Type & Geography: Affiliation / Case Study& RQs & CASP Score \\
		\cite{Bulusu2021demand} & Bulusu et al. (2021) & Data-driven analysis & USA / USA & RQ1.1 & 5 (1/1/1/1/1) \\
		\cite{Causa2023path} & Causa et al. (2023) & Simulation \& Modeling & Italy / Italy & RQ1.2 & 5 (1/1/1/1/1) \\
		\cite{Chae2023vertLocations} & Chae et al. (2023) & Optimization& Korea / Korea & RQ1.1, RQ1.3 & 5 (1/1/1/1/1) \\
		\cite{Chen2024arrivalManag} & Chen et al. (2024) & Simulation \& Modeling & USA / USA & RQ1.3 & 4 (1/1/1/1/0) \\
		\cite{Cho2024lowNoise} & Cho et al. (2024) & Simulation \& Modeling & Korea, USA / - & RQ1.2, RQ1.3 & 5 (1/1/1/1/1) \\
		\cite{Conrad2023trafManag} & Conrad et al. (2023) &Optimization & UK / - & RQ1.3 & 5 (1/1/1/1/1) \\
		\cite{Conrad2024vertManagement} & Conrad et al. (2024) & Simulation \& Modeling & UK / UK & RQ1.3 & 5 (1/1/1/1/1) \\
		\cite{Cotton2019separation} & Cotton et al. (2019) & Analytical \& Conceptual & USA / - & RQ2, RQ3 & 4 (1/1/0/1/1) \\
		\cite{Cummings2024airspace} & Cummings et al. (2024) &  Analytical \& Conceptual  & USA / - & RQ2, RQ3 & 5 (1/1/1/1/1) \\
		\cite{EspejoDiaz2023} & Espejo-Diaz et al. (2023) &Optimization& France, Colombia / - & RQ1.3 &4 (1/1/1/1/0) \\
		\cite{Falck2018acoustic} & Falck et al. (2018) &Optimization& USA / - & RQ1.2 & 5 (1/1/1/1/1) \\
		\cite{Gao2023acoustic} & Gao et al. (2023) & Optimization & USA / - & RQ1.2, RQ1.3, RQ3 & 5 (1/1/1/1/1) \\
		\cite{Garcia2022channel} & Garcia et al. (2022) & Optimization & Spain, Germany / - & RQ1.2 & 5 (1/1/1/1/1) \\
		\cite{Ge2019hierarchical} & Ge et al. (2019) &Optimization & USA, China / - & RQ1.2 & 4 (1/1/1/1/0) \\
		\cite{Geister2018density} & Geister et al. (2018) &  Analytical \& Conceptual  & Germany / - & RQ1, RQ2, RQ3 & 5 (1/1/1/1/1) \\
		\cite{Glaab2019noise} & Glaab et al. (2019) & Data-driven analysis & USA / USA  & RQ1.2, RQ1.3 & 5 (1/1/1/1/1) \\
		\cite{Gomaa2023windWake} & Gomaa et al. (2023) &  Analytical \& Conceptual  & UAE, Australia / - & RQ3 & 5 (1/1/1/1/0) \\
		\cite{Gordo2023layered} & Gordo et al. (2023) & Simulation \& Modeling & Spain, UK, Netherlands / - & RQ3 & 5 (1/1/1/1/1) \\
		\cite{Gray2023airport} & Gray et al. (2023) &  Data-driven analysis & Korea / Korea & RQ1.2, RQ1.3 & 5 (1/1/1/1/1) \\
		\cite{Gray2023obstacles} & Gray et al. (2023) &  Analytical \& Conceptual  & Korea / Korea & RQ3 & 5 (1/1/1/1/1) \\
		\cite{Guo2024vtolSite} & Guo et al. (2024) & Optimization & China, France / China & RQ1.1, RQ1.3 & 5 (1/1/1/1/1) \\
		\cite{Hagag2024energy} & Hagag et al. (2024) &  Analytical \& Conceptual  \& Data-driven analysis & Germany / Germany & RQ1.2, RQ1.3, RQ2 & 5 (1/1/1/1/1) \\
		\cite{Jeong2021kMeansVertiport} & Jeong et al. (2021) &  Data-driven analysis & Korea / Korea & RQ1.1, RQ1.2 & 5 (1/1/1/1/1) \\
		\cite{Jiang2022metric} & Jiang et al. (2022) &  Analytical \& Conceptual  & USA / USA & RQ1.2, RQ1.3 & 4 (1/1/1/1/0) \\
		\cite{Kallies2023Frankfurt} & Kallies et al. (2023) & Simulation \& Modeling & Germany / Germany & RQ1.2, RQ1.3 & 4 (1/1/1/1/0) \\
		\cite{Kim2022Geofence} & Kim et al. (2022) & Simulation \& Modeling & USA / - & RQ2, RQ3 &  4 (1/1/1/1/0) \\
		\cite{Kim2023separation} & Kim et al. (2023) &  Analytical \& Conceptual  & Korea / - & RQ2, RQ3 & 4 (1/1/1/1/0) \\
		\cite{Kleinbekman2018arrScheduling} & Kleinbekman et al. (2018) & Optimization & Netherlands, USA / - & RQ1.3 & 5 (1/1/1/1/1) \\
		\cite{Kleinbekman2020rolling} & Kleinbekman et al. (2020) & Optimization & Netherlands, USA / - & RQ1.3 & 4 (1/1/1/1/0) \\
		\cite{Kotwicz2022restrictions} & Kotwicz et al. (2022) & Optimization \& Data-driven analysis & USA / USA & RQ1.1, RQ1.2 & 5 (1/1/1/1/1) \\
		\cite{Lee2021holding} & Lee et al. (2021) &  Analytical \& Conceptual  & Korea / Korea & RQ1.2 & 5 (1/1/1/1/1) \\
		\cite{Lee2022DCB} & Lee et al. (2022) & Optimization & USA / USA  & RQ1.3 & 5 (1/1/1/1/1)  \\
		\cite{Lim2019Seoul} & Lim et al. (2019) & Data-driven analysis  & Korea / Korea & RQ1.1 & 5 (1/1/1/1/1) \\
		\cite{Lou2021RTT} & Lou et al. (2021) & Optimization & UK / - & RQ1.2, RQ1.3 & 5 (1/1/1/1/1) \\
		\cite{Lu2024vtolTraj} & Lu et al. (2024) & Optimization & Germany, China / - & RQ1.2 & 5 (1/1/1/1/1) \\
		\cite{Neto2022trajEvaluation} & Neto et al. (2022) & Simulation \& Modeling & Brazil / - & RQ1.2 & 4 (1/1/1/1/0) \\
		\cite{Nguyen2023radiation} & Nguyen (2023) & Data-driven analysis  & USA / - & RQ1.2 & 5 (1/1/1/1/1) \\
		\cite{Paradis2022visualizing} & Paradis et al. (2022) & Data-driven analysis  & USA / USA & RQ1.2 & 5 (1/1/1/1/1) \\
		\cite{Park2023comm} & Park et al. (2023) & Optimization & Korea / - & RQ1.2 & 4 (1/1/1/1/0) \\
		\cite{Peng2022hierarchical} & Peng et al. (2022) & Optimization & USA / USA& RQ1.1, RQ1.2 & 5 (1/1/1/1/1) \\
		\cite{Pooladsanj2023VertiSync} & Pooladsanj et al. (2023) & Optimization & USA / USA  & RQ1.3, RQ2, RQ3 & 4 (1/1/1/1/0) \\
		%\cite{Prabhath2023comm4corr_UAV} & Prabhath et al. (2023) & Simulation & USA / - & RQ1.2 & 4 (1/1/1/1/0) \\
		\cite{Qu2024demand} & Qu et al. (2024) & Data-driven analysis  & China / China & RQ1.1 &  4 (1/1/1/1/0) \\
		\cite{Rakas2021fleetSelection} & Rakas et al. (2021) & Data-driven analysis  & USA / - & RQ1.1, RQ1.2 & 5 (1/1/1/1/1) \\
		\cite{Ribeiro2022Geofence} & Ribeiro et al. (2022) & Simulation \& Modeling & Brazil / Brazil & RQ1.2, RQ1.3, RQ3 & 5 (1/1/1/1/1) \\
		\cite{Rimjha2021factorsVertiport} & Rimjha et al. (2021) & Data-driven analysis & USA / USA & RQ1.1, RQ1.3 & 5 (1/1/1/1/1) \\
		\cite{Rimjha2021LA} & Rimjha et al. (2021) & Data-driven analysis & USA / USA & RQ1.1 & 5 (1/1/1/1/1) \\
		\cite{Rimjha22AirRestrict} & Rimjha et al. (2022) & Data-driven analysis & USA / USA  & RQ1.1 & 5 (1/1/1/1/1) \\
		\cite{Schweiger2023wind} & Schweiger et al. (2023) & Data-driven analysis & Germany / Germany & RQ1.3 & 5 (1/1/1/1/1) \\
		\cite{Shin2022Skyport} & Shin et al. (2022) & Optimization & Korea, Canada / Korea & RQ1.1, RQ1.2 & 4 (1/1/1/1/0) \\
		\cite{Slama2022corrRoadmap} & Slama et al. (2022) & Optimization & Czech Republic / Czech Republic & RQ1.2 & 4 (1/1/1/1/0) \\
		\cite{Song2021optSched} & Song et al. (2021) & Optimization & Korea / - & RQ1.2, RQ1.3, RQ3 & 4 (1/1/1/1/0) \\
		\cite{Souza2023confManag} & Souza et al. (2023) & Data-driven analysis & Brazil / Brazil & RQ1.1, RQ1.3 & 5 (1/1/1/1/1) \\
		\cite{Suzuki2022flightReplan} & Suzuki et al. (2022) & Optimization & USA / - & RQ1.3 & 5 (1/1/1/1/1) \\
		\cite{Tang2021autoHDflight} & Tang et al. (2021) & Optimization & USA / USA  & RQ1.2, RQ2 & 4 (1/1/1/1/0) \\
		\cite{Tang2022confPlane} & Tang et al. (2022) & Data-driven analysis & USA / USA  & RQ3 & 4 (1/1/1/1/0) \\
		\cite{Tarafdar2019nortCalifornia} & Tarafdar et al. (2019) & Data-driven analysis & USA / USA& RQ1.1 & 4 (1/1/1/1/0) \\
		\cite{Thompson2023OpVolume} & Thompson et al. (2023) & Optimization & USA, UK / USA  & RQ1.2 & 4 (1/1/1/1/0) \\
		\cite{Vascik2021interoperability} & Vascik et al. (2021) &  Analytical \& Conceptual  & USA / USA & RQ1.1, RQ1.3, RQ3 & 4 (1/1/1/1/0)  \\
		\cite{Verma2022corridors} & Verma et al. (2022) &  Analytical \& Conceptual  & USA / USA & RQ1.2, RQ2 & 4 (1/1/1/1/0) \\
		\cite{Wang2021assignment} & Wang et al. (2021) & Optimization & France, Singapore / - & RQ1.3 & 5 (1/1/1/1/1) \\
		\cite{Wang2023trafAssign} & Wang et al. (2023) & Optimization & France, Singapore / Singapore & RQ1.3 & 5 (1/1/1/1/1) \\
		\cite{Ye2024airRoute} & Ye et al. (2024) & Data-driven analysis & China / China & RQ1.2 & 5 (1/1/1/1/1) \\
	\end{longtblr}
\end{footnotesize}	
		\newpage
		\onecolumn

\section{Enhanced Taxonomies of Corridor Design Factors}\label{Appendix:taxonimies}

		\newpage
		\onecolumn

		\newpage

		\newpage
\begin{footnotesize}
	\renewcommand*{\arraystretch}{0.1}
	\begin{longtblr}[
		caption = {AAM safety-related regulations},
		label = {tab:regulations},
		]{
			colspec = {|m{0.7cm}|m{10cm}|m{2.2cm}|m{1.5cm}|m{0.4cm}|m{0.7cm}|},
			colsep = 2mm,
			rowsep = 0.2px,
			rowhead = 1,
			hlines,
			row{even} = {gray9},
			row{1} = {NavyBlue!30},
		} 
		Org & Title & Code/Note & Country & Year & Ref\#\\
		\SetCell[c=6]{c}{Airspace}\\
		FAA & 14 CFR, Chapter I "Federal Aviation Administration, Department of Transportation" Subchapter E "Airspace" &  & USA & 2024 & \cite{FAA2024_14CFR_I_SubE} \\
		
		FAA & 14 CFR, Chapter I "Federal Aviation Administration, Department of Transportation" Subchapter F "Air Traffic and General Operating Rules" &  & USA & 2024 & \cite{FAA2024_14CFR_I_SubF} \\
		\hline
		\SetCell[c=6]{c}{Aircraft}\\
		FAA & 14 CFR, Chapter I "Federal Aviation Administration, Department of Transportation" Subchapter C "Aircraft" &  & USA & 2024 & \cite{FAA2024_14CFR_I_SubC} \\
		
		EASA & Certification of Aircraft and Related Products, Parts and Appliances, and of Design and Production Organizations (Part 21) & Part 21 & European Union & 2023 & \cite{EASA2023_Part21} \\
		
		EASA & Second Issue of the Special Condition for VTOL-capable Aircraft & SC-VTOL-02 & European Union & 2024 & \cite{EASA2024_SpecialConditionVTOL} \\
		
		ASTM & Standard Specification for Electrical Systems for Aircraft with Electric or Hybrid-Electric Propulsion & F3316/ F3316M-19 & International & 2019 & \cite{ASTM2019_F3316M19} \\
		
		ASTM & Standard Specification for Aircraft Electric Propulsion Systems & F3239-22A & International & 2022 & \cite{ASTM2022_F3239_22A} \\

		ASTM & Standard Practice for Methods to Safely Bound Behavior of Aircraft Systems Containing Complex Functions Using Run-Time Assurance & F3269-21 & International & 2021 & \cite{ASTM2021_F3269} \\
		
		SAE &  Guidelines for Development of Civil Aircraft and Systems & ARP4754B & International & 2023 & \cite{SAE2023_ARP4754B} \\

		ANAC & PROPOSAL FOR SPECIAL CLASS AIRWORTHINESS CRITERIA for the aircraft model EVE-100 from EVE Soluções de Mobilidade Aérea Urbana LTDA & for the EVE-100 eVTOL & Brazil & 2023 & \cite{Anac2023_EVE100} \\
		
		FAA & Airworthiness Criteria: Special Class Airworthiness Criteria for the Archer Aviation, Inc. Model M001 Powered-Lift & for Archer M001 eVTOL & USA & 2024 & \cite{FAA2024_ArcherAviation} \\
		
		FAA & Airworthiness Criteria: Special Class Airworthiness Criteria for the Joby Aero, Inc. Model JAS4-1 Powered-Lift & for the Joby JAS4-1 eVTOL & USA & 2024 & \cite{FAA2024_JobyAero} \\
		\hline
		\SetCell[c=6]{c}{Vertiport}\\
		
		ASTM & Standard Specification for Vertiport Design & F3423/ F3423M-22 & International & 2022 & \cite{ASTM2022f3423m22} \\
		
		ASTM & New Specification for Vertiport Automation Supplemental Data Service Provider (SDSP) Performance & WK85153 & International & 2023 & \cite{ASTM2023wk85153} \\
		
		EASA & Vertiports Prototype Technical Specifications for the Design of VFR Vertiports for Operation with Manned VTOL-Capable Aircraft Certified in the Enhanced Category & PTS-VPT-DSN & European Union & 2022 & \cite{EASA2022PST} \\
		
		ICAO & Heliport Manual, 5th ed. & Doc 9261 & International & 2021 & \cite{ICAO2021_9261} \\
		
		ISO & ISO 5491:2023 Vertiports — Infrastructure and equipment for vertical take-off and landing (VTOL) of electrically powered cargo UAS & ISO 5491:2023 & International & 2023 & \cite{ISO2023_5491} \\
		
		CASA & Guidelines for Vertiport Design—Draft Advisory Circular 139.V-01 v1.0 & AC 139.V-01 v1.0 & Australia & 2023 & \cite{CASA2023_13901} \\
		
		CCAA & Technical Requirements for the Construction of Electric Vertical Take-Off and Landing Aircraft (eVTOL) Landing Fields & CCAATB 000—2024 & China & 2024 & \cite{CCAA2024} \\
		
		MLIT & Vertiport Design Guidelines & & Japan & 2023 & \cite{MLIT2023_VPDesignGuidelines} \\
		
		GACA & AC-140-01 Vertiports Design Specification (VDS) & AC-140-01 & Saudi Arabia & 2024 & \cite{GACA2024_AC14001} \\
		
		GCAA & Heliports (Onshore/Offshore), Vertiports (Onshore) & CAR-HVD & UAE & 2023 & \cite{GCAA2023} \\
		
		CAA & Considerations for Aerodromes and Vertiports Planning to Operate Vertical Take-off and Landing Aircraft (VTOL) & CAP2538 & United Kingdom & 2023 & \cite{CAA2023_2538} \\
		
		FAA & Specification for Runway, Taxiway, Heliport, and Vertiport Light Fixtures & AC 150/5345-46F & USA & 2024 & \cite{FAA2024_ac150-5345-46} \\
		
		FAA & Heliport Design & AC 150/5390-2D & USA & 2023 & \cite{FAA2023_ac150-2D} \\
		
		FAA & Vertiport Design, Supplemental Guidance to Advisory Circular & EB 150A & USA & 2024 & \cite{FAA2024_EB150A} \\
		
		NFPA & Standard for Heliports and Vertiports & NFPA-418 & USA & 2024 & \cite{NFPA2024} \\
	\end{longtblr}
	
	\textbf{Org Acronyms:} ICAO - International Civil Aviation Organization, FAA - Federal Aviation Administration, GCAA - General Civil Aviation Authority, CAA - Civil Aviation Authority, EASA - European Union Aviation Safety Agency, ISO - International Organization for Standardization, ANAC - Agência Nacional de Aviação Civil, ASTM - ASTM International, SAE - SAE International, CASA - Civil Aviation Safety Authority, CCAA - China Civil Airports Association, MLIT - Ministry of Land, Infrastructure, Transport and Tourism, NFPA - National Fire Protection Association.
\end{footnotesize}

	\end{document}